\documentclass[structabstract]{aa}  
\usepackage{graphicx,tabularx,rotating,pdflscape,multirow,lscape,multicol,makecell,xtab,rotating}
\usepackage{dcolumn}
\usepackage{chngcntr}
\usepackage{txfonts}
\usepackage{morefloats}
\usepackage{caption}
\usepackage{placeins}
\usepackage{morefloats}

\def\kms{\ifmmode{\rm km\,s}^{-1}\, \else km\,s$^{-1}$\,\fi}

\def\mujybm{$\mathrm{\muup}$Jy\,beam$^{-1}$}
\def\mjybm{mJy\,beam$^{-1}$}
\def\ltsim{\ifmmode\stackrel{<}{_{\sim}}\else$\stackrel{<}{_{\sim}}$\fi}
\def\gtsim{\ifmmode\stackrel{>}{_{\sim}}\else$\stackrel{>}{_{\sim}}$\fi}
\def\S4195{41.95+575}
\def\S4331{43.31+592}

\def\solmasyr{$\rm{M_\odot yr^{-1}}$}

\def\mdot{$\rm{\dot{M}}\,\,$}

\begin{document}
\title{An ALMA 3mm continuum  census of Westerlund 1}
\author{D.~M.~Fenech\inst{1}
\and J.~S.~Clark\inst{2}
\and R.~K.~Prinja\inst{1}
\and S.~Dougherty\inst{3}
\and F.~Najarro\inst{5}
\and I.~Negueruela\inst{4}
\and A.~Richards\inst{6}
\and B.~W.~Ritchie\inst{2}
\and H.~Andrews\inst{1}}
\institute{
$^1$Dept. of Physics {\&} Astronomy, University College London, 
Gower Street, London WC1E 6BT \\
$^2$School of Physical Science, The Open 
University, Walton Hall, Milton Keynes, MK7 6AA, United Kingdom\\
$^3$Dominion Radio Astrophysical Observatory, National Research Council 
Canada, PO Box 248, Penticton, B.C.  V2A 6J9 \\
$^4$Departamento de Astrof\'{\i}sica, Centro de Astrobiolog\'{\i}a, 
(CSIC-INTA), Ctra. Torrej\'on a Ajalvir, km 4,  28850 Torrej\'on de Ardoz, 
Madrid, Spain
$^5$Departamento de F\'{i}sica, Ingenar\'{i}a de Sistemas y Teor\'{i}a de 
la Se\~{n}al, Universidad de Alicante, Apdo. 99,
E03080 Alicante, Spain\\
$^6$JBCA, Alan Turing Building, University of Manchester, M13 9PL and 
MERLIN/VLBI National Facility, JBO, SK11 9DL, U.K.
}

 \abstract{Massive stars play an important role in both cluster and galactic evolution and the rate at which they lose mass is a key 
 driver of both their own evolution and their interaction with the environment up to and including their terminal SNe explosions. 
 Young  massive clusters provide an ideal opportunity to study a co-eval population of massive stars, 
where both their individual properties and the interaction with their environment can be studied in detail.}{We aim to study the 
constituent stars of the Galactic cluster Westerlund 1 in order to  determine mass-loss rates for the diverse post-main 
sequence population of massive stars.}
{To accomplish this we made 3mm continuum observations with  the Atacama Large Millimetre/submillimetre Array.}
{We detected emission from 50 stars in Westerlund 1, comprising all 21
Wolf-Rayets within the field of view, plus eight cool and 21 OB super-/hypergiants. 
Emission nebulae were associated with a number of the cool hypergiants while, unexpectedly, a  number of hot stars  also appear spatially resolved.}
{We were able to measure the  mass-loss rates for a unique population of massive post-main sequence stars at every stage of evolution, 
confirming a significant increase as stars transitioned from OB supergiant to WR states  via LBV and/or cool hypergiant phases. Fortuitously, the range of spectral types exhibited by the OB supergiants provides a critical test of radiatively-driven  wind theory and in particular the reality of the bi-stability jump. The  extreme mass-loss rate inferred for the interacting binary Wd1-9 in comparison to other cluster members confirmed the key role 
binarity plays in massive  stellar evolution. The presence of compact nebulae around a number of OB and WR stars is
unexpected; by analogy to the cool super-/hypergiants we attribute this to confinement and sculpting of the stellar wind via 
interaction with the  intra-cluster medium/wind. Given the morphology of core collapse SNe depend on the nature  of  the 
pre-explosion circumstellar environment, if this hypothesis is correct then the properties of the explosion depend not just on the 
progenitor, but also the  environment in which it is located.}


\maketitle


\section{Introduction}\label{sect:intro}

Despite their rarity, massive ($>20\,M_{\odot}$) stars are major agents of galactic evolution via the deposition of chemically enriched material, mechanical energy and ionising radiation, while dominating integrated galactic spectra in the UV and IR windows (the latter via re-radiation by hot dust). Despite this central role in galactic astrophysics, their short lives are poorly understood in comparison to stars such as the Sun. Unlike lower-mass stars, heavy mass-loss has long been recognised as a critical factor, along with rotation and the presence (or otherwise) of magnetic fields, in governing their evolutionary pathway \citep{ekstrom} and the nature of their demise, core-collapse or pair-production  supernova (SN), Gamma-ray burst or prompt, quiet collapse to black hole (BH). As might be anticipated from this, the nature of the stellar corpse, either neutron star or BH, is intimately related to the pre-demise mass-loss history, which is of particular interest given the detection of gravitational waves from coalescing black holes \citep{abbott}.

For single stars it has long been thought that line-driven radiative winds serve as the mechanism by which massive stars lose mass. Unfortunately, recent studies have reported observationally derived mass-loss rates of OB stars which are discordant by up to a factor of 10 \citep[e.g. ][]{puls,massa,fullerton,sundqvist}, due to uncertainties related to the degree of structure ('clumping') present in O  and B star winds \citep[e.g. ][]{prinja10,prinja13,sundqvist,surlan}. 
In order to emphasise the far-ranging consequences of this ambiguity, we highlight that it is no longer clear that radiatively-driven winds are able to drive sufficient mass-loss to transition from H-rich main sequence (MS) to H-depleted post-MS Wolf Rayet (WR) star. A popular supposition is that the consequent mass-loss deficit is made up by the short-lived transitional phase between these evolutionary extremes. Observations of ejection nebulae associated with both  hot (luminous blue variable; LBV) and cool (yellow hypergiant and red supergiant; YHG and  RSG) transitional stars imply phases of instability in which extreme, impulsive mass-loss may occur (\.{M}\,$\geq10^{-4}\,M_{\odot}$yr$^{-1}$). However it is not clear which stars encounter such instabilities, nor whether the duration of this phase is sufficient to strip away the H-rich mantle to permit the formation of WRs. Observations of apparently `quiescent' LBVs show that they support dense winds with mass-loss rates comparable to WRs \citep{clark14}  outside of eruption. However direct mass-loss rate determinations for very luminous cool transitional stars are few and far between \citep{dejager}; a critical weakness of current stellar evolutionary codes \citep[e.g. ][]{ekstrom}.

 An alternative evolutionary channel has been suggested by the recent recognition that $\sim70$\% of massive stars are found within 
 binary systems \citep{demink,sana12,sana13}. Interaction between both components may lead to extreme mass-loss from the primary 
\citep[e.g. ][]{petrovic}, allowing for the formation of WRs and favouring the production of NSs over BHs \citep[e.g. ][]{ritchie10},
while the mass gaining undergoes rejuvenation. In extreme cases binary merger my also lead to the production of a massive blue
straggler.

In both mass transfer and merger scenarios for binary evolution, substantial modification of the stellar initial mass function may be anticipated, while binary merger may provide a viable formation route for very massive stars \citep[$>100M_{\odot}$; ][]{schneider14,schneider15}. Moreover, even outside of evolutionary phases dominated by thermal- and nuclear-timescale mass-transfer, massive binaries provide valuable insights of the properties of stellar winds via observations of the resultant wind collision zones. Indeed both the high energy (thermal) X-ray and low energy (non-thermal synchrotron) mm/radio emission that results from shocks within colliding wind binaries (CWBs) provide additional binary identifiers.

Extending this paradigm, when found within stellar aggregates, the collective action of stellar winds and supernovae (SNe) yield powerful cluster winds \citep[e.g. ][]{stevens} that can either disperse or compress their natal giant molecular clouds, respectively inhibiting or initiating subsequent generations of star-formation. Moreover, SNe and their interaction with cluster winds (and the shocks within CWBs) have also been implicated in the production of Galactic cosmic rays and attendant very high energy $\gamma$-ray emission \citep[e.g.][]{abdalla,abra,bykov}.

Given the above considerations, young massive stellar clusters (YMCs) form ideal laboratories for the study and resolution of these issues due to their co-eval stellar populations of a single metallicity. Consequently YMCs have recently received increased attention as (near-IR) Galactic surveys have yielded numerous further examples. Discovered over half a century ago, Westerlund 1 \citep[Wd1; ][]{westerlund} subsequently escaped detailed investigation due to significant interstellar extinction. The serendipitous detection of multiple radio sources associated with  cluster members \citep[][henceforth Do10]{clark98,dougherty} sparked renewed observational efforts which revealed Wd1 to host a  uniquely rich and diverse population  of massive stars \citep{clark02,clark05}. Specifically, Wd1 appears to be co-eval and at an age \citep[$\sim5$Myr;][]{iggy10,kud} where cool supergiants and hypergiants may co-exist with WRs; the uniquely rich population of both \citep{clark05, crowther06}, as well as $>100$ OB supergiants \citep{iggy10,clark17} makes Wd1 a powerful laboratory for the study of massive stellar evolution. 

As such Wd1 has received attention across the electromagnetic spectrum from radio \cite[Do10]{kothes}, IR and optical \cite[e.g.][ and refs. above]{brandner} through to X-ray \citep[][henceforth Cl08]{muno06a,muno06b,clark08} and higher energies \citep[GeV and TeV;][respectively]{ohm, abra}. Multi-epoch spectroscopic radial velocity (RV) surveys have identified a rich population of binaries \citep[][]{ritchie09a,ritchie17}, with tailored modelling of individual systems clearly revealing the influence of binarity on massive stellar evolution  \citep[e.g.][]{ritchie10,clark11,clark14}.

In order to better understand the nature of the massive single and binary stellar populations of Wd1 we undertook Atacama Large Millimetre Array (ALMA) Band 3 (100\,GHz) continuum observations of Wd1 in 2015.The millimetre waveband  is a uniquely powerful diagnostic of mass-loss  and, in conjunction with  observations at other wavelengths to constrain the continuum spectral  energy distribution (SED), may determine the radial run of wind clumping via thermal Bremsstrahlung emission \citep[e.g.][]{blomme}. Moreover, CWBs may be identified by millimetre-radio observations due to either the presence of a non-thermal continuum component from synchrotron emission originating in the wind collision zone, or excess mm emission if such shocks are optically thick \citep{pittard10,pittard11}.

 \cite{fenech} presented results for the brightest continuum source within Wd1, the supergiant B[e] star and interacting binary Wd1-9 \citep{clark13}. In this paper we present a census of the remaining 3-mm sources and discuss their association with stellar counterparts where appropriate. To enable ease of comparison to the radio study of Wd1 by Do10, we choose to mirror the structure of that work in this paper, with departures where necessary due to novel results or synergies between radio and mm datasets. Finally, significant work has been undertaken to establish a distance to Wd1, with most estimates falling between 4-5\,kpc \citep{clark05,crowther06,kothes,brandner,iggy10,kud,clark17}, for the purpose of this paper we adopt a distance, $d=5$\,kpc to Wd1.

\section{Data acquisition, reduction and analysis}\label{sect:data}

\subsection{Observations and imaging}\label{sect:observing}

ALMA was used to observe Wd1 on the 30th June and the 1st July 2015 (Project code: 2013.1.00897.S) covering the central approximately 3.5 sq. arcmin area with 27 pointings. The observations were made at a central frequency of $\sim$100\,GHz with a total  usable bandwidth of 7.5\,GHz over four spectral windows centred on 92.5, 97.5, 102.5 \& 104.5\,GHz respectively. Each spectral window contains 128 channels with a channel frequency width of 15.625\,MHz (only 120 channels are usable). The array consisted of 42 antennas with baselines ranging from 40 to 1500\,m and a total on-source integration time per pointing of 242.3\,secs ($\sim$4\,mins). The data were calibrated using the standard ALMA pipeline procedures in Common Astronomy Software Applications ({\sc {CASA}}; pipeline version 4.3.1: r34044) and included the application of apriori calibration information as well as flagging of erroneous data. Observations of J1617-5848 were used to perform the phase and bandpass calibration and observations of Titan and Pallas were used to amplitude calibrate the data with assumed flux densities of 228.96 and 82.01\,mJy respectively (at 91.495\,GHz). 

As will be discussed in detail in Sect. \ref{sect:mm}-\ref{sect:stellar3} a number of the detected stars in these observations are resolved. In order to ensure that this is not the result of potential phase errors in the data associated with the pipeline water vapour correction performed, the data were re-calibrated applying slightly less flagging than in the original pipeline calibration and also making use of the {\sc{REMCLOUD}}s package \citep{maud17}. This is used to correct for the effect of any water in the form of fog or clouds present in certain weather conditions. As each section of data (30th June and 1st July) were calibrated independently, we found the addition of {\sc{REMCLOUD}} marginally improved the phase calibration for the 30th June but made no discernible improvement for data from the 1st July. We therefore proceeded using re-calibrated 30th June data with {\sc{REMCLOUD}} and re-calibrated 1st July data without {\sc{REMCLOUD}}.

Following initial calibration (both with initial and re-calibrated data), several iterations of phase self-calibration were applied to the data. At each stage images of the full mosaic field were produced. A selection of the brightest compact sources across the field were cleaned to produce the desired model. A selection of pointings, mainly centred around Wd1-9 were used with the model by {\sc {GAINCAL}} to calculate the calibration solutions. {\sc APPLYCAL} was then used to apply the corrections to all pointings. A final round of amplitude and phase calibration was also performed incorporating information from both Wd1-9 and Wd1-26 in the model.

The data were initially imaged using the mosaicing and multi-frequency synthesis functions in the {\sc CASA CLEAN} software. Cleaning was done only on the bright compact component of Wd1-9 (which has a flux density of 153\,mJy) in order to subtract this component from the data and limit high dynamic range problems in subsequent imaging. The final images were produced using the mosaicing and multi-scale capabilities within {\sc CLEAN} including data from all pointings to produce a single wide-field map of Wd1 (see Figs. \ref{fig:forsimage} and \ref{fig:finderimage}). This utilised Cotton-Schwab cleaning and natural weighting with scale sizes set at 1, 14, 12, 37 \& 119. The final image has a fitted beam of 750$\times$570\,mas. 
The final full-field image was primary beam corrected following cleaning to account for the change in sensitivity across the primary beam (PB). Example images of individual sources will be shown in Sect. 4, 5 and 6 and have been taken directly from the PB-corrected wide-field image. A complete catalogue of images of each source is also included in the on-line material. For subsequent data analysis, the images were transferred to the {\sc AIPS} (Astronomical Image Processing Software) environment.

To aid cross-correlation and identification, the absolute astrometry of the final images were checked utilising the 8.6\,GHz radio image and the FORS R-band image (which had previously been aligned to the radio image) as presented in Do10. Following the procedure from Do10, we determined any offset between the ALMA and FORS image by first using the {\sc AIPS} task {\sc HGEOM} to make the geometry of the two images consistent and then performing a Gaussian fit (using the {\sc AIPS} task {\sc JMFIT}) to the compact component of Wd1-9 in both images. A comparison of the fitted peak positions revealed a small offset between the images. As we performed self-calibration on the ALMA data, which causes absolute positional information to be lost, we chose to shift the ALMA image in order to bring the peak positions of Wd1-9 (as determined by the Gaussian fit) into alignment. The aligned FORS and ALMA images are presented in Fig. \ref{fig:forsimage}, the ALMA image showing identified stellar sources is shown in Fig. \ref{fig:finderimage} and the ALMA and radio 8.6\,GHz image is shown in Fig. \ref{fig:radioimage}. Both the original image and the primary-beam corrected image were shifted in this way and used for performing all subsequent analysis.

\subsection{Identification of 3\,mm stars}\label{sect:idents}

 In order to identify and catalogue the sources present in Wd1, the {\sc {SEAC}} source extraction software was used \citep[see][for further details]{peck14,morford}. This utilises a floodfill algorithm to search the map and identify discrete `islands' of emission by locating pixels above a given (seed) threshold and subsequently adding neighbouring pixels to the `island' down to a lower (flood) threshold. Thresholds of 5\,$\sigma$ and 3\,$\sigma$ were set for the initial use of SEAC on the ALMA wide-field image, where $\sigma$ is a regionalised noise level calculated by determining the rms level in individual cells of a 6$\times$6 grid across the whole image. This was necessary to account for the change in noise level across the image especially towards the strong emission from Wd1-26 and Wd1-9. SEAC was used on both the original and primary-beam corrected images. Sources determined by SEAC with a seed threshold of 5\,$\sigma$ in the original image (i.e. not primary beam corrected) are considered as detections. 

This resulted in the detection of 98 sources in the Wd1 field. 
These detections were cross-correlated with previously published catalogues \citep[including][;Do10 and Cl08 as well as the VPHAS point sources catalogue for the region]{clark05,crowther06,iggy10} in order to identify the observed sources. Following this cross-correlation process, the sources identified at 3\,mm fall broadly into three categories; those that are previously known and identified at multiple wavelengths including optical and radio, those that are detected in the radio observations from Do10 though have no other counterparts and those that are detected only in these ALMA observations. Table \ref{tab:srcecounts} lists a summary of the number of sources detected at 3\,mm within these categories along with those for relevant stellar spectral types where possible.

\begin{table}
\begin{center}
\caption{Summary of source categories and type detected in the ALMA 3-mm observations.}
\setlength{\tabcolsep}{0.5em} 
\renewcommand{\arraystretch}{1.5}
\begin{tabular}{l|c}
\hline
Category & Source number \\
\hline
ALMA only & 51 \\
ALMA+optical only & 30 \\
ALMA+optical+radio & 20 \\
\hline
Total & 101 \\
\hline
Wolf Rayets & 21 \\
YHGs & 4 \\
RSGs & 4 \\
BHGs & 3 \\
LBV  & 1 \\
sgB[e]& 1 \\
OB supergiants & 16 \\
\hline
Total & 50 \\
\end{tabular}
\tablefoot{N.B. seven of the ALMA-only sources are associated with the extended nebulae of Wd1-4 and Wd1-20, see Sect. \ref{sect:char} for details.}
\label{tab:srcecounts}
\end{center}
\end{table}

\begin{figure*}
\begin{center}
\includegraphics[width=11cm]{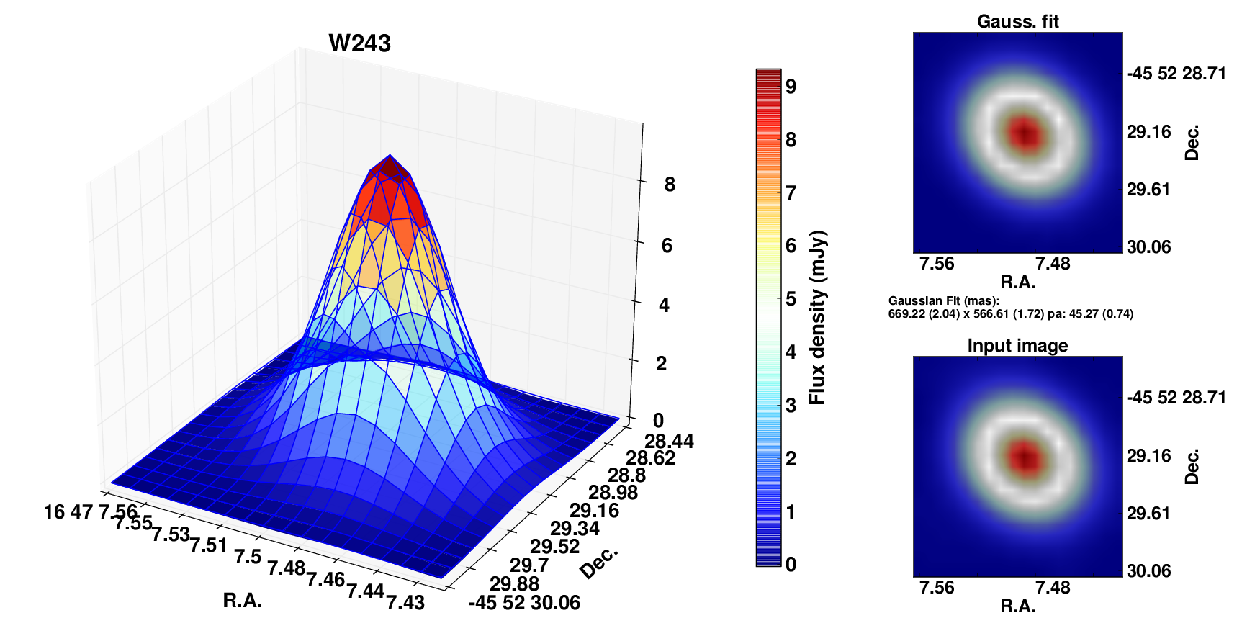}
\caption{An example output of source W243 from the Gaussian fitting process to the sources in the ALMA observations. Left panel in 3D shows the image information in colour-scale with the resulting fit overlaid as a wire-frame. Right shows the image data in the bottom panel and the Gaussian fit in the top panel.}
\label{fig:w243gauss}
\end{center}
\end{figure*}

There are also a number of optically identified sources that do not appear in these ALMA images. In particular there are several sources that have the same or similar spectral classifications to those that have been detected. In order to attempt to locate any weak emission from these sources, further SEAC runs were performed with seed thresholds of 4.5 and 4\,$\sigma$. This identified a further three sources albeit with lower detection thresholds, which have also been included in the final source list. 
The total 3\,mm catalogue list therefore contains 101 sources. Each source has been given an ALMA 3\,mm catalogue number, designated as FCP18, in order of right ascension. All information for these sources is contained in Tables \ref{tab:srcelist} and \ref{tab:srcesizes} for the stellar sources and Table \ref{tab:fcp18_list} for the ALMA-only sources.

\section{The millimetre emission in Wd1}\label{sect:mm}

The millimetre emission observed in Wd1 shows both distinct isolated sources as well as more extended regions of diffuse emission which broadly coincide with similar regions seen in the radio images (see Fig. \ref{fig:radioimage}). 


\subsection{3-mm star characteristics}\label{sect:char}


Of the total 101 objects detected in these observations, 50 have been identified as stellar sources that have previously observed optical and/or radio counterparts. Their identification, flux density as measured from the primary-beam corrected image alongside the radio-mm spectral index are presented in Table \ref{tab:srcelist}. In addition to SEAC, Gaussian fitting was performed on each of the sources (see Sect. \ref{sect:res}). The flux densities presented are either those taken from the Gaussian fitting (for unresolved or Gaussian-like sources) or the integrated flux density of the source `island' returned by SEAC. The errors on the flux density are those from calculating the integrated flux density (either returned from the Gaussian-fitting or SEAC) and the expected absolute amplitude calibration error for ALMA which is taken to be 5\% (see ALMA Cycle 2 Technical Handbook) combined in quadrature. For sources with a clearly defined compact component with extended structure, flux densities and where possible spectral indices have been provided for both the compact and extended components separately.
Radio observations have shown extended nebulae associated with several sources including Wd1-4 and Wd1-20. This extended emission appears in some cases as multiple components within these higher resolution ALMA images presented here. In particular, four of the FCP18 catalogue sources are part of the diffuse extended emission associated with the nebula of Wd1-20 identified by comparison with the radio emission, as well as three sources which are associated with the nebula of Wd1-4. These components have been listed individually in the final catalogue (Table \ref{tab:fcp18_list}) for completeness. However they have been treated as part of Wd1-4 and Wd1-20 for calculating both flux densities and spatial extent.

\subsubsection{Spatial resolution}\label{sect:res}

A large number of the stellar sources appear to be resolved and in order to determine their spatial extent two approaches were used. Primarily a Gaussian fit was performed using the {\sc AIPS} task {\sc JMFIT}. Estimates of the peak flux density and position from SEAC were used and only pixels >3 times the rms local to the source were included in the fit. Three dimensional plots of the source structure and the resulting model fit were used to visually assess the fit (an example of which can be seen in Fig. \ref{fig:w243gauss}). The final convolved and deconvolved source sizes are presented in Table \ref{tab:srcesizes} which lists the spatial information for each source, including the positional offset from the observed ALMA peak to the catalogued source position taken from the literature (see the caption of Table \ref{tab:srcesizes} for details). Where the Gaussian fit was deemed unreasonable i.e. when the source structure was distinctly non-Gaussian, a largest angular size (LAS) was measured and is listed in Table \ref{tab:srcesizes} (the LAS sizes are not deconvolved). The {\sc IRING AIPS} task was also used to perform integrated annular profiles of each source centred on the position of the peak brightness.
The errors included for source sizes measured using Gaussian fitting are those directly returned by the fitting procedure. For the convolved sizes, these represent the errors calculated in the fit. For the deconvolved sizes the errors represent the potential minimum to maximum range the deconvolved size could have (based on the convolved size errors), as a result of subtracting the image restoring beam, and therefore likely overestimate the true error.

A total of 27 of the stellar sources have been determined to be resolved i.e. have sizes larger than the image restoring beam and have convolved and deconvolved sizes listed in Table \ref{tab:srcesizes}. A further 17 appear to be partially resolved (i.e. those that have only one dimension deconvolved listed in Table 3), while the remaining 7 sources are unresolved. The sources which are partially or fully resolved separate into two main categories: those that appear point-like i.e. are generally representable by a Gaussian and those that have extended nebular emission. 
The former include stars of every spectral type found within the cluster. Given that they are in general relatively faint detections, the errors associated with the WRs and OB super-/hypergiants are systematically larger
in relation to source size than those associated with the cool super-/hypergiants. Despite this, the majority of 
WRs and OB super-/hypergiants appear extended, with objects such as Wd1-243 (LBV; Sect. \ref{sect:w243}) and WR A and L (Sect. \ref{sect:wrstars}) robustly so and hence serving as exemplars. The latter mainly appear to be extended nebulae associated with the cool hypergiants e.g. Wd1-26. 

One final feature of note is that a number of compact sources for which it is possible to fit a gaussian appear to show excess emission in the shoulders of their profiles (see Fig. \ref{fig:profiles} and Sect. \ref{sect:resolvedemission}). This is clearly present in the profiles of the compact sources associated with the YHGs Wd1-4, -12a and 16a but also fainter sources such as those associated with e.g. WR D, J, K and Q, albeit with less confidence.

 We defer discussion of these sources for subsequent sections, where their extent can be compared to predictions derived from their varied wind properties as well as the properties of ejection nebulae associated with other examples. 

Comparison of the mm and radio flux densities of those sources associated with the extended emission reveal an apparent flux density deficit in our ALMA observations {\em if}, as suggested by the radio observations, the emission mechanism is optically-thin free-free from ionised circumstellar ejecta. This strongly suggests that some of the mm emission has been resolved out i.e. occupies larger spatial scales than those effectively sampled by ALMA during these observations. As a result we refrain from determining ionised ejecta masses for these sources, favouring the estimates presented in Do10.


\newcolumntype{Y}{>{\centering\arraybackslash}X}
\newcolumntype{L}{>{\arraybackslash}m{3cm}}
\setcellgapes{4pt}
\begin{table*}
\small
\centering
\caption{ALMA 3-mm flux densities of the known stars detected in these observations.}
\setlength{\tabcolsep}{0.2em} 
\renewcommand{\arraystretch}{1.1}
\begin{tabularx}{\linewidth}{|l|Y|Y|Y|Y|Y|Y|Y|L|}
\hline
Source & \multicolumn{1}{c|}{RA} & Dec & \multicolumn{2}{c|}{Flux density (mJy)} & \multicolumn{2}{c|}{$\alpha$} & Sp. Type & Comments \\
& J2000 &J2000& 3\,mm & 3.6\,cm & radio & 3\,mm-3.6\,cm && \\
\hline
\hline
\multicolumn{9}{|l|}{WR stars and hybrids} \\
\hline
WR J & 47  2.472 & 50 59.976 & 0.18\,$\pm$\,0.04 & - & - & >0.03  & WN5h & No RV data, NoB \\
WR R & 47  6.090 & 50 22.535 & 0.35\,$\pm$\,0.07 & - & - & >0.30  & WN5o & No RV data, X-ray suggests binary \\
WR O & 47  7.651 & 52 36.182 & 0.48\,$\pm$\,0.06 & - & - & >0.42  & WN6o & X-ray suggests binary \\
WR U & 47  6.538 & 50 39.184 & 0.40\,$\pm$\,0.05 & - & - & >0.35  & WN6o & X-ray suggests binary \\
WR Q & 46 55.528 & 51 34.608 & 0.40\,$\pm$\,0.06 & - & - & >0.35 & WN6o & No RV data, NoB \\
WR A & 47  8.347 & 50 45.571 & 3.98\,$\pm$\,0.21 & 0.50\,$\pm$\,0.06 & 0.00\,$\pm$\,0.04  & 0.85  & WN7b+OB? & 7.63d binary \\
WR D & 47  6.245 & 51 26.525 & 0.76\,$\pm$\,0.07 & - & - & >0.61 & WN7o & X-ray suggests CWB \\
\multirow{2}{*}{WR B} & \multirow{2}{*}{47  5.367} & \multirow{2}{*}{51 5.016} & 1.82\,$\pm$\,0.12\tablefootmark{c} & & & >-0.35\tablefootmark{c}  & \multirow{2}{*}{WN7o+OB?} & \multirow{2}{*}{3.52d binary} \\
 &  &  & 3.40\,$\pm$\,0.17\tablefootmark{t} & 4.30\,$\pm$\,0.04\tablefootmark{t} & +0.04\,$\pm$\,0.07\tablefootmark{t} & -0.10\,$\pm$\,0.02\tablefootmark{t}  &  &  \\
WR G & 47  4.005 & 51 25.176 & 0.83\,$\pm$\,0.07 & - & - & >0.65  & WN7o & X-ray suggests CWB \\
WR P & 47  1.584 & 51 45.425 & 0.68\,$\pm$\,0.06 & - & - & >0.57  & WN7o & No RV data, NoB \\
WR I & 47  0.878 & 51 20.674 & 2.16\,$\pm$\,0.12 & - & - & >1.04 & WN8o & No RV data, NoB \\
WR V & 47  3.799 & 50 38.916 & 1.25\,$\pm$\,0.08 & 0.40\,$\pm$\,0.06 & 0.00\,$\pm$\,0.50 & 0.46\,$\pm$\,0.07 & WN8o & No RV data, NoB \\
WR L & 47  4.195 & 51 7.356 & 3.65\,$\pm$\,0.19 & 0.40\,$\pm$\,0.06 & >0.5 & 0.90\,$\pm$\,0.07 & WN9h+OB? & RV binary \\
WR S & 47  2.972 & 50 19.836 & 0.51\,$\pm$\,0.06 & 0.30\,$\pm$\,0.06 & >0.0 & 0.22\,$\pm$\,0.10  & WN10-11h/BHG & Runaway, single \\
W13 & 47  6.451 & 50 26.224 & 0.37\,$\pm$\,0.05 & -  & - & >0.32  & WNVL/BHG & Eclipsing binary \\
WR K & 47  3.230 & 50 43.956 & 0.32\,$\pm$\,0.06 & - & - & >0.25 & WC8 & Not obvious binary (NoB) \\
WR E & 47  6.048 & 52  8.465 & 0.91\,$\pm$\,0.07 & - & - & >0.69  & WC9 & RV variable 10-20\,km/s \\
WR F & 47  5.203 & 52 25.116 & 1.44\,$\pm$\,0.09 & 0.30\,$\pm$\,0.06 & +0.66\,$\pm$\,0.09 & 0.64\,$\pm$\,0.09 & WC9d+OB? & 5.05d Binary \\
WR C & 47  4.402 & 51 3.756 & 0.39\,$\pm$\,0.05 & - & - & >0.34  & WC9d & Dust production suggests binary \\
WR H & 47  4.204 & 51 19.956 & 0.49\,$\pm$\,0.06 & - & - & >0.43  & WC9d & Dust suggests binary \\
WR M & 47  3.954 & 51 37.776 & 0.55\,$\pm$\,0.06 & - & - & >0.48  & WC9d & RV shift suggests binary \\
 \hline
\multicolumn{9}{|l|}{Yellow hypergiants and red supergiants} \\
\hline                   
\multirow{2}{*}{W16a} & \multirow{2}{*}{47  6.607} & \multirow{2}{*}{50 42.334} & 0.69\,$\pm$\,0.07\tablefootmark{c} & -  & - & >-0.34\tablefootmark{c}  & \multirow{2}{*}{A5Ia$^+$} &  \\
&  &  & 1.46\,$\pm$\,0.15\tablefootmark{t} & 1.60\,$\pm$\,0.30\tablefootmark{t} & -0.33\,$\pm$\,0.09\tablefootmark{t} & -0.04\,$\pm$\,0.09\tablefootmark{t} &  &  \\
\multirow{2}{*}{W12a} & \multirow{2}{*}{47  2.205} & \multirow{2}{*}{50 59.166} & 1.00\,$\pm$\,0.10\tablefootmark{c} & - & - & >-0.44\tablefootmark{c}  & \multirow{2}{*}{F1Ia$^+$} &  \\
 &  &  & 1.63\,$\pm$\,0.14\tablefootmark{t} & 2.90\,$\pm$\,0.30\tablefootmark{t} & -0.06\,$\pm$\,0.08\tablefootmark{t} & -0.23\,$\pm$\,0.06\tablefootmark{t} &  &  \\
\multirow{2}{*}{W4} & \multirow{2}{*}{47  1.422} & \multirow{2}{*}{50 37.385} & 1.66\,$\pm$\,0.11\tablefootmark{c} & 0.78\,$\pm$\,0.07\tablefootmark{c} & 0.49\,$\pm$\,0.31\tablefootmark{c} & 0.30\,$\pm$\,0.05\tablefootmark{c} & F3Ia$^+$ &  \\
&   &   & 2.03\,$\pm$\,0.25\tablefootmark{r} & 0.80\,$\pm$\,0.08\tablefootmark{r} & -0.23\,$\pm$\,0.07\tablefootmark{r} & 0.00\,$\pm$\,0.07\tablefootmark{r} &  &  \\
W32 & 47  3.678 & 50 43.686 & 0.38\,$\pm$\,0.07 & 0.40\,$\pm$\,0.06 & 0.00\,$\pm$\,0.05 & -0.03\,$\pm$\,0.10 & F5Ia$^+$ &  \\
\multirow{2}{*}{W237} & \multirow{2}{*}{47  3.101} & \multirow{2}{*}{52 19.086} & -  & 1.80\,$\pm$\,0.20\tablefootmark{c} & -0.06\,$\pm$\,0.10\tablefootmark{c} & - & \multirow{2}{*}{M3Ia} &  \\
 &  &  & 1.60\,$\pm$\,0.14\tablefootmark{t} & 7.40\,$\pm$\,2.21\tablefootmark{t} & -0.35\,$\pm$\,0.19\tablefootmark{t} & -0.05\,$\pm$\,0.14\tablefootmark{t} &  &  \\
W75 & 47  8.914 & 49 58.589 & 0.45\,$\pm$\,0.07 & 0.30\,$\pm$\,0.06 & >0.0 & 0.16\,$\pm$\,0.11 & M4Ia &  \\
\multirow{2}{*}{W20} & \multirow{2}{*}{47  4.686} & \multirow{2}{*}{51 24.096} & 1.45\,$\pm$\,0.11\tablefootmark{c} & -  & - & >0.00\tablefootmark{c}  & \multirow{2}{*}{M5Ia} &  \\
 &  &  & 3.66\,$\pm$\,0.19\tablefootmark{t} & 3.80\,$\pm$\,0.40\tablefootmark{t} & -0.11\,$\pm$\,0.07\tablefootmark{t} & -0.02\,$\pm$\,0.05\tablefootmark{t} & &  \\
W26 & 47  5.375 & 50 36.486 & 114.75\,$\pm$\,5.76 & 20.10\,$\pm$\,2.00 & 0.07\,$\pm$\,0.11 & 0.71\,$\pm$\,0.05 & M5-6Ia &  \\
\hline
\multicolumn{9}{|l|}{Blue hypergiants, LBVs, OB supergiants and the sgB[e] star} \\
\hline                    
W25 & 47  5.831 & 50 33.785 & 0.18\,$\pm$\,0.01 & - & - & >0.01  & O9Iab & Faint X-ray, no RV data \\
W17 & 47  6.167 & 50 49.355 & 0.75\,$\pm$\,0.04 & 1.70\,$\pm$\,0.20 & -0.42\,$\pm$\,0.11 & -0.33\,$\pm$\,0.06 & 09.5Ia & Faint X-ray, no RV data  \\
W43a & 47  3.549 & 50 57.816 & 0.17\,$\pm$\,0.05 & - & - & >0.01  & B0Ia & 16.27d binary from RV shifts \\
W61a & 47  2.300 & 51 41.826 & 0.10\,$\pm$\,0.05 & - & - & >-0.22  & B0.5Ia & No RV data, NoB, 4.5$\sigma$ \\
W46a & 47  3.911 & 51 19.866 & 0.35\,$\pm$\,0.06 & - & - & >0.30  & B1Ia &  No RV data, NoB\\
W56a & 46 58.939 & 51 49.110 & 0.11\,$\pm$\,0.05 & - & - & >-0.18  & B1.5Ia & No RV data, NoB\\
W52 & 47  1.843 & 51 29.495 & 0.11\,$\pm$\,0.05 & - & - & >-0.16  & B1.5Ia & 6.7d ellipsoidal modulation \\
W8b & 47  4.953 & 50 26.856 & 0.15\,$\pm$\,0.05 & - & - & >-0.06  & B1.5Ia & No RV data, NoB \\
W243 & 47  7.496 & 52 29.252 & 9.96\,$\pm$\,0.50 & 1.50\,$\pm$\,0.20 & 0.87\,$\pm$\,0.30 & 0.77\,$\pm$\,0.13 & B2Ia (LBV) & RV pulsator, NoB \\
\hline
\end{tabularx}
\tablefoot{3.6\,cm flux densities and spectral indices taken from Do10 are also listed. Radio to mm spectral indices have been calculated using the total flux density unless both core and resolved components are separated in both the radio and ALMA images. Where no flux density is reported for a source in D010, we find a limiting spectral index based on a 3.6\,cm flux density limit of 170\mujybm \,(see Do10 for details). Spectral type identifications have been taken from \cite{crowther06} for WR stars, \cite{clark10} for the YHGs and RSGs, \cite{iggy10} for the OB stars and Do10 for the two D09 sources.\tablefoottext{t}{total flux density}, \tablefoottext{c}{core component flux density}, \tablefoottext{r}{resolved flux density (not including core component)}}
\label{tab:srcelist}
\end{table*}

\begin{table*}
\small
\centering
\caption*{Table \ref{tab:srcelist} --- continued}
\setlength{\tabcolsep}{0.2em} 
\renewcommand{\arraystretch}{1.1}
\begin{tabularx}{\linewidth}{|l|Y|Y|Y|Y|Y|Y|Y|L|}
\hline
Source & RA & Dec & \multicolumn{2}{c|}{Flux density (mJy)} & \multicolumn{2}{c|}{$\alpha$} & Sp. Type & Comments \\
& J2000 &J2000& 3\,mm & 3.6\,cm & radio & 3\,mm-3.6\,cm && \\
\hline
\hline
\multicolumn{9}{|l|}{Blue hypergiants, LBVs, OB supergiants and the sgB[e] star} \\
\hline
W28 & 47  4.660 & 50 38.646 & 0.20\,$\pm$\,0.05 & -  & - & >0.07  & B2Ia & No RV data, NoB \\
W2a & 46 59.707 & 50 51.332 & 0.12\,$\pm$\,0.05 & -  & - & >-0.16  & B2Ia & RV pulsator, NoB, 4.5$\sigma$ \\
W11 & 47  2.231 & 50 47.286 & 0.11\,$\pm$\,0.05 & -  & - & >-0.19  & B2Ia & No RV data or X-ray detection, 4.5$\sigma$ \\
W23a & 47  2.567 & 51 9.066 & 0.23\,$\pm$\,0.05 & -  & - & >0.12  & B2Ia+OB & RV data, NoB \\
W71 & 47  8.450 & 50 49.530 & 0.27\,$\pm$\,0.07 & -  & - & >0.19  & B2.5Ia & RV pulsator, NoB \\
W33 & 47  4.117 & 50 48.636 & 0.23\,$\pm$\,0.05 & - & - & >0.13  & B5Ia$^{+}$ & No RV data, NoB \\
W7 & 47  3.618 & 50 14.526 & 0.37\,$\pm$\,0.05 & - & - & >0.31  & B5Ia$^{+}$ & RV pulsator, NoB \\
W42a & 47  3.239 & 50 52.326 & 0.25\,$\pm$\,0.06 & - & - & >0.16  & B9Ia$^{+}$ & No RV data, NoB \\
\multirow{2}{*}{W9} & \multirow{2}{*}{47  4.140} & \multirow{2}{*}{50 31.430} & 152.92\,$\pm$\,0.08\tablefootmark{c} & 24.90\,$\pm$\,2.50\tablefootmark{c} & 0.68\,$\pm$\,0.70\tablefootmark{c} & 0.74\,$\pm$\,0.04\tablefootmark{c} & \multirow{2}{*}{SgB[e]} & X-ray suggests CWB \\
 & & & 16.15\,$\pm$\,0.90\tablefootmark{r} & 30.50\,$\pm$\,3.00\tablefootmark{r} & 0.16\,$\pm$\,0.17\tablefootmark{r} & -0.26\,$\pm$\,0.05\tablefootmark{r} &  &  \\
\multirow{2}{*}{D09-R1} & \multirow{2}{*}{47  9.071} & \multirow{2}{*}{51 10.139} & \multirow{2}{*}{1.05\,$\pm$\,0.12} & 0.70\,$\pm$\,0.07\tablefootmark{c} & -0.23\,$\pm$\,0.27\tablefootmark{c} & \multirow{2}{*}{0.17\,$\pm$\,0.09\tablefootmark{t}} & \multirow{2}{*}{BSG} &  \\
 & &  &  & 6.5\,$\pm$\,1.2\tablefootmark{r} & -0.61\,$\pm$\,0.11\tablefootmark{r} & & &  \\
D09-R2 & 47  6.908 & 50 37.204 & 0.68\,$\pm$\,0.08 & 0.70\,$\pm$\,0.06 & -0.43\,$\pm$\,0.11 & -0.01\,$\pm$\,0.06 & BSG &  \\
W30 & 47  4.117 & 50 39.456 & 0.17\,$\pm$\,0.06 & - & - & >0.00  & O+O & X-ray suggests CWB \\
\hline
\end{tabularx}
\end{table*}


\begin{table*}[!htbp]
\small
\centering
\caption{Spatial information for the 50 identified stellar sources. }
\newcolumntype{Y}{>{\centering\arraybackslash}X}
\newcolumntype{L}{>{\centering\arraybackslash}m{1cm}}
\newcolumntype{M}{>{\centering\arraybackslash}m{1.7cm}}
\setlength{\tabcolsep}{0.2em} 
\renewcommand{\arraystretch}{1.1}
\begin{tabularx}{\textwidth}{|l|L|Y|Y|M|Y|Y|Y|Y|l|}
\hline
Source & FCP18 & RA & Dec & Offset & \multicolumn{4}{c|}{Size (arcsecs)} & Notes \\
&&&&&  \multicolumn{2}{c|}{Convolved} &  \multicolumn{2}{c|}{Deconvolved} & \\
&&&& arcsecs & Major axis & Minor axis & Major axis & Minor axis &  \\ 
\hline
\hline
\multicolumn{10}{|l|}{WR stars and hybrids} \\
\hline
WR J & 23 & 47 2.472 & 50 59.976 & 0.13 & 0.55\,$\pm$\,0.07 & 0.52\,$\pm$\,0.07 & --- & --- & Unresolved  \\
WR R & 68 & 47 6.090 & 50 22.535 & 0.21 & 0.81\,$\pm$\,0.10 & 0.72\,$\pm$\,0.09 & 0.54\,$\pm$\,0.23 & 0.37\,$\pm$\,0.32 &  \\
WR O & 84 & 47 7.651 & 52 36.182 & 0.30 & 0.74\,$\pm$\,0.06 & 0.64\,$\pm$\,0.05 & 0.37\,$\pm$\,0.20 & 0.28\,$\pm$\,0.24 &  \\
WR U & 78 & 47 6.538 & 50 39.184 & 0.23 & 0.63\,$\pm$\,0.04 & 0.51\,$\pm$\,0.03 & --- & --- & Unresolved  \\
WR Q & 1 & 46 55.528 & 51 34.608 & 0.46 & 0.69\,$\pm$\,0.06 & 0.64\,$\pm$\,0.06 & 0.34\,$\pm$\,0.25 & 0.18\,$\pm$\,0.21 &  \\
WR A & 88 & 47 8.347 & 50 45.571 & 0.29 & 0.665\,$\pm$\,0.005 & 0.579\,$\pm$\,0.004 & 0.16\,$\pm$\,0.03 & 0.09\,$\pm$\,0.07 &  \\
WR D & 71 & 47 6.245 & 51 26.525 & 0.06 & 0.69\,$\pm$\,0.03 & 0.58\,$\pm$\,0.02 & 0.23\,$\pm$\,0.16 & 0.13\,$\pm$\,0.13 &  \\
\multirow{2}{*}{WR B} & \multirow{2}{*}{62} & \multirow{2}{*}{47 5.367} & \multirow{2}{*}{51 5.016} & \multirow{2}{*}{0.07} & 0.97 0.03 & 0.75 0.02 & 0.72\,$\pm$\,0.04 & 0.49\,$\pm$\,0.04 & Compact \\
 &  &  &  &  & 4.43  & LAS & ---  & ---   & Total \\
WR G & 43 & 47 4.005 & 51 25.176 & 0.05 & 0.63\,$\pm$\,0.02 & 0.58\,$\pm$\,0.02 & 0.12\,$\pm$\,0.11 & --- &  \\
WR P & 15 & 47 1.584 & 51 45.425 & 0.10 & 0.68\,$\pm$\,0.03 & 0.57\,$\pm$\,0.03 & 0.20\,$\pm$\,0.15 & 0.07\,$\pm$\,0.12 &  \\
WR I & 10 & 47 0.878 & 51 20.674 & 0.13 & 0.66\,$\pm$\,0.01 & 0.57\,$\pm$\,0.01 & 0.12\,$\pm$\,0.09 & --- &  \\
WR V & 38 & 47 3.799 & 50 38.916 & 0.17 & 0.68\,$\pm$\,0.02 & 0.56\,$\pm$\,0.01 & 0.22\,$\pm$\,0.06 & --- &  \\
WR L & 47 & 47 4.195 & 51 7.356 & 0.07 & 0.661\,$\pm$\,0.005 & 0.573\,$\pm$\,0.005 & 0.13\,$\pm$\,0.04 & 0.08\,$\pm$\,0.06 &  \\
WR S & 29 & 47 2.972 & 50 19.836 & 0.19 & 0.63\,$\pm$\,0.04 & 0.57\,$\pm$\,0.03 & 0.18\,$\pm$\,0.16  & ---  &  \\
W13 & 76 & 47 6.451 & 50 26.224 & 0.23 & 0.64\,$\pm$\,0.05 & 0.55\,$\pm$\,0.04 & 0.20\,$\pm$\,0.18 & --- &  \\
WR K & 31 & 47 3.230 & 50 43.956 & 0.26 & 0.67\,$\pm$\,0.07 & 0.58\,$\pm$\,0.06 & 0.34\,$\pm$\,0.24 & --- &  \\
WR E & 67 & 47 6.048 & 52 8.465 & 0.27 & 0.70\,$\pm$\,0.02 & 0.55\,$\pm$\,0.02 & 0.25\,$\pm$\,0.07 & --- &  \\
WR F (239) & 61 & 47 5.203 & 52 25.116 & 0.21 & 0.69\,$\pm$\,0.02 & 0.59\,$\pm$\,0.01 & 0.24\,$\pm$\,0.06 & 0.15\,$\pm$\,0.10 &  \\
WR C & 50 & 47 4.402 & 51 3.756 & 0.05 & 0.59\,$\pm$\,0.04 & 0.54\,$\pm$\,0.04 & --- & --- & Unresolved \\
WR H & 48 & 47 4.204 & 51 19.956 & 0.30 & 0.66\,$\pm$\,0.04 & 0.58\,$\pm$\,0.04 & 0.21\,$\pm$\,0.18 & --- &  \\
WR M (66) & 41 & 47 3.954 & 51 37.776 & 0.07 & 0.70\,$\pm$\,0.05 & 0.66\,$\pm$\,0.04 & 0.34\,$\pm$\,0.18 & 0.26\,$\pm$\,0.21 &  \\
\hline
\multicolumn{10}{|l|}{Yellow hypergiants and red supergiants} \\
\hline        
\multirow{2}{*}{W16a} & \multirow{2}{*}{79} & \multirow{2}{*}{47 6.607} & \multirow{2}{*}{50 42.334} & \multirow{2}{*}{0.24} & 0.79\,$\pm$\,0.05 & 0.71\,$\pm$\,0.04 & 0.48\,$\pm$\,0.12 & 0.39\,$\pm$\,0.15 & Compact\\
 &  &  &  &  & 3.61  & LAS & ---  & ---  & Total \\
\multirow{2}{*}{W12a} & \multirow{2}{*}{19} & \multirow{2}{*}{47 2.205} & \multirow{2}{*}{50 59.166} & \multirow{2}{*}{0.37} & 1.16 \,$\pm$\,0.08 & 0.77\,$\pm$\,0.05 & 0.97\,$\pm$\,0.10 & 0.52\,$\pm$\,0.09 & Compact \\
 &  &  &  &  & 3.05  & LAS & ---  & ---  & Total \\
\multirow{2}{*}{W4} & \multirow{2}{*}{13} & \multirow{2}{*}{47 1.422} & \multirow{2}{*}{50 37.385} & \multirow{2}{*}{0.29} & 0.95\,$\pm$\,0.04 & 0.78\,$\pm$\,0.02 & 0.69\,$\pm$\,0.05 & 0.54\,$\pm$\,0.04 & Compact \\
 &  &  &  &  & 5.42  & LAS & &  & Total \\
W32 & 36 & 47 3.678 & 50 43.686 & 0.20 & 0.89\,$\pm$\,0.10 & 0.63\,$\pm$\,0.07 & 0.61\,$\pm$\,0.16 & 0.25\,$\pm$\,0.20 &  \\
W237 & 30 & 47 3.101 & 52 19.086 & 0.31 & 1.18\,$\pm$\,0.07 & 1.10\,$\pm$\,0.06 & 1.02\,$\pm$\,0.10 & 0.89\,$\pm$\,0.10 &  \\
W75 & 96 & 47 8.914 & 49 58.589 & 0.25 & 0.85\,$\pm$\,0.01 & 0.71\,$\pm$\,0.07 & 0.56\,$\pm$\,0.16 & 0.41\,$\pm$\,0.20 &  \\
\multirow{2}{*}{W20} & \multirow{2}{*}{55} & \multirow{2}{*}{47 4.686} & \multirow{2}{*}{51 24.096} & \multirow{2}{*}{0.33} & 0.98\,$\pm$\,0.04 & 0.78\,$\pm$\,0.03 & 0.74\,$\pm$\,0.06 & 0.53\,$\pm$\,0.06 & Compact \\
 &  & &   &  & 3.84  & LAS & ---  &  ---  & Total \\
W26 & 63 & 47 5.375 & 50 36.486 & 0.26 & 16.00  & LAS & ---  &  --- &  \\
\hline
\multicolumn{10}{|l|}{Blue hypergiants, LBVs, OB supergiants and the sgB[e] star} \\
\hline
W25 & 65 & 47 5.831 & 50 33.785 & 0.72 & 1.66  & LAS  &  --- &  --- &  \\
W17 & 70 & 47 6.167 & 50 49.355 & 0.88 & 3.37  & LAS  &  --- &  --- & \\
W43a & 33 & 47 3.549 & 50 57.816 & 0.52 & 0.69\,$\pm$\,0.12 & 0.55\,$\pm$\,0.09 & 0.24\,$\pm$\,0.27 & --- &  \\
W61a & 21 & 47 2.300 & 51 41.826 & 0.25 & 0.72\,$\pm$\,0.20 & 0.47\,$\pm$\,0.13 & 0.33\,$\pm$\,0.34 & --- &  \\
W46a & 40 & 47 3.911 & 51 19.866 & 0.37 & 0.73\,$\pm$\,0.07 & 0.58\,$\pm$\,0.05 & 0.32\,$\pm$\,0.24 & 0.12\,$\pm$\,0.15 &  \\
W56a & 4 & 46 58.939 & 51 49.110 & 0.33 & 0.62\,$\pm$\,0.14 & 0.52\,$\pm$\,0.12 & --- & --- & Unresolved \\
W52 & 17 & 47 1.843 & 51 29.495 & 0.30 & 0.62\,$\pm$\,0.13 & 0.48\,$\pm$\,0.10 & --- & --- & Unresolved \\
W8b & 58 & 47 4.953 & 50 26.856 & 0.16 & 0.61\,$\pm$\,0.10 & 0.51\,$\pm$\,0.09 & --- & --- & Unresolved \\
W243 & 83 & 47 7.496 & 52 29.252 & 0.94 & 0.669\,$\pm$\,0.002 & 0.567\,$\pm$\,0.002 & 0.15\,$\pm$\,0.01 & 0.03\,$\pm$\,0.03 &  \\
W28 & 53 & 47 4.660 & 50 38.646 & 0.25 & 0.81\,$\pm$\,0.12 & 0.46\,$\pm$\,0.07 & 0.52\,$\pm$\,0.21 & --- &  \\
W2a & 9 & 46 59.707 & 50 51.332 & 0.23 & 0.72\,$\pm$\,0.18 & 0.50\,$\pm$\,0.12 & 0.37\,$\pm$\,0.34 & --- &  \\
W11 & 20 & 47 2.231 & 50 47.286 & 0.29 & 0.72\,$\pm$\,0.17 & 0.45\,$\pm$\,0.11 & 0.34\,$\pm$\,0.33 & --- &  \\
\hline
\end{tabularx}
\tablefoot{ Reference positions used to derive millimetre offsets for each source are those taken from Do10, \cite{crowther06,clark10,iggy10} and from the recent FLAMES Wd1 survey \cite{clark17}. As described in Sect. \ref{sect:res},the acronym  `LAS' corresponds to `largest angular size' and is used to report dimensions for non-gaussian sources.}
\label{tab:srcesizes}
\end{table*}

\begin{table*}[!htbp]
\small
\centering
\caption*{Table \ref{tab:srcesizes} --- continued}
\newcolumntype{Y}{>{\centering\arraybackslash}X}
\newcolumntype{L}{>{\centering\arraybackslash}m{1cm}}
\newcolumntype{M}{>{\centering\arraybackslash}m{1.7cm}}
\setlength{\tabcolsep}{0.2em} 
\renewcommand{\arraystretch}{1.1}
\begin{tabularx}{\textwidth}{|l|L|Y|Y|M|Y|Y|Y|Y|l|}
\hline
Source & FCP18 & RA & Dec & Offset & \multicolumn{4}{c|}{Size (arcsecs)} & Notes \\
&&&&&  \multicolumn{2}{c|}{Convolved} &  \multicolumn{2}{c|}{Deconvolved} & \\
&&&& arcsecs & Major axis & Minor axis & Major axis & Minor axis &  \\ 
\hline
\hline
\multicolumn{9}{|l|}{Blue hypergiants, LBVs, OB supergiants and the sgB[e] star} \\
\hline
W23a & 24 & 47 2.567 & 51 9.066 & 0.37 & 0.60\,$\pm$\,0.07 & 0.58\,$\pm$\,0.07  & 0.18\,$\pm$\,0.18 & --- &  \\
W71 & 89 & 47 8.450 & 50 49.530 & 0.25 & 0.98\,$\pm$\,0.18 & 0.65\,$\pm$\,0.12 & 0.74\,$\pm$\,0.27 & 0.28\,$\pm$\,0.27 &  \\
W33 & 44 & 47 4.117 & 50 48.636 & 0.34 & 0.62\,$\pm$\,0.07 & 0.56\,$\pm$\,0.07  & 0.18\,$\pm$\,0.20 & --- &  \\
W7 & 34 & 47 3.618 & 50 14.526 & 0.33 & 0.65\,$\pm$\,0.05 & 0.50\,$\pm$\,0.04 & --- & --- & Unresolved \\
W42a & 32 & 47 3.239 & 50 52.326 & 0.25 & 0.67 0.08 & 0.59 0.07 & 0.25\,$\pm$\, 0.24 & --- &  \\
\multirow{2}{*}{W9} & \multirow{2}{*}{46} & \multirow{2}{*}{47 4.136} & \multirow{2}{*}{50 31.400} & \multirow{2}{*}{0.35} & 0.6727\,$\pm$\,0.0002 & 0.5904\,$\pm$\,0.0002 & 0.186\,$\pm$\,0.001 & 0.150\,$\pm$\,0.002 & Compact \\
 &  &  &  &  & 4.50  & LAS &  --- & ---  & Total \\
D09-R1 & 98 & 47 9.071 & 51 10.139 & 0.10 & 2.44  & LAS &  --- &  --- &  \\
D09-R2 & 81 & 47 6.908 & 50 37.204 & 0.43 & 2.28  & LAS & ---  &  --- &  \\
W30 & 45 & 47 4.117 & 50 39.456 & 0.46 & 0.71\,$\pm$\,0.14 & 0.62\,$\pm$\,0.13  & 0.43\,$\pm$\,0.32 & --- &  \\
\hline
\end{tabularx}
\end{table*}

\subsubsection{Spectral indices}\label{sect:specindex}

Where possible the spectral index between the 8.6\,GHz (3.6\,cm) radio and 3\,mm ALMA flux densities are presented in Table \ref{tab:srcelist}. The majority of stellar sources appear to have spectral indices consistent with partially optically-thick free-free emission. However a number appear to show flatter or inverted spectral indices. The divergence of the spectral index from the canonical value of $\alpha=0.6$ expected for a partially optically-thick wind \citep{wright75} can occur from a number of reasons, such as changes in the run of ionisation or wind geometry with radius, or the presence of both optically-thick and thin components \citep[such as optically-thick clumps in a structured wind;][]{ignace}. In long-period ($P_{\rm orb}\gtrsim 1$yr) colliding wind binaries, an additional non-thermal component caused by wind interaction outside the respective radio/mm photospheres can lead to a reduction or flattening of the spectral index \citep[e.g.][Do10]{chapman}. Conversely, in more compact binaries thermal emission from material in the wind collision region (WCR) may come to dominate the spectrum at mm or radio wavelengths \citep{stevens95,pittard06,pittard10,montes}. For instance, optically-thin thermal emission ($\alpha\sim-0.1$) from an adiabatic WCR may dominate emission at long (cm) wavelengths, leading to a flatter than expected spectrum \citep{pittard06}. Alternatively, for shorter period systems, cooling of material in the WCR becomes increasingly important \citep{stevens92} leading to optically-thick thermal emission \citep{pittard10} which could dominate the spectrum at mm wavelengths, resulting in a steeper mm-radio continuum spectral index.

For all eventualities it is important to recall that the time-dependent line of sight through the complex geometry of the WCR and stellar winds imposed by orbital motion may also be reflected in variability in the spectral index. This is of particular concern for very compact (contact) binaries in which interactions occur before appreciable wind acceleration, where computational limitations preclude accurate modelling of the resultant wind and the WCR geometries and interactions, and hence quantitative predictions for the emergent spectrum.

\subsubsection{Mass-loss determination}\label{sect:mdots}

There is considerable advantage to using free-free mm/radio fluxes for determining mass-loss for massive stars in that, unlike H$\alpha$ and UV, the emission due to electron-ion interactions in their ionised winds arises at large radii, where the terminal velocity will have been reached. Therefore the interpretation of the mm/radio flux densities is more straightforward and is not strongly dependent on details of the velocity law, ionization conditions, inner velocity field, or the photospheric profile. Though the greater geometric region and density squared dependence of the free-free flux makes these continuum observations sensitive to clumping in the wind, there is evidence that clumping decreases in the outer wind regions \citep[e.g.][]{runacres,puls}.

The mass-loss rate is related to the observed free-free emitted radiation as

\begin{equation}
\begin{aligned}
S_{\nu} = 2.32 x 10^4 \left( \frac{\dot{M}\sqrt{f_{cl}}}{\mu v_{\infty}} 
\right)^{4/3} \frac{1}{D^2} \left(\gamma g_{ff} \nu \overline{Z^2} 
\right)^{2/3},
\end{aligned}
\label{eqn:mdot} 
\end{equation}

\noindent where, S$_{\nu}$ is our observed radio flux in mJy measured at frequency $\nu$ in Hz; \mdot is in \solmasyr; the terminal velocity v$_{\infty}$ is in kms$^{-1}$; D is the distance in kpc \citep[see e.g.][]{wright75}. The quantities $\mu$, $Z$, and $\gamma$ are the mean molecular weight per ion, ratio of electron to ion density, and mean number of electrons per ion.
The Gaunt factor, $g_{ff}$, can be approximated by

\begin{equation}
\begin{aligned}
g_{ff} \approx 9.77 \left( 1+0.13 \log \left(T_{e}^{3/2}/\nu 
\sqrt{(\overline{Z^2})}\, \right)\right),
\end{aligned} 
\label{eqn:gff}
\end{equation}
 \citep[e.g.][]{leitherer91}.
 
Since the free-free emission process depends on the density-squared, it is affected by wind clumping. Equation \ref{eqn:mdot} includes a simple account of this, such that  all clumps are assumed to have the same clumping factor given by $f_{cl}$ = $<{\rho^2}>  / {<{\rho}>}^2$, where the angle brackets indicate an average over the volume in which continuum emission is formed. A given flux can therefore be interpreted as a certain mass-loss rate for a smooth wind ($f_{cl}$ = 1), or correspondingly as a lower mass-loss rate in a clumped wind ($f_{cl} > 1$). Note, with an assumption of no inter-clump material, $f_{cl}$ here is related to the reciprocal of the volume filling factor. Regarding the remaining terms in the above relations we adopt $T_{e}$ = 0.5 $T_{eff}$ \citep[e.g.][]{drew} and that for OB stars hydrogen is fully ionised, He$^+$ dominates over He$^{2+}$, and a helium abundance of n$_{He}/$n$_{H}$ = 0.1 (i.e. $\mu$ $\sim$ 1.4, Z = 1, $\gamma$ = 1). We note that for hot stars the dominant H ionization stage is controlled essentially by the wind density through mass-loss rate and clumping. However in the case of strong clumping in the mm formation region, recombination would be enhanced and thus favour He$^+$. For the chemistry of the WR star winds we have adopted here the following generalisation for $\mu$ based on \cite{leitherer97}; $\mu$ = 4.0 for WN6 and earlier types; $\mu$ = 2.0 for later-type than WN6; $\mu$ = 4.7 for WC8 and WC9. Once more, He$^+$ is assumed to be the most prevalent in the WR mm-emitting region and we have assumed  Z = 1 and $\gamma$ = 1. Following \cite{leitherer95} we assume the B hypergiant (Ia$^+$) candidates in Wd~1 to have a chemical composition akin to LBVs and adopt $\mu$ = 1.6, $\gamma$=0.8 and  Z=0.9. For the fundamental stellar parameters ($T_{eff}$, terminal velocity $v_\infty$) we have adopted the (observational) compilations of \cite{crowther06b} and \cite{searle} for B0-B5 supergiants; \cite{crowther07} and \cite{sander} for WN stars; and \cite{sander} for WC stars.

Our mass-loss estimates assume the mm fluxes are not affected by binarity, either via synchrotron emission and/or enhanced thermal emission from a (strongly radiative) WCR \citep[e.g.][]{pittard10}. After excluding two sources with spectral indices strongly indicative of non-thermal emission we present the resultant mass-loss rates for a total of 27 stars in Table \ref{tab:masslosstab} and further discuss these in the following sections.

\subsection{Mm sources lacking counterparts}\label{sect:nooptical}

The remaining 51 detected sources which have no catalogued optical or radio counterpart are listed in Table \ref{tab:fcp18_list}. Their positions in Wd1 are shown in Fig. \ref{fig:unknownfinderimage} and images of each source are available in the online material. Seven of these sources have been determined to be knots of emission associated with the extended nebulae of Wd1-4 and Wd1-20. Of the 44 >5\,$\sigma$ detected sources that are unique to these data, at least five appear to be obviously associated with extended emission, in most cases these are clearly low surface-brightness areas with brighter points of emission accounting for the source detections. A further 31 sources are more identifiable as potential sources appearing as well-defined components within diffuse mm emission. However, it is not possible from these observations alone to determine if they are for example stellar sources or again merely brighter components of the more diffuse emission. Five of the ALMA-only sources with no previous counterpart do appear to be bright discrete sources in isolation and not associated with any diffuse emission.

Large surveys of the background galaxy population at mm wavelengths are now making use of high sensitivity observations to probe the faint source population. Whilst some of these surveys are performed at 3-mm (100\,GHz), they tend to be less sensitive \citep[e.g.][ use observations at 95, 150 and 220\,GHz though only probe down to $\sim$4\,mJy]{mocanu13}. The vast majority utilise observations at 1.1-1.3\,mm and the most sensitive of these using ALMA Band 6 or 7 \citep[e.g.][]{ono14,carniani15,oteo16,hatsukade16}. Using the cumulative source counts from these surveys, it is however possible to estimate the number of background sources expected within the field observed for Westerlund 1.

\cite{massardi16} study the spectral evolution of sources from mm to radio wavelengths and show that the majority of the population display a down-turning spectral index with indices ranging from 0 to $\sim -1.8$ for the 100-200\,GHz frequency range (3-1.5\,mm) with an average value around -0.5. The noise level in the central regions of the Westerlund 1 field is $\sim$28\,\mujybm resulting in a 3-$\rm{\sigma}$ value of $\sim$85\,\mujybm. Assuming a -0.5 spectral index is representative, this would equate to $\sim 56-51$\,\mujybm at band 6-7.  \cite{umehata17} used ALMA and the Aztec camera to measure the cumulative source counts and find ${\rm S_{>0.4\,mJy} = 9800^{+5100}_{-2200} deg^{-2}}$. The observations of Wd-1 cover an area of approximately 7.3 ${\rm arcmin^2}$, providing an estimate of approximately 20 background sources within our field. Likewise the theoretical models from \cite{hayward13} give ${\rm S_{ALMA_6 > 0.4\,mJy} = 6897 deg^{-2}}$, estimating of $\sim$ 14 sources in the field. Both of these however, lack some of the faint source population (<0.4\,mJy) which should be detectable in our observations. More recently, \cite{oteo16} use ALMA Band 6/7 calibration survey observations to estimate source counts down to a flux density of 0.20\,mJy, equating to $11^{+10}_{-9}$ sources in our field-of-view above this threshold. They also highlight the increased level of uncertainty in calculating source counts due to potential spurious source contamination, when employing lower detection thresholds often used to probe the fainter source population.

Interestingly, \cite{shimizu12} use hydrodynamical simulations to predict the background submillimetre galaxy population and present results for a number of ALMA bands including 3, 6 and 7. Their results would predict < 1 background sources in our field directly at 3-mm. However assuming the previous correction for spectral index and calculating for Band 6 would instead predict $\sim$20 background sources for this study. As discussed in \cite{carniani15}, without knowing the intrinsic spectral energy distribution of the sampled sources (and their redshift distribution) it is difficult to accurately perform large extrapolations in flux density. Whilst \cite{mocanu13} have suggested that the 3-mm models under-predict the expected source counts, it is likely that there is also large uncertainty in predicting the background source counts in this manner.

\subsection{Extended continuum emission}\label{sect:extended}

There are several areas of low surface brightness emission in the central region of Wd1. These are spatially coincident with the extended regions identified in the radio images presented in Do10 as A1-A8. The higher resolution of these ALMA observations is `resolving-out' some of the more diffuse emission on the larger spatial scales. This results in the significantly more patched or clumped background emission and the breaking down into more discrete components within these regions seen in the ALMA images. This is most clearly evident in Fig. \ref{fig:radioimage} where the 3\,mm emission (in colour-scale) appears to be associated with knots of emission in the more extended structures seen in the radio (blue contours). 
Five of the extended regions are readily associated with individual stars namely the RSGs Wd1-20 and Wd1-26, the BSG D09-R1, the sgB[e] star Wd1-9 and WR B, with these stars showing as more discrete sources in the ALMA observations. The remaining three extended regions are not associated with with any currently known stellar counterpart, however, several of the ALMA-only sources identified in these observations are found in these regions. In some cases, such as around Wd1-20, the ALMA-only sources can be identified as low brightness diffuse emission associated with the larger nebula. In others such as near WR B, the other ALMA-only sources appear bright, compact and very similar to the other known stellar sources.

\begin{figure*}
\begin{minipage}{\textwidth}
\begin{center}
\includegraphics[width=18cm]{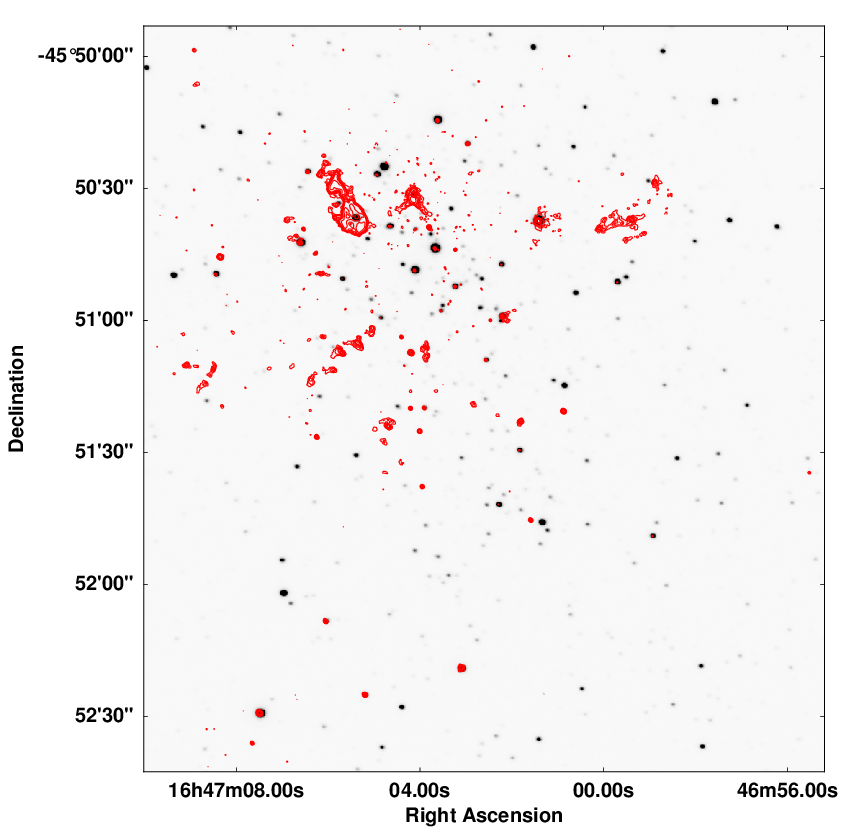}
\caption{ALMA 3-mm contours (from the non-PB corrected image) overlaid on a FORS R-band image with a limiting magnitude of $\sim$17.5\,mag (see Do10). The contours are plotted at -3,3,5,7,9,11,15,30,60,120 $\times$ 33\,\mujybm.}
\label{fig:forsimage}
\end{center}
\end{minipage}
\end{figure*}


\begin{figure*}
\includegraphics[width=18cm]{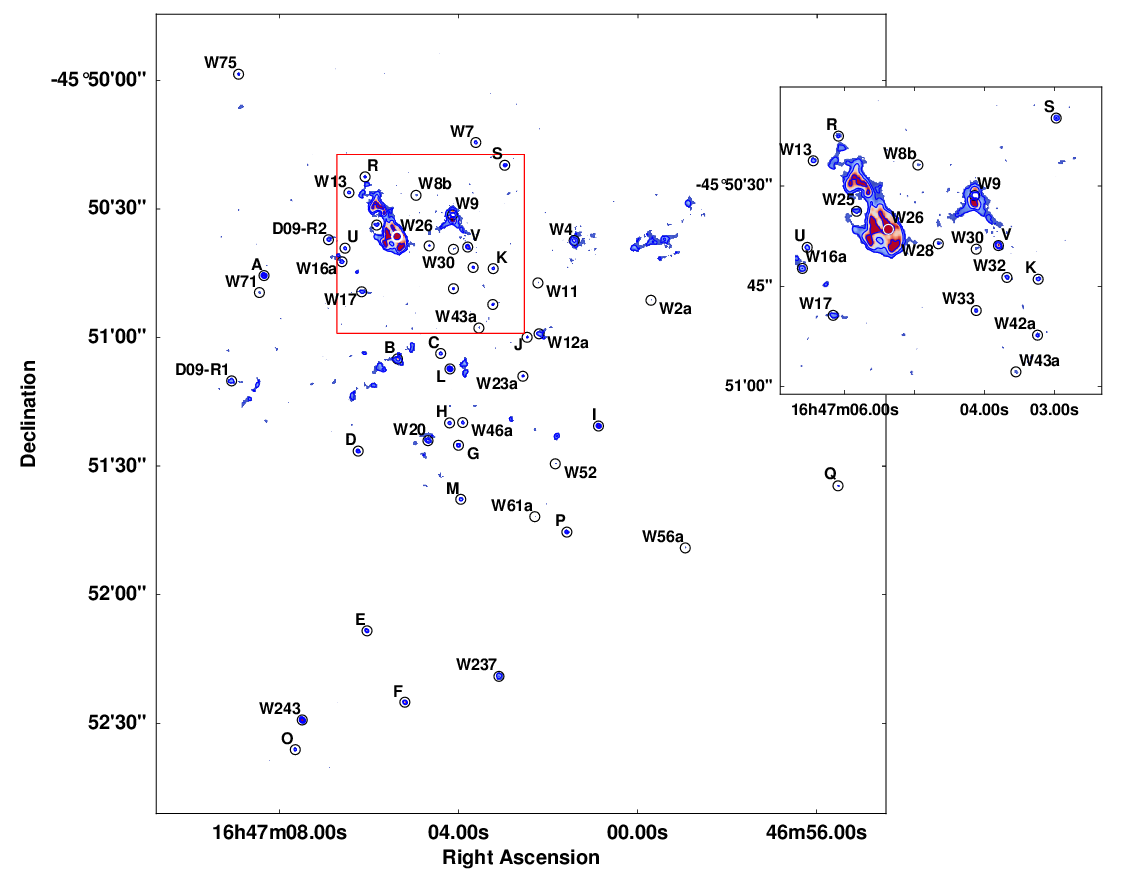}
\caption{ALMA 3-mm colour-scale map with overlaid contours from the non-PB corrected image. The colour-scale ranges from 0.10 to 1.2\,\mjybm and the contours are plotted at levels of 0.16,0.29,0.52,0.96,1.74,3.16\,\mjybm. The stellar sources identified in this study are labelled at the positions observed in the ALMA data. The insert shows a zoom-in of the region defined by the red rectangle.}
\label{fig:finderimage}
\end{figure*}

\begin{figure*}
\includegraphics[width=18cm]{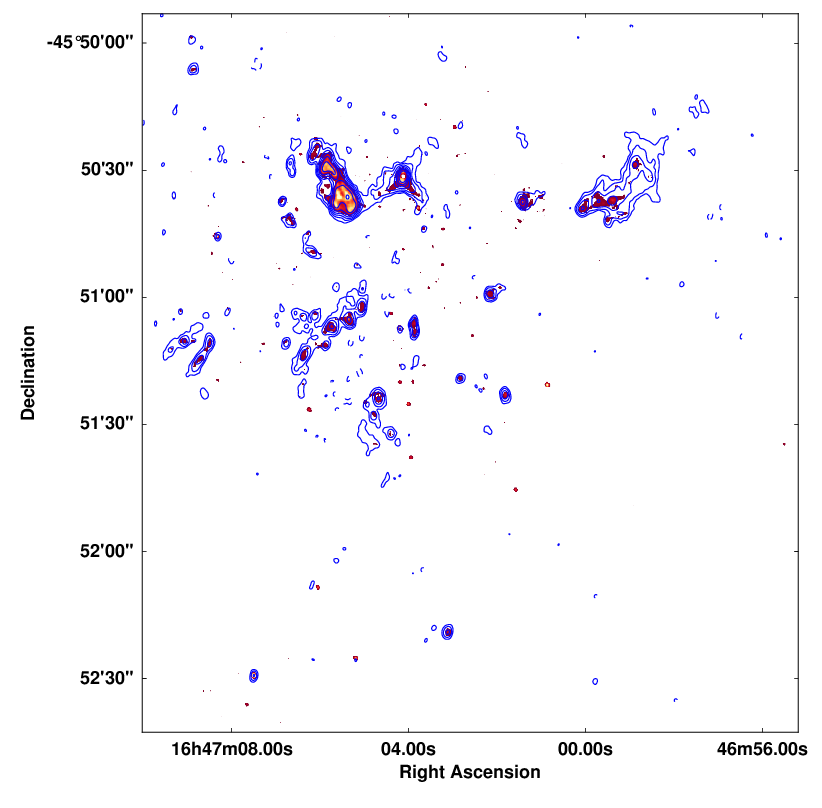}
\caption{ALMA 3-mm colour-scale (from the non-PB corrected image) with overlaid contours from the 8.6\,GHz ATCA observations (Do10). The colour-scale ranges from 0.1 to 2.0\,\mjybm and the radio contours are plotted as those from Fig. 2 Do10, at levels of -3,3,6,8,10,12,18,24,48,96,192 $\times$ 0.06\,\mjybm.}
\label{fig:radioimage}
\end{figure*}

\section{The stellar sources I. The Wolf Rayets}\label{sect:wrstars}

 A key finding of our survey is that almost half of the 50 3-mm continuum sources associated with stellar counterparts within Wd1 are Wolf-Rayet (WR) stars (Table \ref{tab:srcelist}); {\em we detect all such stars within our field-of-view}. This is an important result since it raises the possibility of determining their mass-loss rates. Given the key role of WR-phase mass-loss in stellar evolution, e.g. changes in the mass-loss rates by a factor of 2-3 are sufficient to distinguish between black hole and neutron star formation post-SN \citep{nugis,wellstein}, undertaking such a study for a sample of known distance and for which the properties of the progenitor population may be determined is particularly attractive.

Understandably, given the above, a number of previous studies have investigated the mm- and radio-continuum properties of WR stars. Consequently, in order to place our results in context,  we first briefly review prior findings before discussing our results.

\subsection{Previous mm and radio surveys}\label{sect:wrsurveys}

\citet [][and references therein]{abbott86} \,provide the first comprehensive survey of the radio properties of WR stars, presenting a distance ($\leq$3kpc) and declination  ($\delta > -47^{\rm o}$) limited 4.9-GHz radio survey of 42 of the 43 stars so selected. They report 28 detections, with spectral types  ranging from WN5-8, WC4-9 and a single WN/C star. Complementing these data, 
radio observations of southern hemisphere WRs were obtained by \citet{leitherer95,leitherer97,chapman} who report 16 detections (from 36 targets), with spectral sub-types ranging from WN6h-8h and WC5-8\footnote{We caution  that a number of the WNLh stars within this sample are likely very massive core-H burning objects such as WR24, WR25 and WR89 and hence may not be in the same evolutionary state as other less massive stars such as those found in Wd1.}. \cite{cappa} returned to study WRs north of $\delta > -46^{\rm o}$, detecting 20 of 34 stars and yielding 16 detections of stars missing from the above studies at 8.46\,GHz. Spectral sub-types of detections range from WN5-9h, WC6-9 and a single example of a WN/WCE transitional star. Regarding the nature of sources, \cite{abbott86} reported non-thermal emission for $\sim21$\% of their detections, \cite{cappa} 30\% and \cite{leitherer95,leitherer97,chapman} at least 40\% of their sample. Trivially, while the percentage  of non-thermal emitters at radio wavelengths appears significant,  the differences between these values likely reflect the different methods imposed on the authors by the nature of their respective datasets, making a final fraction difficult to ascertain.

After exclusion of {\em apparent} non-thermal sources, mass-loss rates may be inferred for the remainder of the stars. \cite{willis} re-interpreted the observations of \cite{abbott86} using refined wind terminal velocity and ionisation/abundance values, finding a mean mass-loss rate \.{M} $\sim 5.3\pm2.3 \times 10^{-5}M_{\odot}$ yr$^{-1}$ for the 24 stars considered. \cite{leitherer97} reported a mean \.{M}$\sim4\times10^{-5}M_{\odot}$yr$^{-1}$  for all spectral sub-types with the exception of WC9 stars, for which it appears lower by a factor of $>2$, while \cite{cappa} find  a mean \.{M}$\sim4{\pm3}\times10^{-5}M_{\odot}$yr$^{-1}$ for the WN stars and \.{M}$\sim2{\pm1}\times10^{-5}M_{\odot}$yr$^{-1}$ for the WC8-9 stars.

 Mass-loss rates for individual stars from these samples, broken down by spectral sub-type and assuming $f=1$ (i.e. no wind-clumping), are presented in Fig. \ref{fig:wrmdotsubtype} alongside those from these observations (see Sect. \ref{sect:mdotrates}). Note that a number of {\em confirmed} binary systems\footnote{e.g. WR9, 11, 22,79, 93, 138, 139, 141 and 145} demonstrate apparently thermal radio emission and hence have been included in the plot. In such cases contamination from emission from the WCR (Sect. \ref{sect:specindex}) cannot be excluded, which would lead to an artificially elevated mass-loss rate.

Finally, although not a systematic survey, 42-GHz observations centred on Sgr A$^*$ by \cite{YZ} identified a large number of stars within the Galactic Centre cluster. Given the age of the cluster \citep[6-7\,Myr;][]{martins07} it is expected that these stars will be less massive than those in Wd1, while no cut for non-thermal sources is possible with extant data.  However, given the presence of a large number of the hitherto unrepresented WN9h sub-type we include these values in Fig. \ref{fig:wrmdotsubtype}. In doing so we immediately highlight that the mass-loss rates of the WN9h stars appear systematically lower than those of other spectral sub-types, which we discuss further in Sect. \ref{sect:mdotrates}.

To the best of our knowledge, the only systematic mm continuum survey of WRs was undertaken by \cite{montes}, who report 1.2-mm fluxes for all 17 sources observed. Targets appear to have been selected on the basis of previous radio detections; hence this compilation will automatically be biased towards brighter continuum sources. Combining these detections with radio data revealed that all sources showed a positive (thermal) mm-radio spectral index, even for stars such as WR79a and 105, which \cite{cappa} suggest are possible non-thermal sources. \cite{montes} refrain from determining mass-loss rates from these data and given the substantially larger sample size, we choose to compare mass-loss rated derived for our Wd1 WR cohort to those from radio observations.


\subsection{WRs within Wd1}\label{sect:wrcohort}

\cite{crowther06} provide a census of the WR population of Wd1, finding spectral types spanning WN5 to WN10-11h and WC8-9. We detected all of  the 21 WRs within the  observational field - ranging in sub-type from  the WN5 stars WR J and R through to the WN9-11 stars WR S and Wd1-13 as well as WC stars of sub-types WC8 (WR K) and WC9 (e.g. WR M). Only  WR N, T, W  and X were external to the field. Of the detections, 8 WN  (WR A, B, D, G, L, O, U and Wd1-13) and 5 WC stars (WR C, E, F, H and M) display {\em current} binary signatures (Table \ref{tab:srcelist})\footnote{e.g. spectroscopic radial velocity shifts \citep[e.g.][]{ritchie09a,clark17}, periodic photometric variability \citep{bonanos}, hard and/or over-luminous X-ray emission (Cl08) and/or an IR excess due to emission from hot dust \citep{crowther06}; with the latter two diagnostics indicative of CWBs.}, with the WN10h/BHG star WR S postulated to have evolved via binary channel prior to disruption via a SN \citep{clark14}. We caution that the six WN detections for which no evidence for binarity currently exists (WR I, J, P, Q, R and V) have not been the subject of an RV survey due to a lack of appropriate emission lines that can  function as RV diagnostics. Finally the WC8 star WR K  shows no evidence for RV shifts \citep{ritchie09a}, although this could simply reflect a face-on  inclination or a wide and/or highly eccentric orbit, with the latter also potentially explaining the lack of secondary binary diagnostics, which are known to be transient phenomena in some such systems, in this and other cluster WRs.

\begin{figure}[!htbp]
\begin{minipage}{\linewidth}
\begin{center}
\includegraphics[width=7.5cm]{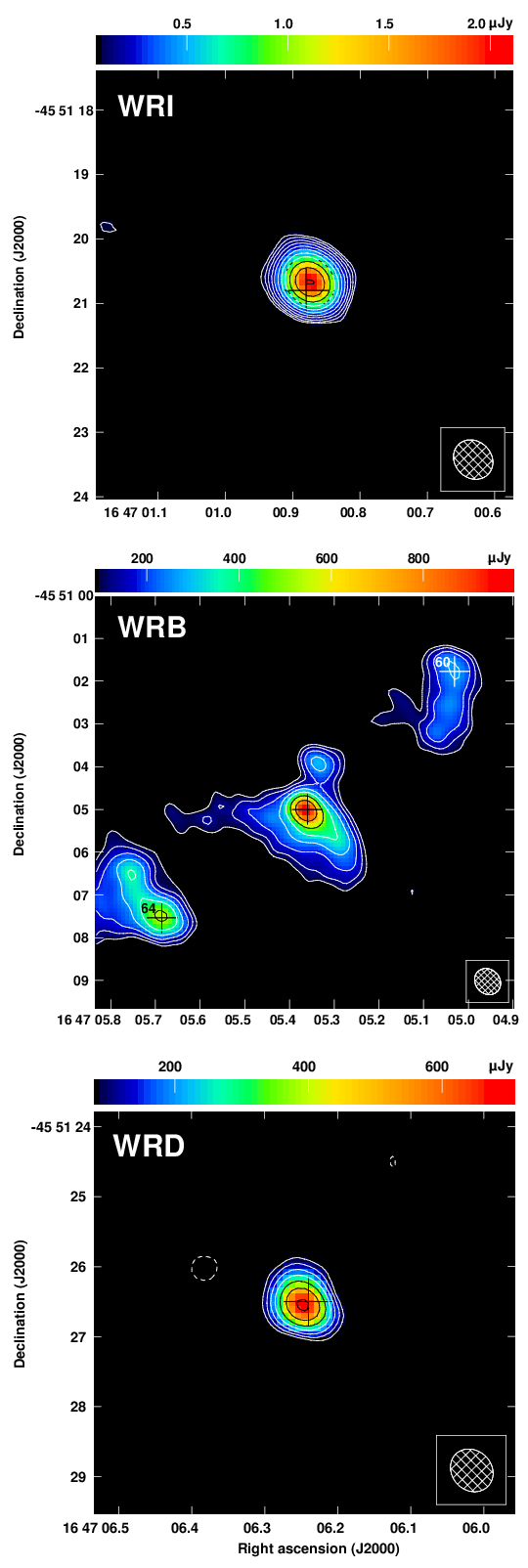}
\captionof{figure}{Wolf-Rayet stars WR C, A and B as seen in the non-primary beam corrected mosaic ALMA image. The contours are plotted at levels of -1,1,1.4,2,2.8,4,5.7,8,11.3,16,22.6,32 $\times$ 3$\sigma$. Where $\sigma$ is 22, 28 and 30\,\mujybm respectively.}
\label{fig:almawr}
\end{center}
\end{minipage}
\end{figure}

\subsubsection{mm-radio spectral indices}\label{sect:mmradioSI}

Employing the 3.6-cm radio observations of Do10 we calculate the two-point mm-radio spectral index, $\alpha$, for each star. Where no flux density is reported for a source in D010, we find a limiting spectral index based on a 3.6\,cm flux density limit of 170\mujybm (see Do10 for details). We first examine those sources with both mm and radio detections since their spectra are most constrained and they provide exemplars for the remaining objects.

WR A (WN7b + OB?, $P_{\rm orb}=7.62$d binary) has a flat radio spectrum ($\alpha_{\rm radio}\sim0.0$) that is best interpreted as a composite of thermal (wind) and non-thermal (WCR) components (Do10), although this is a slightly surprising result given that the short orbital period should place the WCR well within the radio photosphere. The mm-radio spectral index, $\alpha_{\rm mm}\sim0.85$ is unambiguously thermal, suggesting that the stellar wind dominates at mm wavelengths, although this value is slightly steeper than expected for a canonical stellar wind. Following the discussion in Sect. \ref{sect:specindex} this could be the result of a highly structured wind or optically-thin emission from the WCR. The latter would be expected for a short period binary system but would appear difficult to reconcile with the apparent non-thermal emission component at radio wavelengths. 
The canonical $\alpha = 0.6$ spectral index assumes the wind to be at terminal velocity. The spectral index can however be non-linear if the wind is accelerating, with a steeper spectral index at higher frequencies. For WR A (and the other WR stars), the characteristic radius where the free-free optical depth is equal to 1 is $\sim 100\,R_{\star}$) and is therefore significantly beyond the wind acceleration zone for a canonical ($\beta = 1$) velocity law.

WR V (WN8o) likewise has a flat radio spectrum $\alpha_{\rm radio}\sim0.0$) suggestive of a composite thermal+non-thermal origin although, unlike WR A, no corroborative evidence for binarity  exists. As with WR A $\alpha_{\rm mm}\sim0.46$ suggests the increased dominance of the stellar wind at mm-wavelengths.

WR B (WN7b + OB?, $P_{\rm orb}=3.52$d binary) appears comparable to WR A. It too has a flat radio spectrum ($\alpha_{\rm radio}\sim0.04$); however when considering the total flux of the mm source it has a moderately negative spectral index 
($\alpha_{\rm mm}\sim-0.10$). Examination of both radio and mm data shows that it is significantly extended at both wavelengths (Fig. \ref{fig:almawr} and Table \ref{tab:srcesizes}). Following the reasoning in  Do10, we consequently refrain from  interpreting the emission beyond noting that the mm continuum flux is comparable to other cluster WRs, suggesting that it is likely dominated by emission from the stellar wind. Finally we note the twin `lobes' of emission equidistant from the star and apparently aligned upon a single axis (Fig. \ref{fig:almawr}). While this is suggestive of the jet lobes that are sometimes associated with high mass X-ray binaries such as SS433, the double eclipse in the light-curve clearly indicates a normal stellar  companion \citep{bonanos}.

WR L (WN9h +OB?, $P_{\rm orb}=54$d binary) has a radio spectrum ($\alpha_{\rm radio}>0.5$) consistent with the canonical value for a partially optically-thin stellar wind, although current limits would also accommodate an additional contribution from a non-thermal component. The mm-radio spectral index ($\alpha_{\rm mm}\sim0.90$) is directly comparable to that of WR A and so similar conclusions apply here.  

WR F (WC9d +OB?, $P_{\rm orb}=5.05$d binary) has weaker constraints on the radio spectrum ($\alpha_{\rm radio}>0.0$) but a mm-radio spectral index, $\alpha_{\rm mm}\sim0.64$ entirely consistent with emission from a stellar wind. 

WR S (WN10-11h/BHG) is thought to be a single star \citep{clark14} and the radio spectrum ($\alpha_{\rm radio}>0.0$) is consistent with such an hypothesis however, the mm-radio spectral index ($\alpha_{\rm mm}\sim0.22$) is unexpectedly flat for such a scenario. We discuss this below.

The final 15 stars only have lower limits for $\alpha_{\rm mm}$ due to a lack of radio detections (Table \ref{tab:srcelist}). None of these demonstrate spectral indices that imply purely non-thermal emission. Examining the remaining stars by spectral sub-type and all five of the WN5 and  WN6 stars have mm-radio spectra  consistent with thermal emission from a stellar wind, but which could also accommodate contributions from optically-thin thermal or non-thermal emission; in this regard we note that binarity is suggested for WR O, R  and U on the basis of their X-ray properties.  Of the  WN7 stars, WR D, G and P, all appear to have emission dominated by the stellar wind, despite binarity being suggested for WR D and G. 
As with WR L, lower  limits to the  spectral index of WR I (WN8) are higher than expected for a canonical stellar wind, while the WN11h star Wd1-13 replicates the properties of, and conclusions drawn for, the WN5 and WN6 stars. 

Despite it being likely that all four WC9 stars (WR C, E, H and M) are binaries, we find no compelling evidence for a non-thermal emission component in their spectra, with a similar conclusion drawn for the apparently single WC8 star. Unfortunately, the lack of period determinations for the WC9 cohort precludes us commenting on the likelihood of an additional optically-thick continuum contribution from the WCRs \citep[cf.][Sect. \ref{sect:specindex}]{pittard10}.

In summary, we find the mm-radio spectral indices of all the WRs to be consistent with partially optically-thin thermal emission, with no compelling {\em a priori} reason to suspect a significant contribution from a source other than the stellar wind for any star, with the exception of WR V and S. The non-thermally emitting fraction derived from mm observations is therefore clearly lower than derived from extant radio surveys (Sect. \ref{sect:wrsurveys}), which may be expected given the greater flux  expected from the stellar wind at mm wavelengths due to the $\nu^{+0.6}$ dependence.

\begin{table}
\centering
\caption{Mass-loss rates calculated using the observed flux densities at 3\,mm.}
\renewcommand{\arraystretch}{1.1}
\begin{tabular}{|l|c|c|c|c|}
\hline
Source & Spectral type & $T_{eff}$ & $v_{\infty}$ & \mdot$\sqrt{f_{cl}}$ \\
& & kK & (\kms) & (\solmasyr) \\
\hline
WR J    &  WN5h   &  60 & 1500  &   1.62e-05 \\
WR R    &  WN5o   &  60 & 1500  &   2.65e-05 \\
WR O    &  WN6o   &  70 & 1800  &   3.92e-05 \\
WR U    &  WN6o   &  70 & 1800  &   3.41e-05 \\
WR Q    &  WN6o   &  70 & 1800  &   2.73e-05 \\
WR A    &  WN7b+OB?   &  50 & 1300  &   7.19e-05 \\
WR D    &  WN7o   &  50 & 1300  &   2.07e-05 \\
WR B    &  WN7o+OB?   &  50 & 1300  &   3.99e-05 \\
WR G    &  WN7o   &  50 & 1300  &   2.22e-05 \\
WR P    &  WN7o   &  50 & 1300  &   1.92e-05 \\
WR I    &  WN8o   &  45 & 1000  &   3.55e-05 \\
WR V    &  WN8o   &  45 & 1000  &   2.34e-05 \\
WR L    &  WN9h+OB?   &  32 & 700   &   3.82e-05 \\
WR S    &  WN10-11h & 25 & 400  &   5.18e-06 \\
W13     &  WN9-10h  & 28.5 & 500  &   5.52e-06 \\
WR K    &  WC8    &  63 & 1700  &   3.22e-05 \\
WR E    &  WC9    &  45 & 1200  &   5.24e-05 \\
WR F    &  WC9d+OB?   &  45 & 1200  &   7.36e-05 \\
WR C    &  WC9d   &  45 & 1200  &   2.74e-05 \\
WR H    &  WC9d   &  45 & 1200  &   3.29e-05 \\
WR M    &  WC9d   &  45 & 1200  &   3.57e-05 \\
\hline
W25    &   09Iab   & 31.5 & 2100 &  8.25e-06 \\
W43a   &   B0Ia   & 27.5 & 1500  &  5.50e-06 \\
W61a   &   B0.5Ia & 26.5 & 1350  &  3.65e-06 \\
W46a   &   B1 Ia   & 22 & 725   &  3.65e-06 \\
W56a   &   B1.5Ia   & 21 & 500   &  1.45e-06 \\
W52    &   B1.5Ia   & 21 & 500   &  1.50e-06 \\
W8b    &   B1.5Ia   & 21 & 500   &  1.81e-06 \\
W28    &   B2Ia   & 19 & 550   &  2.57e-06 \\
W2a    &   B2Ia   & 19 & 550   &  1.66e-06 \\
W11    &   B2Ia   & 19 & 550   &  1.59e-06 \\
W23a   &   B2Ia+BSG & 18 & 400  &  3.19e-06 \\
W71    &   B2.5Ia   & 19 & 550   &  2.78e-06 \\
W33    &   B5Ia$^{+}$ & 13 & 300  &  2.57e-06 \\
W7     &   B5Ia$^{+}$ & 13 & 300  &  3.64e-06 \\
W42a   &   B9Ia$^{+}$ & 10 & 200  &  1.92e-06 \\
\hline
\end{tabular}
\tablefoot{Adopted parameters are d = 5\,kpc, Z=1, $\gamma$=1, $T_{wind}=0.5\times T_{eff}$. For the WR stars, $\mu$ is taken as 4.0 for WN6 or earlier, 2.0 for later than WN6, 4.7 for WC8 \& WC9. See Sect. \ref{sect:mdotrates} for discussion of the errors.}
\label{tab:masslosstab}
\end{table}

\subsubsection{Mass-loss rates}\label{sect:mdotrates}

\begin{figure}
\begin{center}
\includegraphics[width=6cm,angle=270]{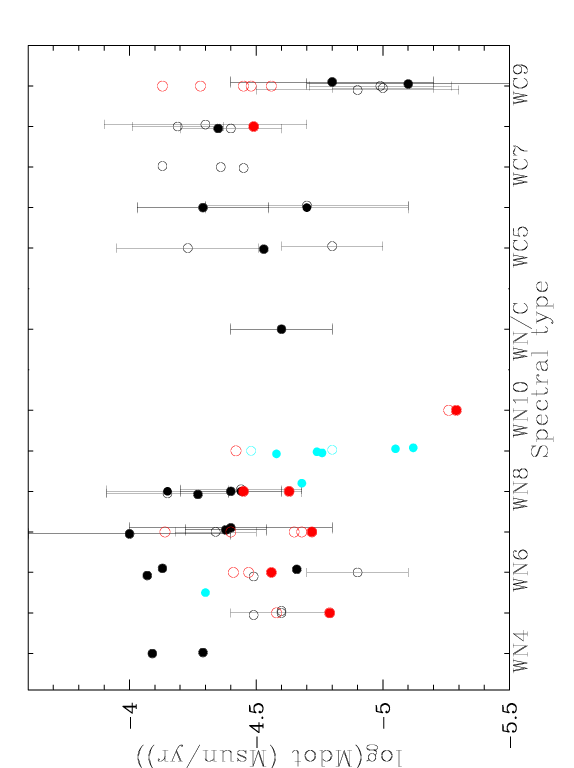}
 \captionof{figure}{Plot of 3-mm mass-loss rate versus spectral sub-type for the WRs within Wd1 (red; open (filled) symbols corresponding to apparent binary (single) stars). Comparable mass-loss rates derived via radio observations of field stars are also presented (symbols in black as before; data from \cite{willis,leitherer97,cappa}). Values for members of the Galactic Centre cluster are given in blue \citep{YZ17}. Error-bars are plotted where given in the source material. 3-mm error-bars for Wd1 stars have not been included for clarity but are estimated to be $\pm0.2$\,dex; please see the text in Sect. \ref{sect:mdotrates} for further details. Note in some instances a slight offset parallel to the x-axis from the spectral-type marker has been applied for reasons of clarity.}
\label{fig:wrmdotsubtype}
\end{center}
\end{figure}

As a consequence of the above discussion we utilise the mm-fluxes in order to infer mass-loss rates for all detections following the methods outlined in Sect. \ref{sect:mdots}. We present the results in Table \ref{tab:masslosstab} and Figs. \ref{fig:wrmdotsubtype} and \ref{fig:OBmdots}, where we also compare our values to previous radio surveys (Sect. \ref{sect:wrsurveys}). While we defer individually tailored non-LTE model-atmosphere analyses for a future work, extant modelling for both WR F and S \citep{clark11,clark14} shows an encouraging consistency in mass-loss rates to within a factor of $\sim2$. Utilising solely the errors in flux density we calculate example ranges for the mass-loss rates for four representative sources. WR O, E, D and U have ranges of $2.90-3.56\times10^{-5}$, $5.09-5.67\times10^{-5}$, $1.89-2.18\times10^{-5}$ and $2.11-2.97\times10^{-5}$\,\solmasyr respectively, implying errors of the order $\pm$0.2\,dex in Fig. \ref{fig:wrmdotsubtype}. The mm mass-loss rates are compared directly with those calculated from radio measurements as published in the literature (see Sect. \ref{sect:wrsurveys}). Utilising directly the radio flux densities and re-calculating the radio mass-loss rates assuming the same chemistry, $\rm{V_{\infty}}$ and  $\rm{T_{eff}}$, as for the mm mass-loss rates, results in a change in the range of $\sim0.05-0.3\times10^{-5}$\,\solmasyr for each object, with the WC mass-loss rates experiencing the smallest shift.

Our survey has a number of advantages over those performed previously. Specifically, our observations detect a much higher percentage of sources (100\% versus 44-65\%; Sect. \ref{sect:wrsurveys}). As a consequence we can be confident that the range of mass-loss reported accurately reflects the underlying distribution, whereas values from previous surveys will be biased due to the non-detection of fainter objects supporting lower mass-loss rates. Moreover, all of our objects are co-located, minimising uncertainties in the distance  estimate that adversely affect surveys of isolated field objects. Additionally, accurate determination of cluster age and  hence progenitor masses for the WRs greatly improves the utility of our mass-loss rates when employed to test theoretical stellar evolutionary predictions and also gives confidence that scatter in the values observed is intrinsic to the stars themselves.

\begin{figure*}[!htbp]
\begin{minipage}{\textwidth}
\begin{center}
\includegraphics[width=18cm]{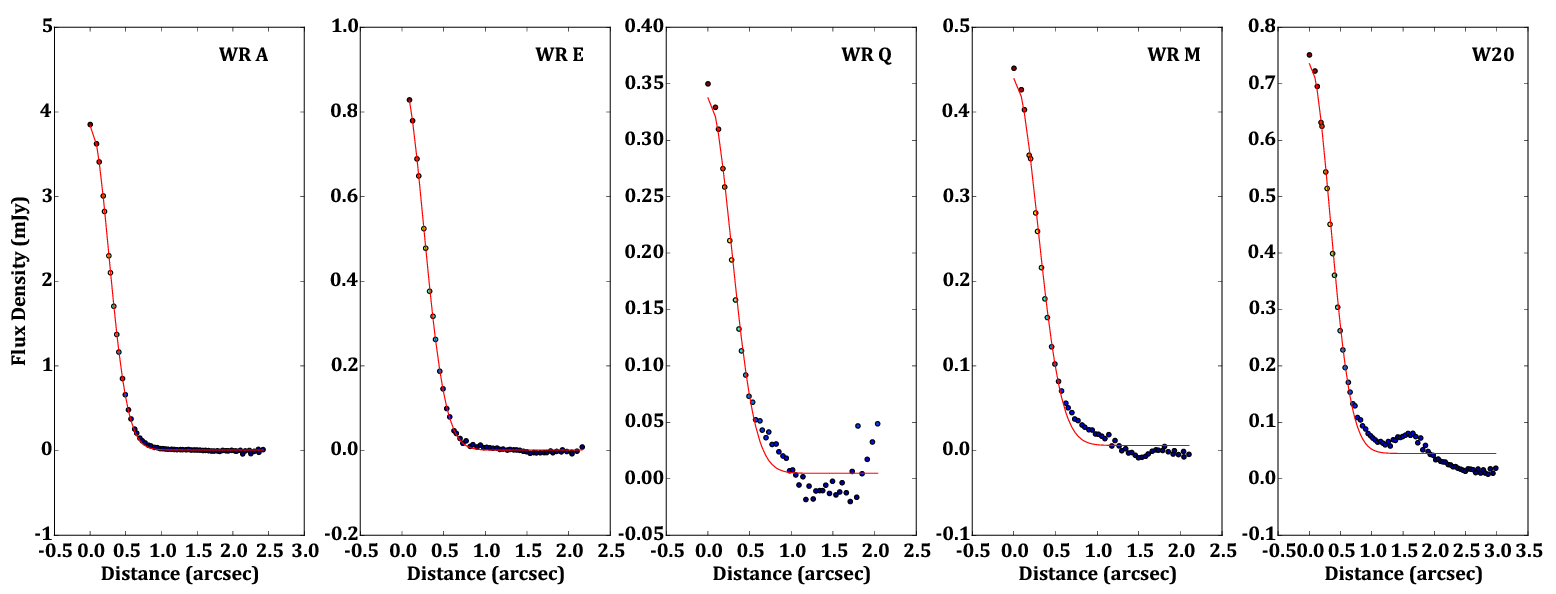}
\caption{Integrated annular profiles for five sources in Wd1. The deviation from a smooth Gaussian profile can be seen for sources WR M and Q, whilst none is visible for sources WR A and E. Wd1-20 shows a clear deviation from a Gaussian courtesy of low surface-brightness emission surrounding the compact core.}
\label{fig:profiles}
\end{center}
\end{minipage}
\end{figure*}

It is immediately obvious that, with the exceptions of the WN10-11h stars WR S and Wd1-13, mass-loss rates for cluster WRs span a limited range of log({\.M})$\,\sim-4.1 \rightarrow -4.8$ or $\sim1.6-7.4\times10^{-5}M_{\odot}$yr$^{-1}$, a scatter which, following the proceeding discussion, is at least in part likely intrinsic. As is apparent from Fig. 5, mean mass-loss rates for the whole ensemble
(\.{M} $\sim 3.6\pm0.4 \times 10^{-5}M_{\odot}$ yr$^{-1}$; excluding WR S and Wd1-13) and both WN  (\.{M} $\sim 3.4\pm0.5 \times 10^{-5}M_{\odot}$ yr$^{-1}$) and WC cohorts (\.{M} $\sim 4.2\pm0.7 \times 10^{-5}M_{\odot}$ yr$^{-1}$) are entirely consistent with previous radio determinations (Sect. \ref{sect:wrsurveys}) as well as values derived from spectroscopic model-atmosphere analysis \citep{hamann,sander}. Strikingly, we find no evidence of the systematically lower mass-loss rate for the WCL stars that had been previously supposed (Sect. \ref{sect:wrsurveys}).

The WNVLH/BHG hybrids WR S and Wd1-13 are (past) members of close binaries for which we hypothesise that binary interaction has resulted in the stripping of their outer layers \citep{ritchie10,clark14}. As such it is not immediately apparent that they are directly comparable to the wider population, appearing intermediate between objects such as the WN9h star WR L and non-cluster early-B hypergiants such as $\zeta^1$ Sco. WR L itself has a mass-loss rate in excess of those reported for the {\em lower-mass} WN9h stars within the Galactic Centre cluster, but which is directly comparable to those of other WRs within Wd1.

Finally in Sect. \ref{sect:specindex} we highlighted the potential blending on WR B with extended emission and the unexpectedly flat mm-radio spectra of WR S and V. Despite this, in each case we find that our mass-loss rate estimate is comparable to those other members of the  same spectral sub-type, supporting  our implicit assumption that the mm-continuum emission in each case remains dominated by the stellar wind.

\subsubsection{Spatially resolved emission}\label{sect:resolvedemission}

As reported in Sect. \ref{sect:res} Gaussian fitting to the sources implies that excluding WR C, J and U the emission associated with the remaining 18 WRs appears spatially extended (Table \ref{tab:srcesizes}; Fig. \ref{fig:almawr})\footnote{No sources were resolved in the surveys summarised in Sect. \ref{sect:wrsurveys} which achieved spatial resolutions of  $\geq 1.2$" for \cite{abbott86},  $>6$" for \cite{cappa} and $>1$" for \cite{leitherer95,leitherer97}.}\footnote{WR B appears qualitatively different from the remaining objects  in terms of source dimensions and hence we do not discuss it further here, since it appears likely that it may simply represent a chance superposition of stellar point source with background optically-thin emission (Sect. \ref{sect:wrcohort}).}.  
As expected the magnitude of the associated errors broadly correlates with source luminosity; as a consequence we cannot comment on any possible relation between source extent and luminosity, WR sub-type or binary status with confidence.  Despite this we emphasise that the errors associated with the more luminous sources are  sufficiently small that the conclusion that a subset of WRs are indeed resolved appears robust, with e.g. WR A, E, F and L resolved at a significance of $\sim5.3\sigma$, $\sim3.6\sigma$, $\sim4\sigma$ and $\sim3.3\sigma$ respectively. 

As described in Sect. \ref{sect:res} an additional emission component, visible as a `shoulder' in the wing of the Gaussian profile, appears present in a number of sources (e.g. WRs J , K and M; see Fig. \ref{fig:profiles}) in a similar manner to the cool super-/hypergiants, although at a lower significance given the fainter nature of the WRs. As a consequence we refrain from quantitative modelling of this feature, although we speculate that these deviations imply a composite structure for these sources, with a bright core  and surrounding low surface-brightness halo.
Pre-empting Sect. \ref{sect:w20}, possible  corroboration of this hypothesis is provided by the structure of the central component of the RSG Wd1-20, which  is sufficiently resolved to directly  observe a similar potential core + halo structure in the image and its corresponding annular profile (also shown in Fig. \ref{fig:profiles}), which displays a similar (albeit more pronounced) deviation from a Gaussian morphology.

\begin{figure*}[!htbp]
\begin{minipage}{\textwidth}
\begin{center}
\includegraphics[width=18cm]{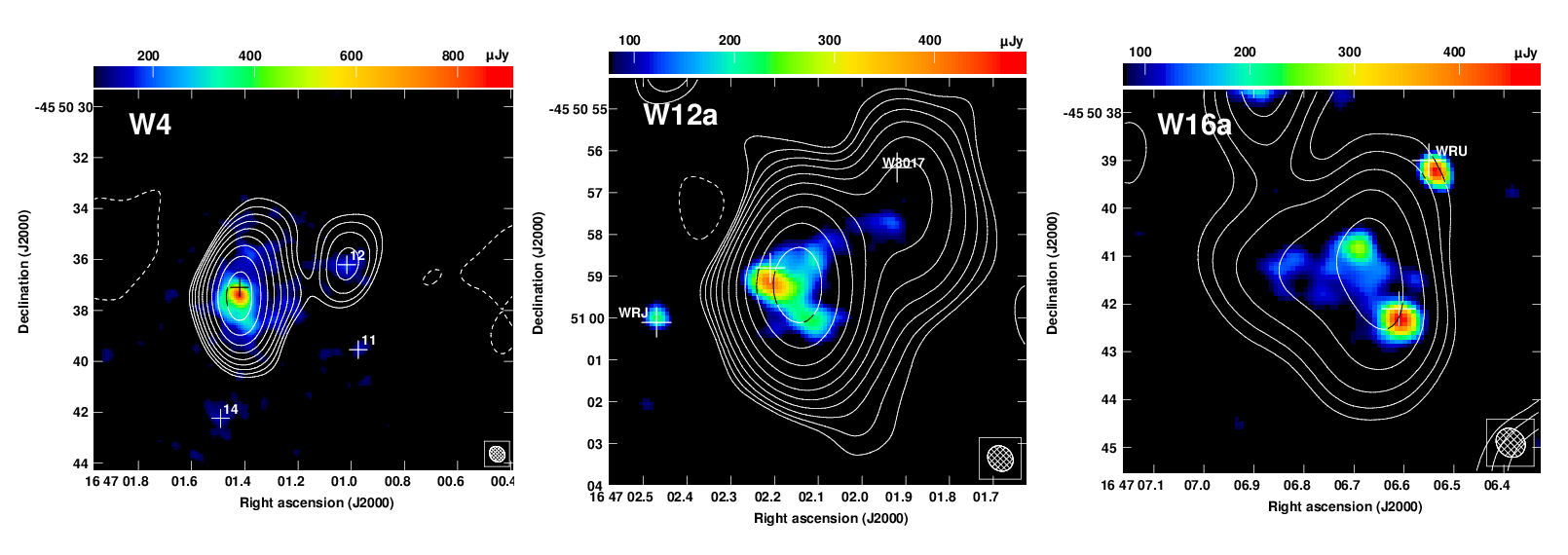}
\captionof{figure}{Yellow hypergiant stars 4a, 12a and 16a as seen by ALMA. The contours are from recent 8.6\,GHz ATCA observations of Wd1 \citep[see][for more details]{andrews18}) and are plotted at -1, 1, 1.41, 2, 2.83, 4, 5.66, 8, 11.31, 16 $\times$ 143, 63 and 242 \mujybm\, respectively.} 
\label{fig:almayhgs}
\end{center}
\end{minipage}
\end{figure*}

Utilising the mm-fluxes and adopted wind and stellar parameters of individual stars (Sect. \ref{sect:mdots})  we may adopt the formalism of \cite{wright75} to infer the radius of the mm-photosphere of each object. These range from 0.1 (WR K) to 3.8 (WR L) milliarcsec; even for WR L this is a factor of $\sim20\times$ smaller than the minor axis (Table \ref{tab:srcesizes}). Comparison of such an analytic estimate to one derived from non-LTE model-atmosphere analysis for WR S \citep{clark14} shows coincidence to within a factor of $\sim2$. As a consequence we may reject the hypothesis that we are resolving the stellar wind in any of these sources. Likewise, while (field) WR stars are commonly associated with wind-blown bubbles, such structures are orders of magnitude larger than  we observe here. We defer further discussion and interpretation of these phenomena until Sect. \ref{sect:discussresn}.

\section{The stellar sources II. The cool super- and hyper-giants}\label{sect:coolsuper}

We next turn to the YHGs and RSGs within Wd1. Such stars are of particular interest since, despite the brevity of the phase, it is thought that they shed mass at sufficient rates to profoundly affect evolutionary pathways.  However current  models \citep[e.g.][]{ekstrom} rely on sparse empirical constraints for mass-loss rates in this phase \citep{dejager}, which often have to be inferred via secondary diagnostics (e.g. via dust emission which then requires the adoption of an uncertain  dust:gas ratio). Wd1 has the potential to be a critical test-bed for understanding mass-loss from stars for three reasons. Firstly it is sufficiently massive that it contains the richest population of RSGs and YHGs known in the Galaxy; moreover we find it at an age ($\sim5$\,Myr) at which such stars first occur and, as a consequence, these are also likely to be the most extreme objects ($M_{initial}\sim35-40\,M_{\odot}$, log$(L/L_{\odot})\sim5.8$) permitted by nature at solar metalicities \citep{ekstrom}.

Radio observations by  Do10 reveal the presence of ionised circumstellar gas around a number of cluster members, with recent, deeper, observations detecting and resolving such ejecta around nine of the ten YHGs and RSGs (missing only Wd-8; Andrews et al. submitted).  Given their photospheric temperatures are insufficient to ionise this material  this leaves the presence of a hot companion or the diffuse cluster radiation field as possible physical agents (cf. discussion in Do10). More importantly this opens the possibility of {\em directly} determining time averaged mass-loss rates for such systems. 
Being more sensitive to extended structure, the radio observations of Do10 provide more accurate constraints on the masses of the  circumstellar nebulae  while, with their enhanced spatial resolution, the ALMA observations provide complementary information on the geometry; both prerequisites if mass-loss estimates are to be determined. Unfortunately, we cannot easily combine both datasets to obtain spectral index information, but Do10 conclude that the {\em spatially extended} emission associated with each star is consistent with an optically-thin thermal origin.

\subsection{The yellow hypergiants}\label{sect:yhg}

We detected and resolved four of the five cluster YHGs within the field of view of our ALMA observations - Wd1-4, 12a, 16a and 32 (Table \ref{tab:srcelist} and Figs. \ref{fig:w26halpha} \& \ref{fig:stargroups}), with Wd1-8 remaining undetected at both mm and radio wavelengths. 

The nebulae associated with Wd1-4 (F2 Ia$^+$), Wd1-12a (A5 Ia$^+$) and W16a (A2 Ia$^+$) are clearly asymmetric at 3-mm, comprising  a dominant (and resolved) point-like source embedded at the apex of more extended, trailing nebulosity, with an arc- or arrow-head like morphology (Figs. \ref{fig:almayhgs}, \label{fig:stargroups} and Table \ref{tab:srcesizes}). These sources are in turn located at the head of elongated radio nebula of greater extent, with an overall cometary morphology comprising a resolved, compact nucleus coincident with the YHG and an  extended tail. While the nebulosity associated with Wd1-32 is too compact to reveal such a configuration, the radio continuum observations of Wd1-265 by \cite{andrews18} show that it, too, is associated with a spectacular cometary nebula. While such behaviour has never before been observed for YHGs it is reminiscent of the nebulae associated with the RSGs 
NML Cyg and GC IRS7 (with the latter being more elongated), which are thought to be shaped via an interaction with the nearby Cyg OB2 association and Galactic Centre cluster respectively \citep{schuster,serabyn,YZ}. Given that the cometary tails of the Wd1 YHGs are also all orientated away from the cluster core we suggest that a similar physical process is in operation here; we return to this below.

What is the origin of the circumstellar material? In interpreting both radio and ALMA data it is important to recognise that individual cluster YHGs may be in either a pre- or post-RSG  phase and hence may have  very different physical properties (e.g. $L/M$ ratio, \.{M} etc.). Nevertheless the comparable angular sizes and fluxes (Tables \ref{tab:srcelist}, \ref{tab:srcesizes}) of the extended emission components and, where available, the radio-determined nebular masses\footnote{$\sim4.5\times10^{-3}M_{\odot}$ for Wd1-4a, $\sim6.1\times10^{-3}M_{\odot}$ for Wd1-12a and $\sim6.4\times10^{-3}M_{\odot}$ for Wd1-265 when scaled to 5\,kpc (Do10).} point to a common physical origin. 
Ejection nebulae have been associated with the post-RSG YHGs IRC +10 420 \citep{tiffany,shenoy} and IRAS 17163-3907 \citep{lagadec,huts,wallstrom} but are physically more extended and contain orders-of-magnitude more mass than those considered here. Likewise, while the YHGs within Wd1 exhibit pulsations \citep{clark10}, none have (yet) demonstrated the long-term secular evolution or giant eruptions that characterise post-RSG examples such as $\rho$ Cas, IRC +10 420 and HR8752 \citep{lobel,nieu,oudmaijer96,oudmaijer98}. Finally, none show the rich emission spectra of both IRC +10 420 and IRAS 18357-0604, which appear to be caused by extreme post-RSG mass-loss rates \citep{oudmaijer98,clark14a}; instead they share a continuous morphological sequence with the mid-late BHGs within Wd1 \citep{clark05} which we argue are in a pre-RSG phase (Sect. \ref{sect:rsg}).

Given the above, if the majority of Wd1 YHGs are indeed in a pre-RSG phase and therefore are not yet prone to the large scale instabilities that characterise post-RSG stars, then the circumstellar material appears most likely to result from the accumulation of a quiescent stellar wind. Adopting a wind velocity of  $\sim200$kms$^{-1}$, we infer a wind-mediated mass-loss rate of \.{M} $\sim 10^{-5}M_{\odot}$ yr$^{-1}$ for the thermal component of Wd1-4a (when scaled to 5kpc; Do10). For a freely expanding wind this would correspond to a dynamical age of $\sim10^2$yr for the nebulae and hence an accumulation of $\sim10^{-3}M_{\odot}$ over this period. Given the nebular morphologies of Wd1-4, 12a 16a and 265 clearly suggest wind confinement in the hemisphere nearest the cluster core this is better interpreted as  a lower limit to nebular mass; indeed estimates from Do10 and \cite{andrews18} are in excess of this value.

\subsection{The red supergiants}\label{sect:rsg}

\begin{figure}[h]
\begin{minipage}{\linewidth}
\begin{center}
\includegraphics[width=0.8\linewidth]{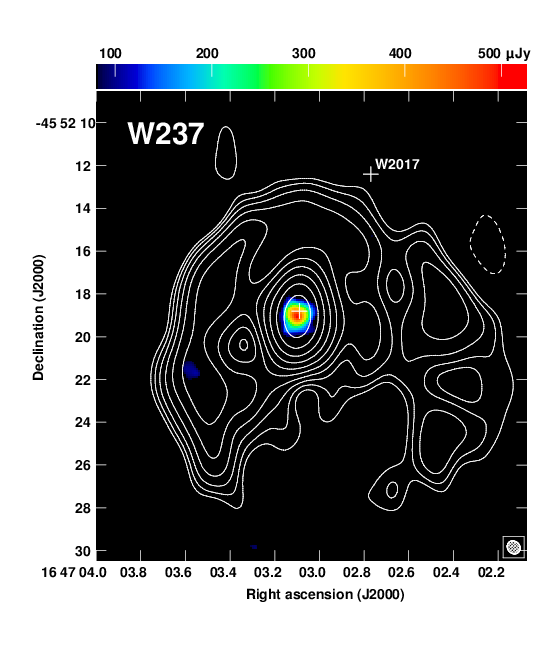}
\captionof{figure}{ALMA image of the red supergiant Wd1-237. The contours are from recent 8.6\,GHz ATCA observations of Wd1 \citep[see][for more details]{andrews18}) and are plotted at -1, 1, 1.41, 2, 2.83, 4, 5.66, 8, 11.31, 16 $\times$ 58 \mujybm.} 
\label{fig:w237}
\end{center}
\end{minipage}
\end{figure}

\begin{figure}
\begin{minipage}{\linewidth}
\begin{center}
\includegraphics[width=\linewidth]{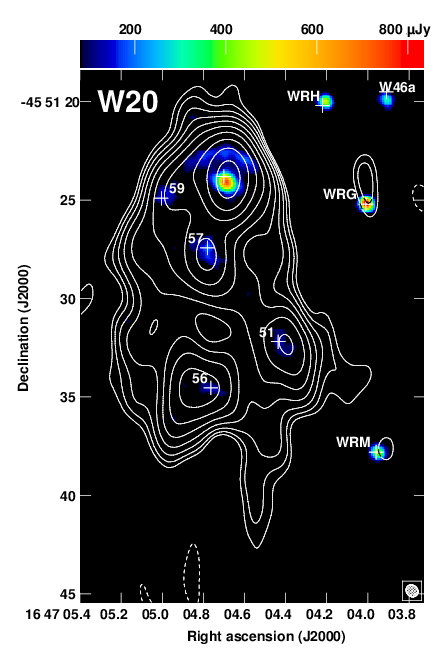}
\captionof{figure}{ALMA image of the red supergiant star Wd1-20 shown in colour-scale. Over-plotted are the 8.6\,GHz radio contours from recent ATCA observations \citep[see][for further details]{andrews18} at levels of -1, 1, 1.41, 2, 2.83, 4, 5.66, 8, 11.31, 16 $\times$ 91\,\mujybm.}
\label{fig:w20}
\end{center}
\end{minipage}
\end{figure}

\begin{figure}
\begin{center}
\includegraphics[width=\linewidth]{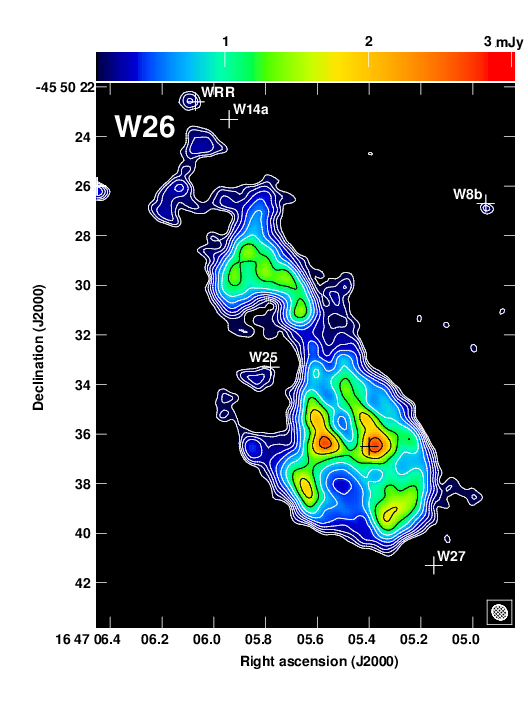}
\captionof{figure}{ALMA image of the red supergiant star Wd1-26 shown in colour-scale. Contours are plotted at -1, 1, 1.41, 2, 2.83, 4, 5.66, 8, 11.31, 16 $\times$ 3$\sigma$ where $\sigma$ is 64 \mujybm. See Appendix B for further images of W26 comparing the mm emission with that seen in the 8.6\,GHz radio emission from Do10 and the H$\alpha$ images presented by \cite{wright}.}
\label{fig:w26}
\end{center}
\end{figure}

The ALMA observations resolved all four RSGs within Wd1, including Wd1-75 for the first time (Figs \ref{fig:w237} - \ref{fig:w26} and Tables \ref{tab:srcelist}, \ref{tab:srcesizes}). To within observational uncertainties, all  extended nebular components have radio spectral indices  consistent with optically-thin thermal emission (Do10). Intriguingly,  their nebular masses are larger than inferred for the YHGs \citep[Do10;][]{andrews18}, suggesting either a longer duration for their mass loss or that a differing physical process may be responsible for their formation. Determination of current mass-loss rates via mid-IR  fluxes is precluded by the saturation of the stars in all available flux-calibrated datasets\footnote{Similarly, their brightness and clear spectral variability \citep[cf.][]{clark10} has, to date, prevented determination of their underlying stellar  parameters; we note that the absolute bolometric luminosities inferred by \cite{fok} appear inconsistent with cluster properties, in terms of both the range spanned and also  the extremely high(low) individual luminosities inferred for individual  objects.}. Nevertheless the detection of  SiO and H$_2$O maser emission associated with both Wd1-26 and -237 argues for very high mass-loss rates which are thought to occur during the latter stages of the RSG phase, when stars have the most extreme $L/M$ ratio \citep{davies,fok}.

Observations of (predominantly lower-luminosity) field RSGs show that, when present, their circumstellar nebulae show a diverse set of properties. Few determinations of  nebular masses are present in the literature, but the nebulae associated with Wd1-20, Wd1-26 and Wd1-237 appear broadly comparable to that surrounding the extreme RSG VY CMa, a star for which transient mass-loss rates in excess of $\sim10^{-3}M_{\odot}$yr$^{-1}$ have been inferred \citep[][]{shenoy,smith01,smith09}. Nebular geometries range from quasi-spherical through to elongated \citep[VX Sgr and S Per respectively;][]{schuster}, highly aspherical and clumpy \citep[(VY CMa;][]{smith01} and cometary \citep[GC IRS7;][]{serabyn,YZ}. A number show evidence for interaction with their immediate environment \citep[e.g. GC IRS7 and Betelgeuse;][]{noriega}, while others provide evidence for time-variable mass-loss rates \citep[e.g. VY CMa and $\mu$ Cep;][]{shenoy}. Given the physical information encoded in the nebular geometries we chose to  group and discuss the cluster RSGs on this basis.

\subsubsection{Wd1-75 and Wd1-237}\label{sect:w237}

At mm-wavelengths Wd1-75 presents as a compact, elongated source (Table \ref{tab:srcesizes} and Figs. \ref{fig:w26halpha} \& \ref{fig:stargroups}). In terms of spatial  extent it is similar to the approximately spherical nebula detected around $\mu$ Cep in the mid-IR by \cite{dewit}\footnote{To within a factor of $\sim3-4$, subject to the uncertainties in distance to $\mu$ Cep.}, although additional structure is observed at both larger and smaller radii at different wavelengths around the latter star \citep{schuster,shenoy,cox}. The new radio continuum observations of Andrews et al. (submitted) show that this emission is embedded within a more extended nebula; however given the compact nature of the nebulosity at both wavelengths no conclusions  regarding morphology may be drawn. 

Located in the southern extremities of the cluster, at mm-wavelengths Wd1-237 appears similar to Wd1-75, albeit with a slightly greater extent (Table \ref{tab:srcesizes}). However radio observations suggest a much more extended circumstellar nebula enveloping this structure ($\sim11.2\times8.5$\,arcsec; Do10). The observations of \citet[][also Fig. \ref{fig:w237} here]{andrews18} clearly resolve this emission, revealing a compact central nebula co-incident with the ALMA source and star itself, which is offset from the centre of a larger, quasi-spherical nebula which appears brighter on the hemisphere facing the cluster core. This morphology is strikingly reminiscent of the mid-IR nebula associated with $\mu$ Cep \citep{shenoy}. As with the YHGs this configuration immediately suggests interaction with the cluster proper. Such nested configurations have been associated with other RSGs (cf. preceding discussion in Sect. \ref{sect:rsg}) and have historically been interpreted as arising from variations in the  mass-loss rate of the star. While this is an obvious hypothesis for Wd1-237 it is not the {\em only} explanation;  an issue we return to below.

 Given the apparent sculpting  of the nebula by the cluster, a robust determination of the  dynamical age will require detailed hydrodynamical simulations. Nevertheless simply taking the  displacement of  the central core from the outer arc of emission facing the cluster core ($\sim0.12$\,pc for a distance of 5\,kpc) and an outflow velocity of 10\,kms$^{-1}$  suggests a minimum age of $\sim10^4$\,yr, which in turn would imply a time-averaged mass loss rate of $\sim10^{-5}\,M_{\odot}$yr$^{-1}$ via the nebular mass \citep[$\sim0.07\,M_{\odot}$; Do10,][]{andrews18}. We regard these as at best  order of magnitude estimates given uncertainties in nebular age, outflow velocity and the possibility of an additional neutral component to the nebula which would be invisible to current observations.

\subsubsection{Wd1-20 and Wd1-26}\label{sect:w20}

Radio observations of the two remaining cluster RSGs have already revealed them to be associated with  pronounced cometary nebula (Do10). Such linearly extended structures have previously only been associated with the Galactic Centre star GC IRS7, although bow shocks have been detected around a further three RSGs; Betelgeuse \citep{noriega}, $\mu$ Cep \citep[e.g.][]{cox} and IRC -10414 \citep{gvar}. Clearly however, they bear close resemblance to the nebulae of the cluster YHGs, while that associated with Wd1-237 appears to be a less collimated analogue.

 We first turn to Wd1-26. Analysis of optical spectroscopy reveals the gaseous element of the nebula to be composed of N-enriched material \citep{mackey15}, while mid-IR imaging reveals a co-spatial dusty component \citep{clark98}. The H$\alpha$ image presented  by \cite{wright}; over-plotted on ALMA data in Fig. \ref{fig:w26halpha}) appears to show Wd1-26 at the centre of a clumpy ring nebula. A second, triangular nebula is seen to the north-east, with both structures potentially connected by a thin, faint bridge of emission. Our ALMA observations clearly replicate this latter feature, implying a physical connection between both components. While the southern  nebular component is obviously clumpy, the higher spatial resolution afforded  by ALMA (and the lack of contamination by background stars that afflicts the H$\alpha$ image) suggests  that Wd1-26 does not sit at the centre of a ring nebula. Rather it is located  at the apex of one of two bright elongated blobs of  emission that run parallel to one another. These  are also aligned with the major axis of the  nebula, which  in turn is directed towards  the cluster core. Leading these two sub-components are a further two emission hot-spots, with all four components in turn embedded  within more extended lower surface-brightness emission. Comparison to the 2-cm radio image of GC IRS7 \citep{YZ} reveals a  striking  similarity, with both nebulae demonstrating `comet-like' morphologies of comparable sizes ($\sim0.38$\,pc for Wd1-26 versus $\sim1$\,pc for GC IRS 7), with significant substructure visible in both (by analogy) `coma' and `tail' components.

As with Wd1-26, radio observations of Wd1-20 reveal the nebular tail to be orientated away from the core  of the cluster and, at shorter wavelengths, it too appears to show similar substructure. Likewise both mass  and spatial extent \citep[Table \ref{tab:srcesizes}; Do10 and][]{andrews18} are comparable. While the nebular tail is not detected as a single coherent structure in the ALMA data, isolated hot-spots coincident with features in the radio data are seen; however these shortcomings are compensated for by the detailed view afforded of the immediate environment of the RSG itself (Fig. \ref{fig:w20}). Specifically a compact aspherical nebula ($\sim0.02 \times \sim0.01$\,pc; Table \ref{tab:srcesizes}) is coincident with the star. To the north of this is a further arc of emission $\sim0.14$\,pc in extent with a projected separation between the apex of this structure and the RSG of $\sim0.036$\,pc. We note that this feature appears absent in the nebula surrounding Wd1-26, it is not clear whether this corresponds to a real physical difference or instead is due to differing orientations/lines-of-sight through otherwise identical nebulae.

 An obvious interpretation for the arc of emission is that it is a bow-shock, with the nebular morphologies of both Wd1-20 and Wd1-26 being shaped by interaction with the wider cluster environment\footnote{While the distance between star and bow-shock for Betelgeuse \citep[0.8pc for a distance of 400\,pc;][]{noriega}, $\mu$ Cep \citep[0.15\,pc;][]{cox} and IRC -10414 \citep[$\sim0.14$\,pc;][]{gvar} are significantly larger than observed for Wd1-20, the cluster environment of the latter object is much more extreme than that experienced by any of the three former objects and so one would not expect equivalence between them.}. Potential  physical agents for this include (i) {\em relative} motion of the RSGs through the intra-cluster medium/wind  \citep[detected at both radio and X-ray wavelengths; Do10,][]{muno06b}, (ii) photoionisation by the hot massive stellar cohort and/or (iii) interaction with the stellar winds of individual cluster members and/or a recent supernova(e). 

Of these,  motion of the stars relative to the cluster as a whole {\em may} be disfavoured by  the observation that  the cometary nebulae associated with both RSGs and YHGs are all orientated {\em away}  from the core region which, under such a scenario, would require all four objects to be falling towards this region. Likewise one might anticipate that velocities significantly higher than the cluster virial velocity would be required for bow shocks to form via interaction with a static intra-cluster medium; essentially all such objects  would be `run-towards' rather than the more normal  `runaways'. As a consequence we favour an interaction between the star and a dynamic stellar/cluster wind(s) as the most likely causal agent for the cometary nebulae.

\cite{mackey14,mackey15} investigated the shaping of RSG winds by external agents providing, respectively, detailed simulations of the  bow-shock around Betelgeuse and the nebula of Wd1-26 . The latter study suggests ram confinement of the stellar wind by the nearby sgB[e] star Wd1-9 as a potential physical origin  for the cometary morphology. More sophisticated  simulations (e.g. incorporating more detailed models of the cluster environment) would be invaluable in order to (i) further test this hypothesis, (ii) constrain the mass-loss rates Wd1-26 (and Wd1-20) and (iii) determine whether the `clumpy'  sub-structure is a result of aspherical mass-loss \citep[cf. VY CMa][]{smith01} or post-ejection hydrodynamical instabilities \citep[cf.][]{cox}. 

Nevertheless, one important insight arising from these simulations is the prediction of a massive static shell {\em interior} to the bow-shock. This forms via the stalling of the neutral RSG wind as a result of external photoionisation and as a consequence of this the accumulation of mass lost by the star. In this regard the observation of such  compact nebulae surrounding both Wd1-20 and -237 is of particular interest.

\section{The stellar sources III. OB stars -  supergiants, hypergiants, LBVs and sgB[e] stars}\label{sect:stellar3}

The last cohort we turn to comprise the hot, early-type stars within Wd1; a somewhat more heterogeneous grouping than those previously considered. We detect a total of 21 mm-continuum sources coincident with cluster members (Tables \ref{tab:srcelist}-\ref{tab:srcesizes} and Figs.\ref{fig:w26halpha} \& \ref{fig:stargroups}). Subject to the classification of two objects  (D09-R1 \& R2)\footnote{Both stars have been classified as generic OB  supergiants on the basis of their photometric magnitudes and colours.} detections range from O9 Iab (Wd1-25) to B9 Ia$^+$ (Wd1-42a) plus  the sgB[e] star Wd1-9 and the LBV Wd1-243. Despite their presence in large numbers, no stars of lower luminosity class were detected (Clark et al. in prep.). In this regard we note that based on radio data (Do10), the O9 Ib star  Wd1-15 should have been detected but was not, implying that it is variable - a potential signature of non-thermal emission. 

Three further stars - D09-R1 and R2 (OB SGs) and Wd1-17 (O9 Iab) - have radio properties suggestive of non-thermal emission (Do10). Our  mm-data are consistent with such a conclusion for Wd1-17 but imply flat or moderately positive mm-radio spectra for D09-R1 and R2 (Table \ref{tab:srcelist}). However, the  radio and mm  point sources associated with each star are further coincident with extended continuum emission; consequently  determination of absolute fluxes and hence the physical nature of the emission is uncertain.  While we currently favour non thermal emission for D09-R2 and Wd1-17 - and hence identification, along with Wd1-15, as CWBs - no corroborative evidence for such a classification is available at optical or X-ray wavelengths \citep[Table \ref{tab:srcelist}; Cl08,][]{ritchie17}. 

Of the remaining objects the emission associated with the LBV Wd1-243 and the central component of Wd1-9 is entirely consistent with arising in a partially optically thick stellar winds. Lower limits to the radio-mm spectral indices for Wd1-21, 8b, 11, 52, 56a and 61a do not constrain the emission mechanism, while those for Wd1-7, 23, 25, 28, 30, 33, 42a, 43a, 46a and 71 are consistent with partially optically thick thermal emission, though the presence of an additional optically thin and/or non-thermal component may not be excluded. For the purposes of this paper we proceed under the assumption that these are indeed thermal wind sources for which mass-loss rates may be determined (Table \ref{tab:masslosstab} \& Fig. \ref{fig:OBmdots}).

Finally, we find a number of sources, including both super- and hypergiants, to be partially resolved (Table \ref{tab:srcesizes}), but with an extent greater than expected on the basis of mass-loss rates derived below; an exemplar being the LBV Wd1-243 ($0.15\pm0.01$\,arcsec); we discuss this phenomenon in Sect. \ref{sect:discussresn}.

\subsection{The OB super- and hypergiants}\label{OBstars}

Post-MS late-O and B stars within Wd1 follow a smooth progression in spectral morphologies leading to a corresponding evolutionary passage from O9 III through O9-9.5 Iab to B0-4 Ia \citep{iggy10,clark15,clark17}. The status of the B5-9 Ia$^+$ stars Wd1-7, 33 and 42a is less  clear-cut since  such classifications have also been applied to post-RSG objects such as HD168607 and HD168625. The simulations of \cite{groh} and \cite{martins17}, suggest massive ($>40\,M_{\odot}$) late O/early B supergiants evolve directly into cooler hypergiants \citep[consistent with the empirical results of ][]{clark12}). Alongside their close spectral similarity to the cluster B supergiants this does however argue for these objects to be a direct extension of this O9III - B0-4 evolutionary sequence, and consequently for all these stars to be in a pre-RSG phase. 

 Accurate  classifications are unavailable for D09-R1 and -R2 and, while the spectral variability and X-ray properties of Wd1-30a reveal that it is a massive O+O star interacting binary, determination of physical parameters is non trivial \citep{clark17}. Following the discussion above both Wd1-15 and Wd1-17 appear likely non-thermal colliding wind sources. No observational constraints on the binary properties of either system are available and consequently it is not possible to disentangle contributions from the components of either system (stars and wind collision zone). Excluding these stars, as well as the LBV Wd1-243 and the sgB[e] star Wd1-9, which we discuss later, leaves a total of 15 cluster members with modern classifications for which mass-loss rate determinations may be made. 


\begin{figure}
\begin{center}
\includegraphics[width=6cm,angle=270]{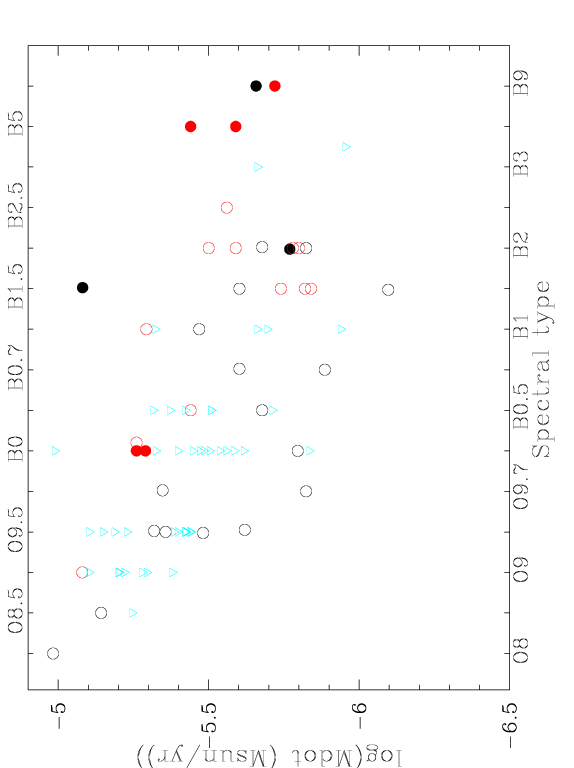}
\captionof{figure}{Plot of mass-loss rate versus spectral type for the population of OB super- (red, open symbols) and hyper-giants (red, filled symbols) within Wd1. Upper limits are given by the inverted blue triangles. Comparable mass-loss rates derived via radio observations of field stars are also presented \citep[symbols in black as before; data from][]{benaglia,leitherer95}. With a (variable) mass-loss rate of log({\.M}$/M_{\odot}(yr^{-1}$))$\sim$-4.89 $\rightarrow$ -4.74, Cyg OB2 \#12 (B3-4 Ia$^+$) lies outside this plot, while the isolated point corresponds to the highly luminous BHG $\zeta^{1}$ Sco. Error bars are not included for clarity, however representative errors are 0.1-0.2\,dex for this work, and around 0.2\,dex for values taken from \cite{benaglia} and \cite{leitherer95}.}
\label{fig:OBmdots}
\end{center}
\end{figure}

\begin{figure*}
\begin{minipage}{\textwidth}
\begin{center}
\includegraphics[width=\linewidth]{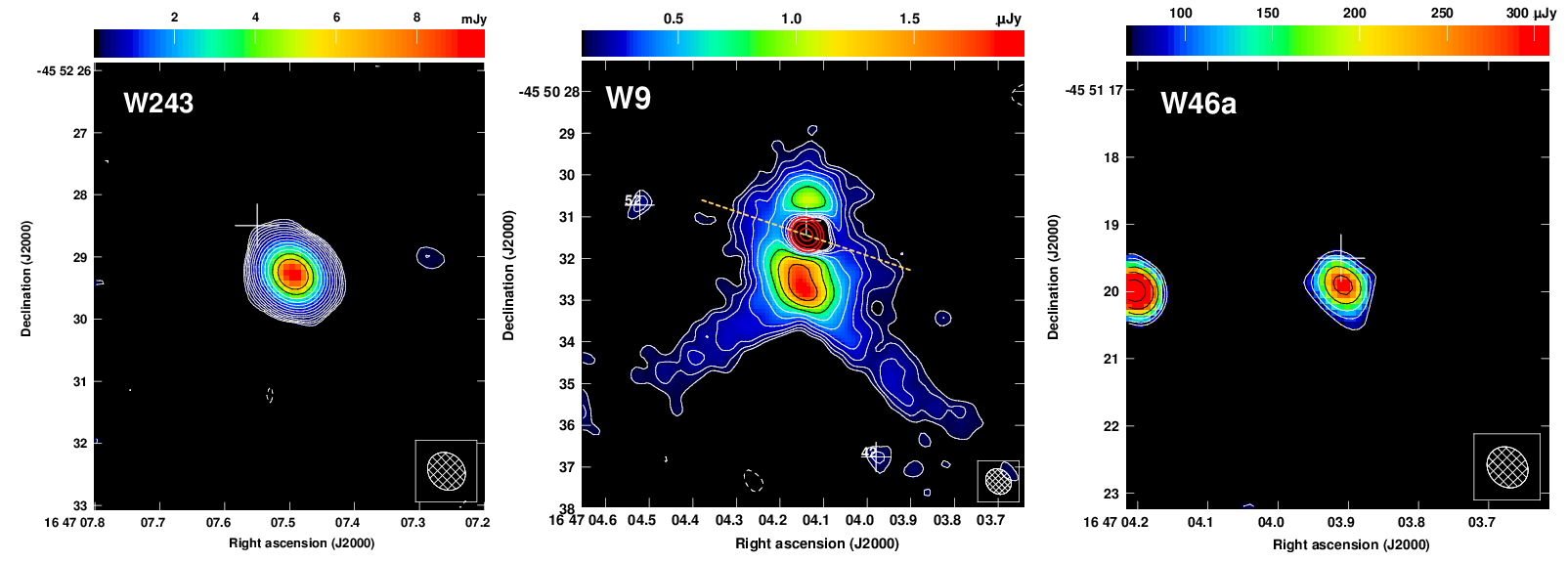}
\captionof{figure}{ALMA images of the LBV Wd1-243 (left), the SgB[e] star Wd1-9 (middle) and the B1 supergiant Wd1-46a (right). For Wd1-243 and Wd1-46a contours are plotted at -1, 1, 1.41, 2, 2.83, 4, 5.66, 8, 11.31, 16 $\times$ 3$\sigma$ where $\sigma$ is 30\,\mujybm. For Wd1-9 the larger white contours are plotted as for  Wd1-243 and -46a with a $\sigma$ of 33\,\mujybm. The red contours show the bright compact source emission and are plotted at levels of 0.1,0.14,0.2,0.28,0.4,0.56,0.8 $\times$ the peak flux density of 411\,mJy. The dashed line bi-secting the central compact source in Wd1-9 reflects the orientation of the apparent bipolar outflow reported in \cite{fenech}. }
\label{fig:w243}
\end{center}
\end{minipage}
\end{figure*}

Of this subset, three stars, Wd1-23a (B2 Ia + B Ia?), 43a (B0 Ia + ?) and 52 (B1.5 Ia + ?) \citep{iggy10}, show compelling evidence for binarity\footnote{RV shifts (Wd1-2a, 7, 33, 42 and 71) and/or variability in H$\alpha$ (Wd1-7, 23, 33, 28, 42, 61 and 71) present in other stars are instead interpreted as originating in pulsations and wind variability respectively \citep{clark10} while none have X-ray properties indicative of CWBs (Cl08).}. Conversely, a number of known and suspected massive OB binaries are not detected within the cluster \citep[e.g. Wd1-27 and 53a;][]{bonanos,clark98} suggesting that their properties do not {\em uniformly} favour discovery at mm-wavelengths (cf. Sect. \ref{sect:specindex}). Indeed, the mass-loss rate inferred for Wd1-52 is entirely comparable to other (apparently single) B1.5Ia stars and those of Wd1-23a and 43a to the wider supergiant cohort and $\sim0.6$\,dex lower than the apparent CWB Wd1-17 (O9.5Ia) {\em if} we were to make the questionable assumption  that its 3mm flux derives solely from wind emission. 

Mass-loss rate estimates for this subset are given in Table \ref{tab:masslosstab}, with upper-limits to the continuum-flux and mass-loss rates for non-detections in Table \ref{tab:obupperlist}; these are presented graphically in Fig. 12\footnote{Unfortunately, the heterogeneous S/N ratio across the field precludes consideration of our results as a flux-limited survey of OB stars within Wd1.}. We supplement these data with the inclusion of the WNVLh/BHG hybrids Wd1-5 and -13\footnote{Consideration of the temperature of Wd1-5 \citep{clark14} leads to indicative spectral classifications of B0 Ia$^+$ for these objects; while these are likely to have considerable uncertainty they are sufficient to illustrate their relation to the other cluster members considered here.}. 

The main finding of our study is the preference for detecting super-/hyper-giants  of spectral type B1.5 and lower (11 from 13 including the B1-1.5 Ia star Wd1-56a) versus earlier spectral types (4 from $>60$ stars), with no cluster O giants detected. Mass-loss rates vary from $\sim 1.5-4.0\times10^{-6}M_{\odot}$yr$^{-1}$ for the B1.5-9 super-/hypergiants and $\sim 4.0-8.3\times10^{-5}M_{\odot}$yr$^{-1}$ for the earlier supergiants. However, given the upper limits inferred from the large population of undetected supergiants of early spectral type we are currently unable to conclude that their mass loss rates are systematically greater than their descendents of  later spectral types. Indeed, 
inspection of Fig. 12 suggests that the $\leq$B1.5 supergiants detected represent the extreme tail of a wider distribution of mass-loss rates for such stars; taken as a whole both detections and non-detections suggest a significant intrinsic scatter in the  mass-loss rates of stars of a given spectral type (e.g. $\sim0.4$\,dex and $\sim0.6$\,dex for B0 and B1 supergiants respectively). 

How do these results compare to other studies? Despite an absence of mm-continuum observations of OB stars in the literature, a number of investigations have been undertaken at radio wavelengths {\citep{benaglia,bieging,lamers,leitherer95,puls,scuderi}. After exclusion of non-thermal sources \cite{benaglia} provide mass-loss rates for 19 O8-B5 supergiants  which we reproduce in Fig. 12, supplemented with the 
late-type BHG HD160529 \citep{leitherer95}; our ALMA observations compare very favourably to these in terms of both numbers of detections and distributions of spectral types. Upon
consideration of both datasets it is apparent that the  mass-loss rates derived by both studies  are broadly comparable. However non-LTE model-atmosphere analysis of spectroscopic data for field B0-9 super-/hypergiants \citep{crowther06b,searle} suggest that the majority of stars so analysed support {\em lower} mass-loss rates than those derived from these radio and mm continuum  observations. Fortuitously seven stars have both radio continuum and spectroscopic mass-loss rate determinations\footnote{HD2905, 30614, 37128, 41117, 152236, 154090} which reveal that the radio detected stars comprise the subset of objects with the highest spectroscopic mass-loss estimates. This implies that the radio detections reported in \cite{benaglia} are biased towards the high radio luminosity subset of a larger population of predominantly fainter objects with lower mass-loss rates, as we infer for our ALMA sample.

How, then, are these observations to be interpreted? A synthesis of the spectroscopic studies suggests that supergiants of later spectral type support lower mass-loss rates \citep[cf. Fig. 9 of ][]{clark14}, but this likely reflects the systematically lower luminosities (and hence masses) of the (field) stars considered; this bias would not be expected to be present in the co-eval population of Wd1. 
Indeed a more complicated relationship between mass-loss rate and terminal wind velocity and spectral type/stellar temperature might be expected as a result of the recombination of Fe\,{\sc iv} to Fe\,{\sc iii} - the so called `bi-stability jump' - around 21kK/spectral type B1 \citep{pp,vink99,vink00,vink,benaglia}. \cite{lamers95} show that the ratio of terminal-wind to escape-velocity $v_{\infty}/v_{\rm escape}$) drops from $\sim$2.6 to $\sim$1.9 around spectral type B1. Given the $v_{\infty}^{-4/3}$ dependence of mm-continuum flux (Eqn 1); in the absence of any other changes to  wind properties one would therefore anticipate that it would be easier to detect stars of later spectral type, as appears to be borne out by our observations. In practice one might also anticipate changes in mass-loss rate and ionisation structure of the wind with spectral type; however quantitative analysis of such affects must await tailored non-LTE analysis of individual stars.

\subsection{The LBV Wd1-243}\label{sect:w243}

The LBV Wd1-243 is a strong,  resolved source in our ALMA data. Spectroscopic observations show that it remained in a long-lived cool-phase between 2002-2009 \citep[approximate spectral type of A3 Ia$^+$;][]{ritchie09b}, with unpublished data suggesting this state persisted until 2015, despite the continued presence of  low level photospheric pulsations. The mm-radio  spectral index suggests a thermal spectrum with mass-loss rates broadly comparable between both radio (log({\.M}/\solmasyr) $\sim-5.0$; data obtained between 1998-2002) and mm determinations (log({\.M}/\solmasyr) $\sim-4.6$; data obtained in 2015). Intriguingly however, both values are an order-of-magnitude greater than that determined from non-LTE model-atmosphere analysis \citep[log({\.M}/\solmasyr) $\sim-6.1$;][]{ritchie09b}, with the {\em caveat} that the latter analysis was unable to simultaneously replicate both the prominent H\,{\sc i} and He\,{\sc i} emission features and photospheric lines evident in the spectra.

Even outside giant eruptions, LBV mass-loss rates are clearly variable \citep{clark09,grohAG}, potentially providing an explanation  the moderate discrepancies between the radio and mm-continuum values. In any event both values are fully consistent with the range  of values quoted for LBVs in the literature \citep[][and refs. therein]{clark14} and in particular the radio derived value  for the post-RSG LBV HD160529 \citep[nominal classification of B8 Ia$^+$ and log({\.M}/\solmasyr) $\sim-4.87$][]{leitherer95}.

\subsection{The sgB[e] star Wd1-9}\label{sect:w9}

While ALMA observations of Wd1-9 have been discussed in detail in \cite{fenech} the improved sensitivity of the mosaic images presented here reveals that the two filaments of emission extending to the south-east and south-west of the source, which were identified with low S/N in \cite{fenech} are much more clearly seen (Fig. \ref{fig:w243}). This strengthens the hypothesis that Wd1-9 shares a comparable nebular morphology to the B[e] star MWC349 \citep{white,gvaram}. However, in the case of Wd1-9 we don't see filamentary structure at mm-wavelengths to the north of the nebular. Intriguingly, initial results of an analysis of Very Large Telescope Interferometer/MIDI data suggest the presence of compact, elongated mid-IR continuum source aligned along the N-S axis of the radio and mm-nebula \citep{hummel17}, with a size comparable to that of the circumbinary torus inferred from the IR spectral energy distribution of Wd1-9 \citep{clark13}. If interpreted as the circumbinary torus this would provide powerful corroboration of the hypothesis that the mm-continuum emission extended in the NS direction resides in the orbital plane of the binary, with the fainter filamentary structure resulting from the disc wind.

\section{The magnetar CXO J164710.2-455216}\label{sect:CXO}

Only a handful of out-bursting magnetars have been found to exhibit (transient, pulsed) radio emission \citep{halpern,rea,szary}. Unlike normal pulsars they appear to have shallow or flat spectral indices, making detection at mm wavelengths viable, with \cite{torne} demonstrating that the magnetar SGR 1734-2900 is visible up to 225\,GHz despite a distance of 8.3\,kpc. Although the origin of radio emission from magnetars is still unclear, both \cite{rea} and \cite{szary} predict that given the physical properties of CXO J164710.2-455216 ({\em viz.} a ratio of quiescent X-ray luminosity  to spin down power greater than unity and its residual temperature) we should not expect to  detect it at mm- or radio-wavelengths. Moreover CXO J164710.2-455216 has yet to exhibit the intense flares that were associated with the transient radio emission of SGR1900+14 and SGR1806-20 \citep[][and refs. therein]{halpern}. Unsurprisingly, therefore, our ALMA observations yield only an upper limit to the continuum flux of 65\,\mujybm.

\section{Discussion}\label{sect:discussion}

\subsection{Post-MS mass-loss rates and stellar evolution}\label{sect:evolution}

 Following the discussion in Sect. \ref{sect:intro} a key omission in our description of the  physics of massive stars is an accurate determination of mass-loss rates as a function of initial mass, metallicity and age. These ALMA observations of Wd1 provide multiple detections of {\em every} post-MS phase and offer the {\em potential} of uniformly determined, unbiased mass-loss rate determinations for stars of $M_{\rm init}\sim30-50M_{\odot}$.

 In the absence of tailored modelling of individual stars, detailed comparison of mass-loss rates to predictions from stellar-evolution models \citep[e.g.][]{brott,ekstrom,martins17} are premature, although we may make limited qualitative comparisons. Each of these codes utilises the theoretical mass-loss recipe of \cite{vink} for OB stars (although \cite{martins17} scales this by a factor of a third to account for recent observational findings) and the empirical relation of \cite{nugis} for field WRs, which together imply a significant increase in mass-loss rates as stars transition between these phases. Such behaviour is indeed observed for Wd1, with a simple comparison of  Figs. \ref{fig:wrmdotsubtype} and \ref{fig:OBmdots} indicating that WRs span log({\.M}/$M_{\odot}$(yr$^{-1})\sim-4.1 \rightarrow -4.8$ and B super-/hypergiants log({\.M}/$M_{\odot}$(yr$^{-1})\sim-5.1 \rightarrow -5.8$  (with the two WNVLh/BHG hybrids intermediate between both groups). However we caution that the large number of OB supergiants non-detections currently biases the sample to more extreme stars  and so one might expect an extension of this range to lower mass-loss rates; deeper observations to  detect the remaining cohort would be of great value. This is especially true since the range of spectral types exhibited by early super-/hypergiants within Wd1 (O9-B9) spans the predicted location ($\sim$B1/21\,kK) of the bi-stability jump due to the transition from Fe\,{\sc iv} to Fe\,{\sc iii} as the dominant ion in the stellar wind (Sect. 6.1).

 Mass-loss rates for the cool super-/hypergiants utilised by codes are again empirical in origin and for RSGs span multiple orders of magnitude \citep{dejager,nieu90,sylvester,vanloon}. Do10 report log({\.M}$/M_{\odot}(yr^{-1})\sim-5$ for the partially optically-thick thermal component associated with the YHG Wd1-4a and infer a similar time-averaged mass-loss rate from the nebular properties of the  RSG Wd1-237. Such values are in excess of those exhibited by the OB supergiants within Wd1 and are comparable to the quiescent mass-loss rates  of the  handful of other similarly extreme RSGs \citep[log({\.M}$/M_{\odot}($yr$^{-1})\sim-5$ (quiescent) $\rightarrow -3$ (outburst); ][]{blocker99,blocker01,smith01}. However they are not as high as those assumed in stellar evolution codes which, for the Geneva models for a $M_{\rm init}\sim40\,M_{\odot}$ star, may exceed log({\.M}$/M_{\odot}($yr$^{-1})\sim-4$. However we {\em caveat} this with the observation that the mass-loss histories of such stars within Wd1 are clearly complex, as evidenced by their nebular morphologies, and further detailed modelling will be required to provide more accurate mass-loss rates. Moreover, observations to investigate the possibility of a reservoir of neutral material within the circumstellar environment, to which our  observations would be insensitive, would be particularly valuable since the presence of such gas would render current estimates of time-averaged mass-loss rates from nebular properties as lower limits.

Finally we turn to the apparent binary Wd1-9. Current evolutionary codes do not incorporate binary interaction, but simulations of potentially comparative systems \citep{petrovic}\footnote{$M_{\rm primary}\sim41M_{\odot}$, $M_{\rm secondary}\sim20-30\,M_{\odot}$ and $P_{\rm orb}\sim3-6$\,d.} suggest  {\em time averaged} mass-loss rates of log({\.M}$/M_{\odot}($yr$^{-1})\geq-4$ over the $4-6\times10^4$\,yr period of fast case-A mass-transfer. The current  mass-loss rate inferred for Wd1-9 \citep[log({\.M}$/M_{\odot}($yr$^{-1})\sim-4.2$][]{fenech} is indeed broadly consistent with such predictions. Moreover it is directly comparable to the highest mass-loss rates determined for the cluster WRs, greater than that of the `quiescent' LBV Wd1-243 and provisional estimates for the cool super-/hypergiants and most striking of all,  a factor of $\gtrsim20\times$ larger than found for the most `extreme' B super/hypergiants within the cluster. Placing Wd1-9 into such an evolutionary context therefore  provides powerful confirmation of the hypothesis that binary interaction plays a central role in the lifecycle of massive stars.

Similar comparative studies of the evolution of mass-loss rates through distinct post-MS phases are possible for both the Arches and Galactic Centre clusters. \cite{martins07,martins08} present the results of a comprehensive model-atmosphere analysis of near-IR spectro-photometric data for members of both clusters, from which we find qualitative agreement with our results in the  sense that mass-loss rates are also found to increase with evolutionary state. Specifically the  mass-loss rates for the mid-O supergiants within the Arches cluster are approximately an order-of-magnitude lower than those of the more massive and evolved WN7-9h stars, with the mid-O hypergiants intermediate between both extremes \citep{martins08}. Likewise the mass-loss rate obtained from analysis of a mean spectrum derived from observations of ten late-O/early-B supergiants within the Galactic centre cluster is over a magnitude lower than the WNLh and WC groups \citep{martins07}.

Unfortunately, we may not quantitatively compare mass-loss rates between Wd1 and the two galactic centre clusters since results from the two different methods employed appear prone to a systematic discrepancy. Specifically, \cite{martins08} demonstrate that the clumping-corrected (\.{M}$/f_{cl}^{0.5}$) spectro-photometric mass-loss estimates for the Arches cluster members are significantly larger than the radio determinations of \cite{lang} for the subset of stars for which such a comparison is possible. We arrive at a similar conclusion after comparison of the results of \cite{YZ,martins07} for the Galactic Centre cluster\footnote{Of the seven stars in common between the two studies, the four determined to have clumped winds via spectro-photometric analysis all show significant discrepancies when compared to radio-continuum mass-loss rates - the WN9Lh stars GC IRS33E  ($\sim 7\times$ greater than the radio determination) and AF ($\sim 2.2\times$), the WN8 star AF NW ($\sim5\times$) and the WN5/6 star  GC 16SE2 ($\sim 5\times$); by contrast the mass-loss estimates for the remaining three stars, for which clumping was not inferred, were  consistent to within a factor of $<1.4\times$.}.

\cite{martins08} suggests that the discrepancy may in part arise due to differences in the clumping fractions in the regions of the inner and outer wind responsible for, respectively, the near-IR spectral diagnostics and mm-/radio-continuum. Such an evolution in the run of clumping with radius has also been suggested by \cite{puls,runacres}; we return to this issue in Sect. \ref{sect:discussresn}.

\subsection{Resolved emission associated with the  OB super/hyper-giant  and WR cohorts}\label{sect:discussresn}

As detailed in Sect. \ref{sect:resolvedemission} and \ref{sect:stellar3} a significant number of the hot stars within Wd1 (both WRs, and OB super-/hyper-giants) appear to be moderately extended in our continuum observations. Stellar wind sources have been spatially resolved at mm and radio wavelengths in the past; examples being the BHG Cyg OB2\#12 \citep{morford} and the LBV P Cygni \citep{skinner}, in both cases facilitated by the combination of relative proximity of the star, low terminal wind velocity and high mass-loss rate. However as discussed in Sect. \ref{sect:resolvedemission}, analytic and numerical estimates for the sizes of the mm-/radio photospheres of the WRs indicate they should not be resolved with the current observations, with an identical conclusion following for the OB super-/hypergiants.

High resolution radio observations of the CWB WR140 has resolved the WCR between both stellar components \citep{dougherty05} and for WR146 both binary components and the WCR \citep{dougherty00b}. However despite the distance to both being lower than to Wd1 and their orbital separation much greater than e.g. WR A and B, the resolved emission has a smaller angular scale than observed for the majority of stars within Wd1; we  therefore conclude that it unlikely that we have resolved the (putative) WCRs of these objects.

Both binary driven mass-loss (cf. Wd1-9; Sect. \ref{sect:w9}) and impulsive ejection events associated with the LBV phase \citep[e.g.][]{duncan} can give rise to resolved ejection nebulae. However not all of the resolved stars within Wd1 currently show evidence for binarity while, to the best of our knowledge, LBV-like instabilities have not been associated with early OB supergiants nor, with only a couple of exceptions, with WN stars earlier than WN9 or WC stars. Moreover, given the similarity in size of the resolved sources (Table \ref{tab:srcesizes}) one would require multiple stars to synchronise the timing and duration of any such interaction or instability in order to replicate this observational finding. 

Similarly, we note that a number of resolved sources demonstrate mm-/radio spectral indices inconsistent with optically-thin thermal emission, as might be expected for ejection nebulae. Additionally, mass-loss rates derived for the sources are also fully consistent with expectations for WR and OB super-/hypergiants respectively (Sects. 4.2.2 and 6.1). Both observations suggest sources dominated by emission from stellar winds; it seems highly improbable that either binary or LBV ejection mechanisms could conspire to replicate the fluxes expected from the stellar winds of such stars as a function of their spectral type.

Given the results of profile fitting (Sects. 3.1.1 and 4.2.3), it appears possible that these sources are composite, with an unresolved core that dominates the emergent flux due to the stellar wind and an extended, low surface brightness halo surrounding this. If confirmed via higher-resolution observations, what would be the physical origin of such a configuration? After eliminating the above possibilities, and following our findings for the cool super-/hyper-giant cohort, an obvious explanation would be the confinement of the stellar winds leading to the formation of a compact wind-blown bubble.

Since, to the best of our knowledge, such a phenomenon has not been observed for isolated massive stars it is attractive to attribute this to their membership of a YMC and we  highlight that \cite{lang} report spatially resolved radio-continuum emission of comparable extent associated with a number of stars in the Quintuplet cluster. Prospective physical agents for this would include pressure confinement by the  intra-cluster medium or wind/wind interaction with the stellar cohort and/or one or more recent SNe. Both X-ray \citep{muno06b}, mm and radio continuum observations (Do10; Sect \ref{sect:cluster}) argue for the presence of an intra-cluster medium or wind associated with Wd1. \cite{mackey15} conclude that  pressure confinement is unlikely to explain the properties of the nebula surrounding Wd1-26; given the greater wind momenta of, for example, the WR cohort such a conclusion would appear to apply to such stars as well. This would appear to favour interaction between stellar and/or a cluster wind as a physical mechanism. However, given the highly structured nature of the intra-cluster medium/wind within Wd1 (as revealed by mm and radio continuum emission; Sect. \ref{sect:mm}) and the asymmetrical distribution of massive stars within Wd1, an assessment of both possibilities would required  detailed hydrodynamical simulation which are clearly beyond the scope of this work.

\subsection{Stellar and cluster wind interaction}\label{sect:cluster}

It has long been understood that feedback from clusters via a stellar-wind- and SN-driven cluster-wind contributes to chemical evolution of the interstellar medium \citep{krumholz}. However the properties and the mechanisms by which such winds are initiated and maintained are currently uncertain, the later due to uncertainties regarding  the efficiency by which stellar kinetic energy is converted to thermal energy, which in turn  drives the outflow \citep[e.g.][]{stevens}. Previous observations of Wd1 have shown that the stars are embedded in a highly complex cluster medium, with both hot \citep{muno06b} and cool components present (Do10). The resolution of our observations confirm that the cluster medium is highly structured, with multiple `clumps' of characteristic scale present in the core regions (Sect. \ref{sect:extended}), while also allowing us to study the interaction and entrainment of the winds of individual stars in unprecedented detail (Sect. \ref{sect:wrstars},\ref{sect:coolsuper} and \ref{sect:stellar3}).

In particular the pronounced cometary nebulae of the cool super-/hypergiant cohort and the compact `bubbles' associated with both OB super-/hypergiant and Wolf-Rayet stars imply that the circumstellar  environments of stars embedded within  young massive clusters may be profoundly different from their isolated brethren. One important consequence of this interaction may be in modifications to the phenomenology of the resultant SNe. 

There is a wealth of observational evidence for a subset ($\sim10$\%) of SNe where the explosive ejecta interacts  with pre-existing circumstellar ejecta/material \citep{smartt}. Although comparatively infrequent, such  type IIn events can be highly luminous \citep[e.g.][]{smith07}, potentially allowing their detection in the early Universe. Consequently the nature of the progenitors has been subject to considerable interest. The apparent requirement for the ejection of large amounts of  circumstellar material in the past  has led to the suggestion that  LBVs may be the direct progenitors of such SNe \citep{smith07,gal}. Multiwavelength studies by \cite{fox} show a diversity of behaviour amongst type IIn SNe, suggesting a range of progenitor systems, while \cite{smith09} suggested that extreme RSGs such as VY CMa may also be viable progenitor systems \citep[see also][]{yoon}. Moreover spectroscopic signatures revealing the presence of circumstellar material has also been observed in a wide range of other SNe sub-types\footnote{Superluminous SN \citep{galyam12}, Type IIP \citep{mauerhan}, IIL \citep{liu} and Ibc \cite{foley}.}. 

However eruptions supporting greatly enhanced mass-loss rates just prior to SNe are not the only route for the production of dense circumstellar shells. The possibility of the retention of significant circumstellar material close to irradiated red supergiants {\em in the absence of impulsive mass-ejection events} has already been highlighted by \cite{mackey14,mackey15}; our observations corroborate this hypothesis, while potentially extending it to yellow hypergiants. Equally  exciting is the identification of compact emission halos surrounding the WRs which, given the age of Wd1, are expected to be the immediate progenitors of core-collapse SNe. Flash spectroscopy of the type II event SN2013fs \citep{yaron} and the type IIb SN2013cu \citep{galyam14} suggest circumstellar material confined to within $<10^{15}$cm of the progenitor; directly comparable to the sizes of the extended emission associated with the cluster WRs. While SN2013fs has been attributed to the explosion of a RSG, the wind signatures identified for SN2013cu suggest instead a WN(h) progenitor \citep{galyam14}. Further observations to quantify the amount of material contained within these structures to compare to the estimate obtained for SN2013cu would be of obvious interest.  

Finally we note that multiple lines of evidence suggest that the pre-SN circumstellar environment may be aspherical, with both clumpy \citep{smith09b} or disc-like configurations proposed \citep{levesque} for a subset of SNe, of which SN1987A is the most obvious exemplar. Clearly the circumstellar nebula surrounding e.g. Wd1-26 satisfies the former condition but we also highlight the apparent toroidal geometry of the supergiant B[e] star Wd1-9, which in this case is  thought to arise from binary interaction. Nevertheless, taken as a whole the properties of the circumstellar environments of the massive  stars within Wd1 suggest that they, and potential SNe arising from them,  may not be fully understood in isolation.

\section{Concluding remarks}\label{conclusion}

In this work and that of \cite{fenech} we have presented the results of a Band 3 (3mm) continuum survey of the Galactic YMC Wd1. This target was chosen due to a combination of proximity, co-evality, age and mass; the combination of the latter three parameters resulting in a rich population of post-MS stars characterised by the co-existence of both cool super-/hypergiants and Wolf Rayets. An unprecedented total of 50 stellar sources were detected, comprising examples of every phase of massive post-MS evolution present within the cluster. The full exploitation of this exceptional dataset in order to constrain the physics of (radiatively-driven) stellar winds - e.g. the presence of  `clumping'  -  requires quantitative analysis (e.g. non-LTE model-atmosphere analysis and/or hydrodynamical simulations of nebular emission) which is beyond the scope of this work. Nevertheless we may immediately draw two broad conclusions from the provisional analysis included here.

Firstly we were able to utilise canonical stellar- and wind-properties and the formulation of \cite{wright75} to infer mass-loss rates for both the hot (WR, LBV, sgB[e] and OB Ia/Ia$^+$) and cool (YHG and RSG) cohorts, with additional estimates for the latter objects being drawn from their bulk nebular properties (Do10). As a consequence we were able to follow the evolution of the mass-loss  rate through all post-MS phases to the brink of SN; an essential observational  constraint for quantifying the physics of massive stars. As expected we found that the mass-loss rates  of the (brighter) OB super-/hypergiants are approximately an order of magnitude lower than those of the WRs, with both appearing broadly consistent with determinations from field populations. Where available, estimates for the sole LBV and the YHGs/RSGs within the cluster were comparable to the WR population, but not as extreme as measured for other such examples (e.g. \.{M}$>10^{-4}M_{\odot}$yr$^{-1}$ for such stars during outburst).
We highlight that the distribution of spectral types of the OB supergiants is particularly fortuitous since  it spans the predicted location of the bi-stability jump induced by the recombination of Fe\,{\sc iv} to Fe\,{\sc iii}; Wd1 therefore offers a powerful test of predictions for wind properties either side of this discontinuity.

In general we find no evidence that  massive binaries are systematically brighter than single stars due to an additional contribution from the wind-collision zone. However the mass-loss rate inferred for the sgB[e] star Wd1-9 is exceptionally high \citep{fenech}, being directly comparable to the most extreme WRs 
present (WR F (WC9) and WR A (WN7b)). Multiple lines of evidence point to Wd1-9 being a massive interacting binary currently exhibiting rapid (case A) mass transfer/loss. Our observations suggest this has led to the formation of a massive circumstellar torus which drives a bipolar wind. Since Wd1-9 is the  only star within Wd1 to exhibit this phenomenon it suggests that the duration of this violent phase is comparatively short, although if the current mass-loss rate were maintained for only the $\sim5-10{\times}10^4$\,yr predicted by theory \citep{petrovic} its influence on the subsequent evolution of both components would be profound.

Secondly, we highlight the degree to which the environment in which the stars are located affects their circumstellar properties. This is most evident in the manner in which the ejecta associated with both the  RSGs and YHGs appears to be sculpted during interaction with the wider cluster, although the mechanism(s) involved - cluster wind and/or radiation field - remain uncertain. While this finding was anticipated the resolution of the sources associated with a number of the OB supergiants and WRs was not. Intriguingly, while such a phenomenon has not been associated with isolated stars, \cite{lang} report a similar extension to the radio sources associated with a subset of the O supergiants and WRs within the Quintuplet. Following Sect. \ref{sect:discussresn} the least objectionable hypothesis is that these represent compact wind blown bubbles confined via one of the preceding physical effects. 

If correct this would have important ramifications both for the initiation and driving of the cluster wind via the combined action of individual stellar winds and SNe and also for the nature of the SNe themselves. We are long accustomed to the idea that the binary nature of a SN progenitor  will affect the nature of the explosive endpoint via stripping of the outer H-rich stellar layers and/or the subsequent shaping of the pre-explosion  circumstellar material - with Wd1-9 providing a case-in-point. However if the circumstellar envelope  of a SN progenitor may be shaped by its immediate environment, potentially leading to the formation of a dense envelope in the absence of an episode(s) of enhanced impulsive mass-loss, then one must accept that the diversity of SNe morphologies does not solely result from the properties of the progenitor and instead originates in part from the wider context of explosion. Given the expectation that the majority of massive stars are born in a clustered environment, this is potentially a far reaching conclusion. 

  To conclude; ALMA appears a uniquely powerful tool for the quantitative analysis of mass-loss rates for massive stars, as well as 
 their interaction with their wider environment. As such it has the potential to revolutionise our understanding of such objects, 
 providing a window into short-lived, violent phases of their evolution and and opening a window onto quantitatively new phenomena. 

\begin{acknowledgements}
This paper makes use of the following ALMA data: ADS/JAO.ALMA\#2013.1.00897.S. ALMA is a partnership of ESO (representing its member states), NSF (USA) and NINS (Japan), together with NRC (Canada), NSC and ASIAA (Taiwan), and KASI (Republic of Korea), in cooperation with the Republic of Chile. The Joint ALMA Observatory is operated by ESO, AUI/NRAO and NAOJ. D. Fenech wishes to acknowledge funding from a STFC consolidated grant (ST/M001334/1). IN is partially supported by the Spanish Government Ministerio de Econom\'{i}a y Competitivad (MINECO/FEDER) under grant AYA2015-68012-C2-2-P. The authors would also like to thank Dr. Roger Wesson (UCL) for providing the VPHAS data for comparison.

\end{acknowledgements}

\bibliographystyle{aa} 
\bibliography{Wd1_census} 

\newcommand{\noop}[1]{}
\begin{thebibliography}{157}
\expandafter\ifx\csname natexlab\endcsname\relax\def\natexlab#1{#1}\fi

\bibitem[{{Abbott} {et~al.}(2016){Abbott}, {Abbott}, {Abbott}, {fake}, {fake},
  {fake}, {fake}, {fake}, \& {fake}}]{abbott}
{Abbott}, B.~P., {Abbott}, R., {Abbott}, T.~D., {et~al.} 2016, Physical Review
  Letters, 116, 241103

\bibitem[{{Abbott} {et~al.}(1986){Abbott}, {Beiging}, {Churchwell}, \&
  {Torres}}]{abbott86}
{Abbott}, D.~C., {Beiging}, J.~H., {Churchwell}, E., \& {Torres}, A.~V. 1986,
  \apj, 303, 239

\bibitem[{{Abdalla} {et~al.}(\noop{2016}in press){Abdalla}, {Abramowski},
  {Aharonian}, {fake}, {fake}, {fake}, {fake}, {fake}, \& {fake}}]{abdalla}
{Abdalla}, H., {Abramowski}, A., {Aharonian}, F., {et~al.} \noop{2016}in press,
  \aap

\bibitem[{{Abramowski} {et~al.}(2012){Abramowski}, {Acero}, {Aharonian},
  {fake}, {fake}, {fake}, {fake}, {fake}, \& {fake}}]{abra}
{Abramowski}, A., {Acero}, F., {Aharonian}, F., {et~al.} 2012, \aap, 537, A114

\bibitem[{{Andrews} {et~al.}(2018){Andrews}, {Fenech}, {Prinja}, {Clark}, \&
  {Hindson}}]{andrews18}
{Andrews}, H., {Fenech}, D., {Prinja}, R.~K., {Clark}, J.~S., \& {Hindson}, L.
  2018, \mnras [\eprint[arXiv]{1803.07008}]

\bibitem[{{Benaglia} {et~al.}(2007){Benaglia}, {Vink}, {Mart{\'{\i}}},
  {Ma{\'{\i}}z Apell{\'a}niz}, {Koribalski}, \& {Crowther}}]{benaglia}
{Benaglia}, P., {Vink}, J.~S., {Mart{\'{\i}}}, J., {et~al.} 2007, \aap, 467,
  1265

\bibitem[{{Bieging} {et~al.}(1989){Bieging}, {Abbott}, \&
  {Churchwell}}]{bieging}
{Bieging}, J.~H., {Abbott}, D.~C., \& {Churchwell}, E.~B. 1989, \apj, 340, 518

\bibitem[{{Bl{\"o}cker} {et~al.}(1999){Bl{\"o}cker}, {Balega}, {Hofmann},
  {Lichtenth{\"a}ler}, {Osterbart}, \& {Weigelt}}]{blocker99}
{Bl{\"o}cker}, T., {Balega}, Y., {Hofmann}, K.-H., {et~al.} 1999, \aap, 348,
  805

\bibitem[{{Bl{\"o}cker} {et~al.}(2001){Bl{\"o}cker}, {Balega}, {Hofmann}, \&
  {Weigelt}}]{blocker01}
{Bl{\"o}cker}, T., {Balega}, Y., {Hofmann}, K.-H., \& {Weigelt}, G. 2001, \aap,
  369, 142

\bibitem[{{Blomme} {et~al.}(2003){Blomme}, {van de Steene}, {Prinja},
  {Runacres}, \& {Clark}}]{blomme}
{Blomme}, R., {van de Steene}, G.~C., {Prinja}, R.~K., {Runacres}, M.~C., \&
  {Clark}, J.~S. 2003, \aap, 408, 715

\bibitem[{{Bonanos}(2007)}]{bonanos}
{Bonanos}, A.~Z. 2007, \aj, 133, 2696

\bibitem[{{Brandner} {et~al.}(2008){Brandner}, {Clark}, {Stolte}, {Waters},
  {Negueruela}, \& {Goodwin}}]{brandner}
{Brandner}, W., {Clark}, J.~S., {Stolte}, A., {et~al.} 2008, \aap, 478, 137

\bibitem[{{Brott} {et~al.}(2011){Brott}, {de Mink}, {Cantiello}, {Langer}, {de
  Koter}, {Evans}, {Hunter}, {Trundle}, \& {Vink}}]{brott}
{Brott}, I., {de Mink}, S.~E., {Cantiello}, M., {et~al.} 2011, \aap, 530, A115

\bibitem[{{Bykov} {et~al.}(2015){Bykov}, {Ellison}, {Gladilin}, \&
  {Osipov}}]{bykov}
{Bykov}, A.~M., {Ellison}, D.~C., {Gladilin}, P.~E., \& {Osipov}, S.~M. 2015,
  \mnras, 453, 113

\bibitem[{{Cappa} {et~al.}(2004){Cappa}, {Goss}, \& {van der Hucht}}]{cappa}
{Cappa}, C., {Goss}, W.~M., \& {van der Hucht}, K.~A. 2004, \aj, 127, 2885

\bibitem[{{Carniani} {et~al.}(2015){Carniani}, {Maiolino}, {De Zotti},
  {Negrello}, {Marconi}, {Bothwell}, {Capak}, {Carilli}, {Castellano},
  {Cristiani}, {Ferrara}, {Fontana}, {Gallerani}, {Jones}, {Ohta}, {Ota},
  {Pentericci}, {Santini}, {Sheth}, {Vallini}, {Vanzella}, {Wagg}, \&
  {Williams}}]{carniani15}
{Carniani}, S., {Maiolino}, R., {De Zotti}, G., {et~al.} 2015, \aap, 584, A78

\bibitem[{{Chapman} {et~al.}(1999){Chapman}, {Leitherer}, {Koribalski},
  {Bouter}, \& {Storey}}]{chapman}
{Chapman}, J.~M., {Leitherer}, C., {Koribalski}, B., {Bouter}, R., \& {Storey},
  M. 1999, \apj, 518, 890

\bibitem[{{Clark} {et~al.}(2009){Clark}, {Crowther}, {Larionov}, {Steele},
  {Ritchie}, \& {Arkharov}}]{clark09}
{Clark}, J.~S., {Crowther}, P.~A., {Larionov}, V.~M., {et~al.} 2009, \aap, 507,
  1555

\bibitem[{{Clark} {et~al.}(1998){Clark}, {Fender}, {Waters}, {Dougherty},
  {Koornneef}, {Steele}, \& {van Blokland}}]{clark98}
{Clark}, J.~S., {Fender}, R.~P., {Waters}, L.~B.~F.~M., {et~al.} 1998, \mnras,
  299, L43

\bibitem[{{Clark} {et~al.}(2008){Clark}, {Muno}, {Negueruela}, {Dougherty},
  {Crowther}, {Goodwin}, \& {de Grijs}}]{clark08}
{Clark}, J.~S., {Muno}, M.~P., {Negueruela}, I., {et~al.} 2008, \aap, 477, 147

\bibitem[{{Clark} {et~al.}(2012){Clark}, {Najarro}, {Negueruela}, {Ritchie},
  {Urbaneja}, \& {Howarth}}]{clark12}
{Clark}, J.~S., {Najarro}, F., {Negueruela}, I., {et~al.} 2012, \aap, 541, A145

\bibitem[{{Clark} \& {Negueruela}(2002)}]{clark02}
{Clark}, J.~S. \& {Negueruela}, I. 2002, \aap, 396, L25

\bibitem[{{Clark} {et~al.}(2005){Clark}, {Negueruela}, {Crowther}, \&
  {Goodwin}}]{clark05}
{Clark}, J.~S., {Negueruela}, I., {Crowther}, P.~A., \& {Goodwin}, S.~P. 2005,
  \aap, 434, 949

\bibitem[{{Clark} {et~al.}(2014{\natexlab{a}}){Clark}, {Negueruela}, \&
  {Gonz{\'a}lez-Fern{\'a}ndez}}]{clark14a}
{Clark}, J.~S., {Negueruela}, I., \& {Gonz{\'a}lez-Fern{\'a}ndez}, C.
  2014{\natexlab{a}}, \aap [\eprint[arXiv]{1311.3956}]

\bibitem[{{Clark} {et~al.}(2014{\natexlab{b}}){Clark}, {Ritchie}, {Najarro},
  {Langer}, \& {Negueruela}}]{clark14}
{Clark}, J.~S., {Ritchie}, B.~W., {Najarro}, F., {Langer}, N., \& {Negueruela},
  I. 2014{\natexlab{b}}, \aap [\eprint[arXiv]{1405.3109}]

\bibitem[{{Clark} {et~al.}(2010){Clark}, {Ritchie}, \& {Negueruela}}]{clark10}
{Clark}, J.~S., {Ritchie}, B.~W., \& {Negueruela}, I. 2010, \aap, 514, A87

\bibitem[{{Clark} {et~al.}(2013){Clark}, {Ritchie}, \& {Negueruela}}]{clark13}
{Clark}, J.~S., {Ritchie}, B.~W., \& {Negueruela}, I. 2013, \aap, 560, A11

\bibitem[{{Clark} {et~al.}(2011){Clark}, {Ritchie}, {Negueruela}, {Crowther},
  {Damineli}, {Jablonski}, \& {Langer}}]{clark11}
{Clark}, J.~S., {Ritchie}, B.~W., {Negueruela}, I., {et~al.} 2011, \aap, 531,
  A28

\bibitem[{{Clark} {et~al.}(\noop{2017}in prep.){Clark}, {Negueruela},
  {Ritchie}, {fake}, {fake}, {fake}, {fake}, {fake}, \& {fake}}]{clark17}
{Clark}, S., {Negueruela}, I., {Ritchie}, {et~al.} \noop{2017}in prep.

\bibitem[{{Clark} {et~al.}(2015){Clark}, {Negueruela}, {Ritchie}, {Najarro},
  {Langer}, {Crowther}, {Bartlett}, {Fenech}, {Gonz{\'a}lez-Fern{\'a}ndez},
  {Goodwin}, {Lohr}, \& {Prinja}}]{clark15}
{Clark}, S., {Negueruela}, I., {Ritchie}, B., {et~al.} 2015, The Messenger,
  159, 30

\bibitem[{{Cox} {et~al.}(2012){Cox}, {Kerschbaum}, {van Marle}, {Decin},
  {Ladjal}, {Mayer}, {Groenewegen}, {van Eck}, {Royer}, {Ottensamer}, {Ueta},
  {Jorissen}, {Mecina}, {Meliani}, {Luntzer}, {Blommaert}, {Posch},
  {Vandenbussche}, \& {Waelkens}}]{cox}
{Cox}, N.~L.~J., {Kerschbaum}, F., {van Marle}, A.-J., {et~al.} 2012, \aap,
  537, A35

\bibitem[{{Crowther}(2007)}]{crowther07}
{Crowther}, P.~A. 2007, \araa, 45, 177

\bibitem[{{Crowther} {et~al.}(2006{\natexlab{a}}){Crowther}, {Hadfield},
  {Clark}, {Negueruela}, \& {Vacca}}]{crowther06}
{Crowther}, P.~A., {Hadfield}, L.~J., {Clark}, J.~S., {Negueruela}, I., \&
  {Vacca}, W.~D. 2006{\natexlab{a}}, \mnras [\eprint{astro-ph/0608356}]

\bibitem[{{Crowther} {et~al.}(2006{\natexlab{b}}){Crowther}, {Lennon}, \&
  {Walborn}}]{crowther06b}
{Crowther}, P.~A., {Lennon}, D.~J., \& {Walborn}, N.~R. 2006{\natexlab{b}},
  \aap [\eprint{astro-ph/0509436}]

\bibitem[{{Davies} {et~al.}(2008){Davies}, {Figer}, {Law}, {Kudritzki},
  {Najarro}, {Herrero}, \& {MacKenty}}]{davies}
{Davies}, B., {Figer}, D.~F., {Law}, C.~J., {et~al.} 2008, \apj, 676, 1016

\bibitem[{{de Jager} {et~al.}(1988){de Jager}, {Nieuwenhuijzen}, \& {van der
  Hucht}}]{dejager}
{de Jager}, C., {Nieuwenhuijzen}, H., \& {van der Hucht}, K.~A. 1988, \aaps,
  72, 259

\bibitem[{{de Mink} {et~al.}(2014){de Mink}, {Sana}, {Langer}, {Izzard}, \&
  {Schneider}}]{demink}
{de Mink}, S.~E., {Sana}, H., {Langer}, N., {Izzard}, R.~G., \& {Schneider},
  F.~R.~N. 2014, \apj, 782, 7

\bibitem[{{de Wit} {et~al.}(2008){de Wit}, {Oudmaijer}, {Fujiyoshi}, {Hoare},
  {Honda}, {Kataza}, {Miyata}, {Okamoto}, {Onaka}, {Sako}, \&
  {Yamashita}}]{dewit}
{de Wit}, W.~J., {Oudmaijer}, R.~D., {Fujiyoshi}, T., {et~al.} 2008, \apjl,
  685, L75

\bibitem[{{Dougherty} {et~al.}(2005){Dougherty}, {Beasley}, {Claussen},
  {Zauderer}, \& {Bolingbroke}}]{dougherty05}
{Dougherty}, S.~M., {Beasley}, A.~J., {Claussen}, M.~J., {Zauderer}, B.~A., \&
  {Bolingbroke}, N.~J. 2005, \apj, 623, 447

\bibitem[{{Dougherty} {et~al.}(2010){Dougherty}, {Clark}, {Negueruela},
  {Johnson}, \& {Chapman}}]{dougherty}
{Dougherty}, S.~M., {Clark}, J.~S., {Negueruela}, I., {Johnson}, T., \&
  {Chapman}, J.~M. 2010, \aap, 511, A58

\bibitem[{{Dougherty} {et~al.}(2000b){Dougherty}, {Williams}, \&
  {Pollacco}}]{dougherty00b}
{Dougherty}, S.~M., {Williams}, P.~M., \& {Pollacco}, D.~L. 2000b, \mnras, 316,
  143

\bibitem[{{Drew}(1989)}]{drew}
{Drew}, J.~E. 1989, \apjs, 71, 267

\bibitem[{{Duncan} \& {White}(2002)}]{duncan}
{Duncan}, R.~A. \& {White}, S.~M. 2002, \mnras, 330, 63

\bibitem[{{Ekstr{\"o}m} {et~al.}(2012){Ekstr{\"o}m}, {Georgy}, {Eggenberger},
  {Meynet}, {Mowlavi}, {Wyttenbach}, {Granada}, {Decressin}, {Hirschi},
  {Frischknecht}, {Charbonnel}, \& {Maeder}}]{ekstrom}
{Ekstr{\"o}m}, S., {Georgy}, C., {Eggenberger}, P., {et~al.} 2012, \aap, 537,
  A146

\bibitem[{{Fenech} {et~al.}(2017){Fenech}, {Clark}, {Prinja}, {Morford},
  {Dougherty}, \& {Blomme}}]{fenech}
{Fenech}, D.~M., {Clark}, J.~S., {Prinja}, R.~K., {et~al.} 2017, \mnras, 464,
  L75

\bibitem[{{Fok} {et~al.}(2012){Fok}, {Nakashima}, {Yung}, {Hsia}, \&
  {Deguchi}}]{fok}
{Fok}, T.~K.~T., {Nakashima}, J.-i., {Yung}, B.~H.~K., {Hsia}, C.-H., \&
  {Deguchi}, S. 2012, \apj, 760, 65

\bibitem[{{Foley} {et~al.}(2007){Foley}, {Smith}, {Ganeshalingam}, {Li},
  {Chornock}, \& {Filippenko}}]{foley}
{Foley}, R.~J., {Smith}, N., {Ganeshalingam}, M., {et~al.} 2007, \apjl, 657,
  L105

\bibitem[{{Fox} {et~al.}(2013){Fox}, {Filippenko}, {Skrutskie}, {Silverman},
  {Ganeshalingam}, {Cenko}, \& {Clubb}}]{fox}
{Fox}, O.~D., {Filippenko}, A.~V., {Skrutskie}, M.~F., {et~al.} 2013, \aj, 146,
  2

\bibitem[{{Fullerton} {et~al.}(2006){Fullerton}, {Massa}, \&
  {Prinja}}]{fullerton}
{Fullerton}, A.~W., {Massa}, D.~L., \& {Prinja}, R.~K. 2006, \apj, 637, 1025

\bibitem[{{Gal-Yam}(2012)}]{galyam12}
{Gal-Yam}, A. 2012, Science, 337, 927

\bibitem[{{Gal-Yam} {et~al.}(2014){Gal-Yam}, {Arcavi}, {Ofek}, {Ben-Ami},
  {Cenko}, {Kasliwal}, {Cao}, {Yaron}, {Tal}, {Silverman}, {Horesh}, {De Cia},
  {Taddia}, {Sollerman}, {Perley}, {Vreeswijk}, {Kulkarni}, {Nugent},
  {Filippenko}, \& {Wheeler}}]{galyam14}
{Gal-Yam}, A., {Arcavi}, I., {Ofek}, E.~O., {et~al.} 2014, \nat, 509, 471

\bibitem[{{Gal-Yam} \& {Leonard}(2009)}]{gal}
{Gal-Yam}, A. \& {Leonard}, D.~C. 2009, \nat, 458, 865

\bibitem[{{Groh} {et~al.}(2009){Groh}, {Hillier}, {Damineli}, {Whitelock},
  {Marang}, \& {Rossi}}]{grohAG}
{Groh}, J.~H., {Hillier}, D.~J., {Damineli}, A., {et~al.} 2009, \apj, 698, 1698

\bibitem[{{Groh} {et~al.}(2014){Groh}, {Meynet}, {Ekstr{\"o}m}, \&
  {Georgy}}]{groh}
{Groh}, J.~H., {Meynet}, G., {Ekstr{\"o}m}, S., \& {Georgy}, C. 2014, \aap,
  564, A30

\bibitem[{{Gvaramadze} \& {Menten}(2012)}]{gvaram}
{Gvaramadze}, V.~V. \& {Menten}, K.~M. 2012, \aap, 541, A7

\bibitem[{{Gvaramadze} {et~al.}(2014){Gvaramadze}, {Menten}, {Kniazev},
  {Langer}, {Mackey}, {Kraus}, {Meyer}, \& {Kami{\'n}ski}}]{gvar}
{Gvaramadze}, V.~V., {Menten}, K.~M., {Kniazev}, A.~Y., {et~al.} 2014, \mnras,
  437, 843

\bibitem[{{Halpern} {et~al.}(2005){Halpern}, {Gotthelf}, {Becker}, {Helfand},
  \& {White}}]{halpern}
{Halpern}, J.~P., {Gotthelf}, E.~V., {Becker}, R.~H., {Helfand}, D.~J., \&
  {White}, R.~L. 2005, \apjl, 632, L29

\bibitem[{{Hamann} {et~al.}(2006){Hamann}, {Gr{\"a}fener}, \&
  {Liermann}}]{hamann}
{Hamann}, W.-R., {Gr{\"a}fener}, G., \& {Liermann}, A. 2006, \aap, 457, 1015

\bibitem[{{Hatsukade} {et~al.}(2016){Hatsukade}, {Kohno}, {Umehata},
  {Aretxaga}, {Caputi}, {Dunlop}, {Ikarashi}, {Iono}, {Ivison}, {Lee},
  {Makiya}, {Matsuda}, {Motohara}, {Nakanishi}, {Ohta}, {Tadaki}, {Tamura},
  {Wang}, {Wilson}, {Yamaguchi}, \& {Yun}}]{hatsukade16}
{Hatsukade}, B., {Kohno}, K., {Umehata}, H., {et~al.} 2016, \pasj, 68, 36

\bibitem[{{Hayward} {et~al.}(2013){Hayward}, {Narayanan}, {Kere{\v s}},
  {Jonsson}, {Hopkins}, {Cox}, \& {Hernquist}}]{hayward13}
{Hayward}, C.~C., {Narayanan}, D., {Kere{\v s}}, D., {et~al.} 2013, \mnras,
  428, 2529

\bibitem[{{Hummel et al.}(\noop{2017}in prep.)}]{hummel17}
{Hummel et al.} \noop{2017}in prep.

\bibitem[{{Hutsem{\'e}kers} {et~al.}(2013){Hutsem{\'e}kers}, {Cox}, \&
  {Vamvatira-Nakou}}]{huts}
{Hutsem{\'e}kers}, D., {Cox}, N.~L.~J., \& {Vamvatira-Nakou}, C. 2013, \aap,
  552, L6

\bibitem[{{Ignace} \& {Churchwell}(2004)}]{ignace}
{Ignace}, R. \& {Churchwell}, E. 2004, \apj, 610, 351

\bibitem[{{Kothes} \& {Dougherty}(2007)}]{kothes}
{Kothes}, R. \& {Dougherty}, S.~M. 2007, \aap, 468, 993

\bibitem[{{Krumholz} {et~al.}(2014){Krumholz}, {Bate}, {Arce}, \&
  et~al.}]{krumholz}
{Krumholz}, M.~R., {Bate}, M.~R., {Arce}, H.~G., \& et~al. 2014, Protostars and
  Planets VI, 243

\bibitem[{{Kudryavtseva} {et~al.}(2012){Kudryavtseva}, {Brandner}, {Gennaro},
  {Rochau}, {Stolte}, {Andersen}, {Da Rio}, {Henning}, {Tognelli}, {Hogg},
  {Clark}, \& {Waters}}]{kud}
{Kudryavtseva}, N., {Brandner}, W., {Gennaro}, M., {et~al.} 2012, \apjl, 750,
  L44

\bibitem[{{Lagadec} {et~al.}(2011){Lagadec}, {Zijlstra}, {Oudmaijer},
  {Verhoelst}, {Cox}, {Szczerba}, {M{\'e}karnia}, \& {van Winckel}}]{lagadec}
{Lagadec}, E., {Zijlstra}, A.~A., {Oudmaijer}, R.~D., {et~al.} 2011, \aap, 534,
  L10

\bibitem[{{Lamers} \& {Leitherer}(1993)}]{lamers}
{Lamers}, H.~J.~G.~L.~M. \& {Leitherer}, C. 1993, \apj, 412, 771

\bibitem[{{Lamers} {et~al.}(1995){Lamers}, {Snow}, \& {Lindholm}}]{lamers95}
{Lamers}, H.~J.~G.~L.~M., {Snow}, T.~P., \& {Lindholm}, D.~M. 1995, \apj, 455,
  269

\bibitem[{{Lang} {et~al.}(2005){Lang}, {Johnson}, {Goss}, \&
  {Rodr{\'{\i}}guez}}]{lang}
{Lang}, C.~C., {Johnson}, K.~E., {Goss}, W.~M., \& {Rodr{\'{\i}}guez}, L.~F.
  2005, \aj, 130, 2185

\bibitem[{{Leitherer} {et~al.}(1995){Leitherer}, {Chapman}, \&
  {Koribalski}}]{leitherer95}
{Leitherer}, C., {Chapman}, J.~M., \& {Koribalski}, B. 1995, \apj, 450, 289

\bibitem[{{Leitherer} {et~al.}(1997){Leitherer}, {Chapman}, \&
  {Koribalski}}]{leitherer97}
{Leitherer}, C., {Chapman}, J.~M., \& {Koribalski}, B. 1997, \apj, 481, 898

\bibitem[{{Leitherer} \& {Robert}(1991)}]{leitherer91}
{Leitherer}, C. \& {Robert}, C. 1991, \apj, 377, 629

\bibitem[{{Levesque} {et~al.}(2014){Levesque}, {Stringfellow}, {Ginsburg},
  {Bally}, \& {Keeney}}]{levesque}
{Levesque}, E.~M., {Stringfellow}, G.~S., {Ginsburg}, A.~G., {Bally}, J., \&
  {Keeney}, B.~A. 2014, \aj, 147, 23

\bibitem[{{Liu} {et~al.}(2000){Liu}, {Hu}, {Hang}, {Qiu}, {Zhu}, \&
  {Qiao}}]{liu}
{Liu}, Q.-Z., {Hu}, J.-Y., {Hang}, H.-R., {et~al.} 2000, \aaps, 144, 219

\bibitem[{{Lobel} {et~al.}(2003){Lobel}, {Dupree}, {Stefanik}, {Torres},
  {Israelian}, {Morrison}, {de Jager}, {Nieuwenhuijzen}, {Ilyin}, \&
  {Musaev}}]{lobel}
{Lobel}, A., {Dupree}, A.~K., {Stefanik}, R.~P., {et~al.} 2003, \apj, 583, 923

\bibitem[{{Mackey} {et~al.}(2015){Mackey}, {Castro}, {Fossati}, \&
  {Langer}}]{mackey15}
{Mackey}, J., {Castro}, N., {Fossati}, L., \& {Langer}, N. 2015, \aap, 582, A24

\bibitem[{{Mackey} {et~al.}(2014){Mackey}, {Mohamed}, {Gvaramadze}, {Kotak},
  {Langer}, {Meyer}, {Moriya}, \& {Neilson}}]{mackey14}
{Mackey}, J., {Mohamed}, S., {Gvaramadze}, V.~V., {et~al.} 2014, \nat, 512, 282

\bibitem[{{Martins} {et~al.}(2007){Martins}, {Genzel}, {Hillier}, {Eisenhauer},
  {Paumard}, {Gillessen}, {Ott}, \& {Trippe}}]{martins07}
{Martins}, F., {Genzel}, R., {Hillier}, D.~J., {et~al.} 2007, \aap, 468, 233

\bibitem[{{Martins} {et~al.}(2008){Martins}, {Hillier}, {Paumard},
  {Eisenhauer}, {Ott}, \& {Genzel}}]{martins08}
{Martins}, F., {Hillier}, D.~J., {Paumard}, T., {et~al.} 2008, \aap, 478, 219

\bibitem[{{Martins} \& {Palacios}(2017)}]{martins17}
{Martins}, F. \& {Palacios}, A. 2017, \aap, 598, A56

\bibitem[{{Massa} {et~al.}(2003){Massa}, {Fullerton}, {Sonneborn}, \&
  {Hutchings}}]{massa}
{Massa}, D., {Fullerton}, A.~W., {Sonneborn}, G., \& {Hutchings}, J.~B. 2003,
  \apj, 586, 996

\bibitem[{{Massardi} {et~al.}(2016){Massardi}, {Bonaldi}, {Bonavera}, {De
  Zotti}, {Lopez-Caniego}, \& {Galluzzi}}]{massardi16}
{Massardi}, M., {Bonaldi}, A., {Bonavera}, L., {et~al.} 2016, \mnras, 455, 3249

\bibitem[{{Maud} {et~al.}(2017){Maud}, {Tilanus}, {van Kempen}, {Hogerheijde},
  {Schmalzl}, {Yoon}, {Contreras}, {Toribio}, {Asaki}, {Dent}, {Fomalont}, \&
  {Matsushita}}]{maud17}
{Maud}, L.~T., {Tilanus}, R.~P.~J., {van Kempen}, T.~A., {et~al.} 2017, \aap,
  605, A121

\bibitem[{{Mauerhan} {et~al.}(2013){Mauerhan}, {Smith}, {Silverman},
  {Filippenko}, {Morgan}, {Cenko}, {Ganeshalingam}, {Clubb}, {Bloom},
  {Matheson}, \& {Milne}}]{mauerhan}
{Mauerhan}, J.~C., {Smith}, N., {Silverman}, J.~M., {et~al.} 2013, \mnras, 431,
  2599

\bibitem[{{Mocanu} {et~al.}(2013){Mocanu}, {Crawford}, {Vieira}, {Aird},
  {Aravena}, {Austermann}, {Benson}, {B{\'e}thermin}, {Bleem}, {Bothwell},
  {Carlstrom}, {Chang}, {Chapman}, {Cho}, {Crites}, {de Haan}, {Dobbs},
  {Everett}, {George}, {Halverson}, {Harrington}, {Hezaveh}, {Holder},
  {Holzapfel}, {Hoover}, {Hrubes}, {Keisler}, {Knox}, {Lee}, {Leitch},
  {Lueker}, {Luong-Van}, {Marrone}, {McMahon}, {Mehl}, {Meyer}, {Mohr},
  {Montroy}, {Natoli}, {Padin}, {Plagge}, {Pryke}, {Rest}, {Reichardt}, {Ruhl},
  {Sayre}, {Schaffer}, {Shirokoff}, {Spieler}, {Spilker}, {Stalder},
  {Staniszewski}, {Stark}, {Story}, {Switzer}, {Vanderlinde}, \&
  {Williamson}}]{mocanu13}
{Mocanu}, L.~M., {Crawford}, T.~M., {Vieira}, J.~D., {et~al.} 2013, \apj, 779,
  61

\bibitem[{{Montes} {et~al.}(2015){Montes}, {Alberdi}, {P{\'e}rez-Torres}, \&
  {Gonz{\'a}lez}}]{montes}
{Montes}, G., {Alberdi}, A., {P{\'e}rez-Torres}, M.~A., \& {Gonz{\'a}lez},
  R.~F. 2015, \rmxaa, 51, 209

\bibitem[{{Morford} {et~al.}(2017){Morford}, {Fenech}, {Prinja}, {Blomme}, \&
  {Yates}}]{morford}
{Morford}, J.~C., {Fenech}, D.~M., {Prinja}, R.~K., {Blomme}, R., \& {Yates},
  J.~A. 2017, \mnras, 463, 763

\bibitem[{{Muno} {et~al.}(2006{\natexlab{a}}){Muno}, {Clark}, {Crowther},
  {Dougherty}, {de Grijs}, {Law}, {McMillan}, {Morris}, {Negueruela}, {Pooley},
  {Portegies Zwart}, \& {Yusef-Zadeh}}]{muno06a}
{Muno}, M.~P., {Clark}, J.~S., {Crowther}, P.~A., {et~al.} 2006{\natexlab{a}},
  \apjl, 636, L41

\bibitem[{{Muno} {et~al.}(2006{\natexlab{b}}){Muno}, {Law}, {Clark},
  {Dougherty}, {de Grijs}, {Portegies Zwart}, \& {Yusef-Zadeh}}]{muno06b}
{Muno}, M.~P., {Law}, C., {Clark}, J.~S., {et~al.} 2006{\natexlab{b}}, \apj,
  650, 203

\bibitem[{{Negueruela} {et~al.}(2010){Negueruela}, {Clark}, \&
  {Ritchie}}]{iggy10}
{Negueruela}, I., {Clark}, J.~S., \& {Ritchie}, B.~W. 2010, \aap, 516, A78

\bibitem[{{Nieuwenhuijzen} \& {de Jager}(1990)}]{nieu90}
{Nieuwenhuijzen}, H. \& {de Jager}, C. 1990, \aap, 231, 134

\bibitem[{{Nieuwenhuijzen} {et~al.}(2012){Nieuwenhuijzen}, {De Jager}, {Kolka},
  {Israelian}, {Lobel}, {Zsoldos}, {Maeder}, \& {Meynet}}]{nieu}
{Nieuwenhuijzen}, H., {De Jager}, C., {Kolka}, I., {et~al.} 2012, \aap, 546,
  A105

\bibitem[{{Noriega-Crespo} {et~al.}(1997){Noriega-Crespo}, {van Buren}, {Cao},
  \& {Dgani}}]{noriega}
{Noriega-Crespo}, A., {van Buren}, D., {Cao}, Y., \& {Dgani}, R. 1997, \aj,
  114, 837

\bibitem[{{Nugis} \& {Lamers}(2000)}]{nugis}
{Nugis}, T. \& {Lamers}, H.~J.~G.~L.~M. 2000, \aap, 360, 227

\bibitem[{{Ohm} {et~al.}(2013){Ohm}, {Hinton}, \& {White}}]{ohm}
{Ohm}, S., {Hinton}, J.~A., \& {White}, R. 2013, \mnras, 434, 2289

\bibitem[{{Ono} {et~al.}(2014){Ono}, {Ouchi}, {Kurono}, \& {Momose}}]{ono14}
{Ono}, Y., {Ouchi}, M., {Kurono}, Y., \& {Momose}, R. 2014, \apj, 795, 5

\bibitem[{{Oteo} {et~al.}(2016){Oteo}, {Zwaan}, {Ivison}, {Smail}, \&
  {Biggs}}]{oteo16}
{Oteo}, I., {Zwaan}, M.~A., {Ivison}, R.~J., {Smail}, I., \& {Biggs}, A.~D.
  2016, \apj, 822, 36

\bibitem[{{Oudmaijer}(1998)}]{oudmaijer98}
{Oudmaijer}, R.~D. 1998, \aaps, 129, 541

\bibitem[{{Oudmaijer} {et~al.}(1996){Oudmaijer}, {Groenewegen}, {Matthews},
  {Blommaert}, \& {Sahu}}]{oudmaijer96}
{Oudmaijer}, R.~D., {Groenewegen}, M.~A.~T., {Matthews}, H.~E., {Blommaert},
  J.~A.~D.~L., \& {Sahu}, K.~C. 1996, \mnras, 280, 1062

\bibitem[{{Pauldrach} \& {Puls}(1990)}]{pp}
{Pauldrach}, A.~W.~A. \& {Puls}, J. 1990, \aap, 237, 409

\bibitem[{Peck(2014)}]{peck14}
Peck, L.~W. 2014, PhD thesis, University College London

\bibitem[{{Petrovic} {et~al.}(2005){Petrovic}, {Langer}, \& {van der
  Hucht}}]{petrovic}
{Petrovic}, J., {Langer}, N., \& {van der Hucht}, K.~A. 2005, \aap, 435, 1013

\bibitem[{{Pittard}(2011)}]{pittard11}
{Pittard}, J. 2011, Bulletin de la Societe Royale des Sciences de Liege, 80,
  555

\bibitem[{{Pittard}(2010)}]{pittard10}
{Pittard}, J.~M. 2010, \mnras, 403, 1633

\bibitem[{{Pittard} {et~al.}(2006){Pittard}, {Dougherty}, {Coker}, {O'Connor},
  \& {Bolingbroke}}]{pittard06}
{Pittard}, J.~M., {Dougherty}, S.~M., {Coker}, R.~F., {O'Connor}, E., \&
  {Bolingbroke}, N.~J. 2006, \aap, 446, 1001

\bibitem[{{Prinja} {et~al.}(2010){Prinja}, {Hodges}, {Urbaneja}, \&
  {Massa}}]{prinja10}
{Prinja}, R.~K., {Hodges}, S.~E., {Urbaneja}, M.~A., \& {Massa}, D.~L. 2010,
  \mnras, 402, 641

\bibitem[{{Prinja} \& {Massa}(2013)}]{prinja13}
{Prinja}, R.~K. \& {Massa}, D.~L. 2013, \aap, 559, A15

\bibitem[{{Puls} {et~al.}(2006){Puls}, {Markova}, {Scuderi}, {Stanghellini},
  {Taranova}, {Burnley}, \& {Howarth}}]{puls}
{Puls}, J., {Markova}, N., {Scuderi}, S., {et~al.} 2006, \aap, 454, 625

\bibitem[{{Rea} {et~al.}(2012){Rea}, {Pons}, {Torres}, \& {Turolla}}]{rea}
{Rea}, N., {Pons}, J.~A., {Torres}, D.~F., \& {Turolla}, R. 2012, \apjl, 748,
  L12

\bibitem[{{Ritchie} {et~al.}(2009{\natexlab{a}}){Ritchie}, {Clark},
  {Negueruela}, \& {Crowther}}]{ritchie09a}
{Ritchie}, B.~W., {Clark}, J.~S., {Negueruela}, I., \& {Crowther}, P.~A.
  2009{\natexlab{a}}, \aap, 507, 1585

\bibitem[{{Ritchie} {et~al.}(2010){Ritchie}, {Clark}, {Negueruela}, \&
  {Langer}}]{ritchie10}
{Ritchie}, B.~W., {Clark}, J.~S., {Negueruela}, I., \& {Langer}, N. 2010, \aap,
  520, A48

\bibitem[{{Ritchie} {et~al.}(2009{\natexlab{b}}){Ritchie}, {Clark},
  {Negueruela}, \& {Najarro}}]{ritchie09b}
{Ritchie}, B.~W., {Clark}, J.~S., {Negueruela}, I., \& {Najarro}, F.
  2009{\natexlab{b}}, \aap, 507, 1597

\bibitem[{{Ritchie} {et~al.}(\noop{2017}in prep.){Ritchie}, {Clark},
  {Negueruela}, {fake}, {fake}, {fake}, {fake}, {fake}, \& {fake}}]{ritchie17}
{Ritchie}, B.~W., {Clark}, S., {Negueruela}, I., {et~al.} \noop{2017}in prep.

\bibitem[{{Runacres} \& {Owocki}(2002)}]{runacres}
{Runacres}, M.~C. \& {Owocki}, S.~P. 2002, \aap, 381, 1015

\bibitem[{{Sana} {et~al.}(2013){Sana}, {de Koter}, {de Mink}, {Dunstall},
  {Evans}, {H{\'e}nault-Brunet}, {Ma{\'{\i}}z Apell{\'a}niz},
  {Ram{\'{\i}}rez-Agudelo}, {Taylor}, {Walborn}, {Clark}, {Crowther},
  {Herrero}, {Gieles}, {Langer}, {Lennon}, \& {Vink}}]{sana13}
{Sana}, H., {de Koter}, A., {de Mink}, S.~E., {et~al.} 2013, \aap, 550, A107

\bibitem[{{Sana} {et~al.}(2012){Sana}, {de Mink}, {de Koter}, {Langer},
  {Evans}, {Gieles}, {Gosset}, {Izzard}, {Le Bouquin}, \& {Schneider}}]{sana12}
{Sana}, H., {de Mink}, S.~E., {de Koter}, A., {et~al.} 2012, Science, 337, 444

\bibitem[{{Sander} {et~al.}(2012){Sander}, {Hamann}, \& {Todt}}]{sander}
{Sander}, A., {Hamann}, W.-R., \& {Todt}, H. 2012, \aap, 540, A144

\bibitem[{{Schneider} {et~al.}(2014){Schneider}, {Izzard}, {de Mink}, {Langer},
  {Stolte}, {de Koter}, {Gvaramadze}, {Hu{\ss}mann}, {Liermann}, \&
  {Sana}}]{schneider14}
{Schneider}, F.~R.~N., {Izzard}, R.~G., {de Mink}, S.~E., {et~al.} 2014, \apj,
  780, 117

\bibitem[{{Schneider} {et~al.}(2015){Schneider}, {Izzard}, {Langer}, \& {de
  Mink}}]{schneider15}
{Schneider}, F.~R.~N., {Izzard}, R.~G., {Langer}, N., \& {de Mink}, S.~E. 2015,
  \apj, 805, 20

\bibitem[{{Schuster} {et~al.}(2006){Schuster}, {Humphreys}, \&
  {Marengo}}]{schuster}
{Schuster}, M.~T., {Humphreys}, R.~M., \& {Marengo}, M. 2006, \aj, 131, 603

\bibitem[{{Scuderi} {et~al.}(1998){Scuderi}, {Panagia}, {Stanghellini},
  {Trigilio}, \& {Umana}}]{scuderi}
{Scuderi}, S., {Panagia}, N., {Stanghellini}, C., {Trigilio}, C., \& {Umana},
  G. 1998, \aap, 332, 251

\bibitem[{{Searle} {et~al.}(2008){Searle}, {Prinja}, {Massa}, \&
  {Ryans}}]{searle}
{Searle}, S.~C., {Prinja}, R.~K., {Massa}, D., \& {Ryans}, R. 2008, \aap, 481,
  777

\bibitem[{{Serabyn} {et~al.}(1991){Serabyn}, {Lacy}, \& {Achtermann}}]{serabyn}
{Serabyn}, E., {Lacy}, J.~H., \& {Achtermann}, J.~M. 1991, \apj, 378, 557

\bibitem[{{Shenoy} {et~al.}(2016){Shenoy}, {Humphreys}, {Jones}, {Marengo},
  {Gehrz}, {Helton}, {Hoffmann}, {Skemer}, \& {Hinz}}]{shenoy}
{Shenoy}, D., {Humphreys}, R.~M., {Jones}, T.~J., {et~al.} 2016, \aj, 151, 51

\bibitem[{{Shimizu} {et~al.}(2012){Shimizu}, {Yoshida}, \&
  {Okamoto}}]{shimizu12}
{Shimizu}, I., {Yoshida}, N., \& {Okamoto}, T. 2012, \mnras, 427, 2866

\bibitem[{{Skinner} {et~al.}(1997){Skinner}, {Exter}, {Barlow}, {Davis}, \&
  {Bode}}]{skinner}
{Skinner}, C.~J., {Exter}, K.~M., {Barlow}, M.~J., {Davis}, R.~J., \& {Bode},
  M.~F. 1997, \mnras, 288, L7

\bibitem[{{Smartt}(2009)}]{smartt}
{Smartt}, S.~J. 2009, \araa, 47, 63

\bibitem[{{Smith} {et~al.}(2009{\natexlab{a}}){Smith}, {Hinkle}, \&
  {Ryde}}]{smith09}
{Smith}, N., {Hinkle}, K.~H., \& {Ryde}, N. 2009{\natexlab{a}}, \aj
  [\eprint[arXiv]{0811.3037}]

\bibitem[{{Smith} {et~al.}(2001){Smith}, {Humphreys}, {Davidson}, {Gehrz},
  {Schuster}, \& {Krautter}}]{smith01}
{Smith}, N., {Humphreys}, R.~M., {Davidson}, K., {et~al.} 2001, \aj, 121, 1111

\bibitem[{{Smith} {et~al.}(2007){Smith}, {Li}, {Foley}, {Wheeler}, {Pooley},
  {Chornock}, {Filippenko}, {Silverman}, {Quimby}, {Bloom}, \&
  {Hansen}}]{smith07}
{Smith}, N., {Li}, W., {Foley}, R.~J., {et~al.} 2007, \apj, 666, 1116

\bibitem[{{Smith} {et~al.}(2009{\natexlab{b}}){Smith}, {Silverman}, {Chornock},
  {Filippenko}, {Wang}, {Li}, {Ganeshalingam}, {Foley}, {Rex}, \&
  {Steele}}]{smith09b}
{Smith}, N., {Silverman}, J.~M., {Chornock}, R., {et~al.} 2009{\natexlab{b}},
  \apj [\eprint[arXiv]{0809.5079}]

\bibitem[{{Stevens}(1995)}]{stevens95}
{Stevens}, I.~R. 1995, \mnras, 277, 163

\bibitem[{{Stevens} {et~al.}(1992){Stevens}, {Blondin}, \&
  {Pollock}}]{stevens92}
{Stevens}, I.~R., {Blondin}, J.~M., \& {Pollock}, A.~M.~T. 1992, \apj, 386, 265

\bibitem[{{Stevens} \& {Hartwell}(2003)}]{stevens}
{Stevens}, I.~R. \& {Hartwell}, J.~M. 2003, \mnras, 339, 280

\bibitem[{{Sundqvist} {et~al.}(2011){Sundqvist}, {Puls}, {Feldmeier}, \&
  {Owocki}}]{sundqvist}
{Sundqvist}, J.~O., {Puls}, J., {Feldmeier}, A., \& {Owocki}, S.~P. 2011, \aap,
  528, A64

\bibitem[{{Sylvester} {et~al.}(1998){Sylvester}, {Skinner}, \&
  {Barlow}}]{sylvester}
{Sylvester}, R.~J., {Skinner}, C.~J., \& {Barlow}, M.~J. 1998, \mnras, 301,
  1083

\bibitem[{{Szary} {et~al.}(2015){Szary}, {Melikidze}, \& {Gil}}]{szary}
{Szary}, A., {Melikidze}, G.~I., \& {Gil}, J. 2015, \apj, 800, 76

\bibitem[{{Tiffany} {et~al.}(2010){Tiffany}, {Humphreys}, {Jones}, \&
  {Davidson}}]{tiffany}
{Tiffany}, C., {Humphreys}, R.~M., {Jones}, T.~J., \& {Davidson}, K. 2010, \aj,
  140, 339

\bibitem[{{Torne} {et~al.}(2015){Torne}, {Eatough}, {Karuppusamy}, {Kramer},
  {Paubert}, {Klein}, {Desvignes}, {Champion}, {Wiesemeyer}, {Kramer},
  {Spitler}, {Thum}, {G{\"u}sten}, {Schuster}, \& {Cognard}}]{torne}
{Torne}, P., {Eatough}, R.~P., {Karuppusamy}, R., {et~al.} 2015, \mnras, 451,
  L50

\bibitem[{{Umehata} {et~al.}(2017){Umehata}, {Tamura}, {Kohno}, {Ivison},
  {Smail}, {Hatsukade}, {Nakanishi}, {Kato}, {Ikarashi}, {Matsuda}, {Fujimoto},
  {Iono}, {Lee}, {Steidel}, {Saito}, {Alexander}, {Yun}, \& {Kubo}}]{umehata17}
{Umehata}, H., {Tamura}, Y., {Kohno}, K., {et~al.} 2017, \apj, 835, 98

\bibitem[{{{\v S}urlan} {et~al.}(2012){{\v S}urlan}, {Hamann}, {Kub{\'a}t},
  {Oskinova}, \& {Feldmeier}}]{surlan}
{{\v S}urlan}, B., {Hamann}, W.-R., {Kub{\'a}t}, J., {Oskinova}, L.~M., \&
  {Feldmeier}, A. 2012, \aap, 541, A37

\bibitem[{{van Loon} {et~al.}(1999){van Loon}, {Groenewegen}, {de Koter},
  {Trams}, {Waters}, {Zijlstra}, {Whitelock}, \& {Loup}}]{vanloon}
{van Loon}, J.~T., {Groenewegen}, M.~A.~T., {de Koter}, A., {et~al.} 1999,
  \aap, 351, 559

\bibitem[{{Vink} {et~al.}(1999){Vink}, {de Koter}, \& {Lamers}}]{vink99}
{Vink}, J.~S., {de Koter}, A., \& {Lamers}, H.~J.~G.~L.~M. 1999, \aap, 350, 181

\bibitem[{{Vink} {et~al.}(2000){Vink}, {de Koter}, \& {Lamers}}]{vink00}
{Vink}, J.~S., {de Koter}, A., \& {Lamers}, H.~J.~G.~L.~M. 2000, \aap, 362, 295

\bibitem[{{Vink} {et~al.}(2001){Vink}, {de Koter}, \& {Lamers}}]{vink}
{Vink}, J.~S., {de Koter}, A., \& {Lamers}, H.~J.~G.~L.~M. 2001, \aap, 369, 574

\bibitem[{{Wallstr{\"o}m} {et~al.}(2017){Wallstr{\"o}m}, {Lagadec}, {Muller},
  \& et~al.}]{wallstrom}
{Wallstr{\"o}m}, S.~H.~J., {Lagadec}, E., {Muller}, S., \& et~al. 2017, \aap,
  597, A99

\bibitem[{{Wellstein} \& {Langer}(1999)}]{wellstein}
{Wellstein}, S. \& {Langer}, N. 1999, \aap, 350, 148

\bibitem[{{Westerlund}(1961)}]{westerlund}
{Westerlund}, B. 1961, \pasp, 73, 51

\bibitem[{{White} \& {Becker}(1985)}]{white}
{White}, R.~L. \& {Becker}, R.~H. 1985, \apj, 297, 677

\bibitem[{{Willis}(1991)}]{willis}
{Willis}, A.~J. 1991, in IAU Symposium, Vol. 143, Wolf-Rayet Stars and
  Interrelations with Other Massive Stars in Galaxies, ed. K.~A. {van der
  Hucht} \& B.~{Hidayat}, 265

\bibitem[{{Wright} \& {Barlow}(1975)}]{wright75}
{Wright}, A.~E. \& {Barlow}, M.~J. 1975, \mnras, 170, 41

\bibitem[{{Wright} {et~al.}(2014){Wright}, {Wesson}, {Drew}, \&
  et~al.}]{wright}
{Wright}, N.~J., {Wesson}, R., {Drew}, J.~E., \& et~al. 2014, \mnras, 437, L1

\bibitem[{{Yaron} {et~al.}(2017){Yaron}, {Perley}, {Gal-Yam}, \&
  et~al.}]{yaron}
{Yaron}, O., {Perley}, D.~A., {Gal-Yam}, A., \& et~al. 2017, Nature Physics,
  13, 510

\bibitem[{{Yoon} \& {Cantiello}(2010)}]{yoon}
{Yoon}, S.-C. \& {Cantiello}, M. 2010, \apjl, 717, L62

\bibitem[{{Yusef-Zadeh} \& {Morris}(1991)}]{YZ}
{Yusef-Zadeh}, F. \& {Morris}, M. 1991, \apjl, 371, L59

\bibitem[{{Yusef-Zadeh} {et~al.}(2017){Yusef-Zadeh}, {Sch{\"o}del}, {Wardle},
  {Bushouse}, {Cotton}, {Royster}, {Kunneriath}, {Roberts}, \&
  {Gallego-Cano}}]{YZ17}
{Yusef-Zadeh}, F., {Sch{\"o}del}, R., {Wardle}, M., {et~al.} 2017, \mnras, 470,
  4209

\end{thebibliography}


\appendix

\counterwithin{table}{section}

\onecolumn

\section{Tables}

\setcellgapes{4pt}
\begin{table*}[ht!]
\small
\centering
\caption{ALMA 3mm source properties for all SEAC detected sources without catalogued identification in the literature.}
\newcolumntype{Y}{>{\centering\arraybackslash}X}
\newcolumntype{L}{>{\arraybackslash}m{3cm}}
\setlength{\tabcolsep}{0.2em} 
\renewcommand{\arraystretch}{1.1}
\begin{tabularx}{\linewidth}{|l|Y|Y|Y|Y|Y|Y|Y|L|}
\hline
Source & RA & Dec & Flux density (mJy) & \multicolumn{4}{c|}{Size} & Source type \\
&&&& \multicolumn{2}{c|}{Convolved} & \multicolumn{2}{c|}{Deconvolved} & \\
& J2000 & J2000 & & Major axis & Minor axis & Major axis & Minor axis & Indicator \\
\hline
2 & 46 58.528 &   50 31.259 & 0.21\,$\pm$\,0.06 & 1.65  & LAS  &---&--- & Extended/diffuse \\
3 & 46 58.890  &  50 28.560 & 1.83\,$\pm$\,0.17 & 3.93  & LAS   &---&--- & Isolated/diffuse  \\
3A & 46 58.890  &  50 28.560 & 1.67\,$\pm$\,0.16 & 1.31\,$\pm$\,0.08 & 1.11\,$\pm$\,0.07 & 1.14\,$\pm$\,0.10 & 0.95\,$\pm$\,0.10 &  \\
5 & 46 59.337  &  50 33.781 & 0.22\,$\pm$\,0.06 & 0.97  & LAS   &---&--- & Isolated/diffuse \\
6 & 46 59.397  &  50 37.111 & 8.98\,$\pm$\,0.55 & 11.87  & LAS   &---&--- & Extended/diffuse  \\
6A & 46 59.397  &  50 37.111 & 2.35\,$\pm$\,0.19 & 1.43\,$\pm$\,0.07 & 1.14\,$\pm$\,0.06 & 1.28\,$\pm$\,0.09 & 0.98\,$\pm$\,0.07 &  \\
6B & 46 59.613  &  50 37.922 & 0.77\,$\pm$\,0.11 & 1.05 \,$\pm$\,0.10 & 0.98\,$\pm$\,0.10 & 0.85\,$\pm$\,0.18 & 0.77\,$\pm$\,0.19 &  \\
6C & 46 59.716 &   50 37.562 & 1.39\,$\pm$\,0.18 & 1.68\,$\pm$\,0.17 & 1.14\,$\pm$\,0.11 & 1.56\,$\pm$\,0.19 & 0.96\,$\pm$\,0.14 &  \\
6D & 47 0.086  &  50 39.093 & 1.91\,$\pm$\,0.18 & 1.53\,$\pm$\,0.11 & 1.20\,$\pm$\,0.08 & 1.40\,$\pm$\,0.12 & 1.03\,$\pm$\,0.11 &  \\
7 & 46 59.509  &  50 41.611 & 1.09\,$\pm$\,0.12 & 5.79  & LAS   &---&--- & Extended/diffuse  \\
7A & 46 59.148  &  50 40.260 & 0.63\,$\pm$\,0.12 & 1.34\,$\pm$\,0.19 & 0.90\,$\pm$\,0.13 & 1.20\,$\pm$\,0.22 & 0.63\,$\pm$\,0.20 &  \\
7B & 46 59.268  &  50 41.071 & 0.22\,$\pm$\,0.07 & 0.82\,$\pm$\,0.17 & 0.73\,$\pm$\,0.15 & 0.59\,$\pm$\,0.41 & 0.32\,$\pm$\,0.38 &  \\
7C & 46 59.509  &  50 41.611 & 0.66\,$\pm$\,0.11 & 1.12\,$\pm$\,0.13 & 0.94\,$\pm$\,0.11 & 0.95\,$\pm$\,0.18 & 0.70\,$\pm$\,0.18 &  \\
8 & 46 59.604 &   50 26.762 & 0.28\,$\pm$\,0.08 & 0.92\,$\pm$\,0.20 & 0.87\,$\pm$\,0.19 & 0.72\,$\pm$\,0.29 & 0.57\,$\pm$\,0.35 & Isolated  \\
11 & 47 0.974  &  50 39.544 & 0.06\,$\pm$\,0.03 & 0.94  & LAS   &---&--- & W4 nebula emission \\
12 & 47 1.017 &   50 36.214 & 0.38\,$\pm$\,0.08 & 1.77  & LAS   &---&--- & W4 nebula emission \\
14 & 47 1.490  &  50 42.245 & 0.23\,$\pm$\,0.06 & 1.86  & LAS   &---&--- & W4 nebula emission \\
16 & 47 1.809 &   51 22.925 & 1.35\,$\pm$\,0.12 & 1.13\,$\pm$\,0.06 & 0.86\,$\pm$\,0.05 & 0.98\,$\pm$\,0.07 & 0.57\,$\pm$\,0.08 & Isolated \\
18 & 47 2.050  &  51 38.945 & 0.10\,$\pm$\,0.05 & 0.59\,$\pm$\,0.16 & 0.54\,$\pm$\,0.14 & 0.15\,$\pm$\,0.24 & 0.00\,$\pm$\,0.19 & Isolated/diffuse  \\
22 & 47 2.308 &   51 21.486 & 0.18\,$\pm$\,0.04 & 1.21  & LAS   &---&--- & Isolated \\
25 & 47 2.722  &  50 26.046 & 0.18\,$\pm$\,0.06 & 0.83\,$\pm$\,0.18 & 0.60 0.13 & 0.59\,$\pm$\,0.33 & 0.00\,$\pm$\,0.22 & Isolated/diffuse  \\
26 & 47 2.722  &  50 5.616 & 0.06\,$\pm$\,0.02 & 0.93  & LAS   &---&--- & Isolated/diffuse \\
27 & 47 2.842  &  51 18.966 & 0.77\,$\pm$\,0.12 & 1.26\,$\pm$\,0.14 & 0.95\,$\pm$\,0.11 & 1.08\,$\pm$\,0.17 & 0.76\,$\pm$\,0.14 & Isolated \\
28 & 47 2.912  &  50 26.676 & 0.13\,$\pm$\,0.05 & 1.05  & LAS   &---&--- & Isolated/diffuse \\
35 & 47 3.644  &  51 16.086 & 0.36\,$\pm$\,0.09 & 1.24\,$\pm$\,0.23 & 0.72\,$\pm$\,0.13 & 1.10\,$\pm$\,0.27 & 0.33\,$\pm$\,0.29 & Isolated/diffuse \\
37 & 47 3.764 &   50 24.336 & 0.14\,$\pm$\,0.05 & 0.69\,$\pm$\,0.15 & 0.53\,$\pm$\,0.11 & 0.32\,$\pm$\,0.31 & 0.00\,$\pm$\,0.27 & Isolated/diffuse \\
39 & 47 3.868  &  51 6.186 & 2.46\,$\pm$\,0.18 & 5.00  & LAS   &---&--- & Extended/potential sources \\
39A & 47 3.868 &   51 6.186 & 1.42\,$\pm$\,0.14 & 1.26\,$\pm$\,0.09 & 1.05\,$\pm$\,0.07 & 1.10\,$\pm$\,0.11 & 0.87\,$\pm$\,0.10 &  \\
39B & 47 3.833 &   51 7.986 & 1.21\,$\pm$\,0.15 & 1.42\,$\pm$\,0.13 & 1.04\,$\pm$\,0.09 & 1.30\,$\pm$\,0.14 & 0.82\,$\pm$\,0.13 &  \\
42 & 47 3.980 &   50 36.756 & 0.06\,$\pm$\,0.08 & 0.53  & LAS   &---&--- & Isolated/diffuse \\
49 & 47 4.204  &  51 16.086 & 0.23\,$\pm$\,0.07 & 0.87\,$\pm$\,0.19 & 0.74\,$\pm$\,0.16 & 0.66\,$\pm$\,0.45 & 0.35\,$\pm$\,0.43 & Isolated/diffuse \\
51 & 47 4.436  &  51 32.196 & 0.33\,$\pm$\,0.06 & 1.79  & LAS   &---&--- & W20 nebula emission \\
52 & 47 4.522  &  50 30.726 & 0.17\,$\pm$\,0.05 & 0.88\,$\pm$\,0.15 & 0.40\,$\pm$\,0.07 & 0.67\,$\pm$\,0.20 & 0.00\,$\pm$\,0.00 & Isolated/diffuse \\
54 & 47 4.660  &  50 35.676 & 0.18\,$\pm$\,0.06 & 1.15  & LAS   &---&--- & Isolated/diffuse, possibly associated with W9 filament \\
56 & 47 4.764  &  51 34.536 & 0.19\,$\pm$\,0.04 & 1.46  & LAS   &---&--- & W20 nebula emission \\
57 & 47 4.781 &   51 27.426 & 0.49\,$\pm$\,0.06 & 2.22  & LAS  &---&---  & W20 nebula emission \\
59 & 47 5.005 &   51 24.906 & 0.32\,$\pm$\,0.06 & 1.49  & LAS   &---&--- & W20 nebula emission \\
60 & 47 5.031  &  51 1.776 & 1.27\,$\pm$\,0.13 & 2.43  & LAS   &---&--- & Isolated/diffuse \\
64 & 47 5.685  &  51 7.535 & 2.63\,$\pm$\,0.18 & 3.29  & LAS   &---&--- & Isolated/diffuse \\
66 & 47 5.883  &  51 11.135 & 0.97\,$\pm$\,0.11 & 3.23  & LAS   &---&--- & Isolated/diffuse \\
69 & 47 6.107  &  51 3.755 & 0.65\,$\pm$\,0.10 & 1.26\,$\pm$\,0.14 & 0.78\,$\pm$\,0.08 & 1.09\,$\pm$\,0.17 & 0.50\,$\pm$\,0.16 & Isolated/diffuse \\
72 & 47 6.271  &  50 44.765 & 0.42\,$\pm$\,0.07 & 0.97\,$\pm$\,0.11 & 0.65\,$\pm$\,0.07 & 0.78\,$\pm$\,0.14 & 0.04\,$\pm$\,0.17 & Isolated/diffuse \\
73 & 47 6.383  &  51 13.474 & 1.16\,$\pm$\,0.13 & 2.98  & LAS   &---&--- & Isolated/diffuse \\
74 & 47 6.392 &   51 20.494 & 0.11\,$\pm$\,0.28 & 1.05  & LAS  &---&---  & Isolated/diffuse \\

\hline
\end{tabularx}
\tablefoot{ Sources positions, flux densities and sizes are listed. Where the sources are small and compact major and minor axis measurements and flux densities are taken from Gaussian fitting of the source. Where the source structure is more extended, largest angular sizes (LAS) are included and the flux densities as measured from SEAC (see Sect.). Where extended sources contain knots of emission, Gaussian fitting has been performed on the individual knots in addition to the whole source. For the relevant sources these are referred as A-D. Source type descriptions are as follows: Isolated sources show bright emission clearly isolated from other features and represent the most likely candidates for potential emission associated with stellar sources. Isolated/diffuse sources are similar to the isolated sources but are in the presence of low surface brightness emission. These could therefore represent stellar emission or brighter knots in diffuse background emission. Extended/diffuse represent sources that appear to be extended background emission.}
\label{tab:fcp18_list}
\end{table*}

\setcellgapes{4pt}
\begin{table*}
\small
\centering
\caption*{Table \ref{tab:fcp18_list}---continued}
\newcolumntype{Y}{>{\centering\arraybackslash}X}
\newcolumntype{L}{>{\arraybackslash}m{3cm}}
\setlength{\tabcolsep}{0.2em} 
\renewcommand{\arraystretch}{1.1}
\begin{tabularx}{\linewidth}{|l|Y|Y|Y|Y|Y|Y|Y|L|}
\hline
Source & RA & Dec & Flux density (mJy) & \multicolumn{4}{c|}{Size} & Source type \\
&&&& \multicolumn{2}{c|}{Convolved} & \multicolumn{2}{c|}{Deconvolved} & \\
& J2000 & J2000 & & Major axis & Minor axis & Major axis & Minor axis & Indicator \\
\hline
75 & 47 6.409 &   51 4.384 & 0.28\,$\pm$\,0.07 & 1.05\,$\pm$\,0.19 & 0.62\,$\pm$\,0.11 & 0.88\,$\pm$\,0.23 & 0.00\,$\pm$\,0.17 & Isolated/diffuse \\
77 & 47 6.469  &  51 23.374 & 0.09\,$\pm$\,0.02 & 1.10  & LAS   &---&--- & Isolated/diffuse \\
80 & 47 6.805  &  51 10.684 & 0.24\,$\pm$\,0.04 & 1.43  & LAS  & --- & --- & Isolated/diffuse \\
82 & 47 7.279 &   51 10.953 & 0.31\,$\pm$\,0.08 & 1.02\,$\pm$\,0.17 & 0.69\,$\pm$\,0.12 & 0.83\,$\pm$\,0.23 & 0.26\,$\pm$\,0.26 & Isolated/diffuse \\
85 & 47 7.805  &  51 15.272 & 0.06\,$\pm$\,0.02 & 1.21  & LAS  &--- &---& Isolated/diffuse \\
86 & 47 8.295  &  50 43.321 & 0.16\,$\pm$\,0.06 & 0.71\,$\pm$\,0.16 & 0.66\,$\pm$\,0.15 & 0.39\,$\pm$\,0.33 & 0.21\,$\pm$\,0.32 & Isolated/diffuse \\
87 & 47 8.304  &  51 19.501 & 0.35\,$\pm$\,0.08 & 1.00\,$\pm$\,0.16 & 0.74\,$\pm$\,0.12 & 0.78\,$\pm$\,0.23 & 0.45\,$\pm$\,0.30 & Isolated/diffuse \\
90 & 47 8.494  &  51 11.040 & 1.33\,$\pm$\,0.13 & 3.61  & LAS  &--- &---& Isolated/diffuse \\
90A & 47 8.494  &  51 11.040 & 1.01\,$\pm$\,0.11 & 1.34\,$\pm$\,0.10 & 0.78\,$\pm$\,0.06 & 1.21\,$\pm$\,0.12 & 0.44\,$\pm$\,0.11 &  \\
91 & 47 8.527  &  50 38.370 & 0.12\,$\pm$\,0.05 & 0.63\,$\pm$\,0.15 & 0.59\,$\pm$\,0.14 & 0.20\,$\pm$\,0.22 & 0.00\,$\pm$\,0.26 & Isolated/diffuse \\
92 & 47 8.642  &  52 32.940 & 0.21\,$\pm$\,0.07 & 0.81\,$\pm$\,0.17 & 0.69\,$\pm$\,0.14 & 0.57\,$\pm$\,0.40 & 0.24\,$\pm$\,0.37 & Diffuse/Isolated \\
93 & 47 8.666  &  51 14.280 & 1.24\,$\pm$\,0.15 & 3.56  & LAS  & && Isolated/diffuse \\
93A & 47 8.666 &   51 14.280 & 0.82\,$\pm$\,0.12 & 1.16\,$\pm$\,0.12 & 0.10\,$\pm$\,0.10 & 0.96\,$\pm$\,0.16 & 0.82\,$\pm$\,0.16 &  \\
94 & 47 8.847  &  51 10.499 & 0.24\,$\pm$\,0.06 & 1.39  & LAS  &--- &---& Isolated/diffuse \\
95 & 47 8.871  &  50 6.419 & 0.45\,$\pm$\,0.06 & 1.58  & LAS  &--- &---& Isolated \\
97 & 47 8.915  &  50 30.359 & 0.15\,$\pm$\,0.06 & 0.78\,$\pm$\,0.18 & 0.55\,$\pm$\,0.13 & 0.45\,$\pm$\,0.37 & 0.00\,$\pm$\,0.19 & Diffuse/Isolated \\
99 & 47 9.148   & 51 3.208 & 0.16\,$\pm$\,0.04 & 1.12  & LAS  & ---&---& Isolated/diffuse \\
100 & 47 9.355  &  51 12.118 & 0.20\,$\pm$\,0.06 & 0.78\,$\pm$\,0.15 & 0.59\,$\pm$\,0.11 & 0.43\,$\pm$\,0.34 & 0.18\,$\pm$\,0.24 & Diffuse/Isolated \\
101 & 47 10.741  &  50 30.083 & 0.22\,$\pm$\,0.06 & 0.80\,$\pm$\,0.14 & 0.63\,$\pm$\,0.11 & 0.47\,$\pm$\,0.36 & 0.26\,$\pm$\,0.34 & Isolated/diffuse \\
\hline
\end{tabularx}
\end{table*}

\begin{table*}
\small
\centering
\caption{ALMA 3-mm upper limits for stars that are undetected.}
\begin{tabular}{|l|l|c|c|}
\hline
Spectral type & Source & \multicolumn{2}{|c|}{Upper limits}  \\
& & Flux Density (Jy) & Mass-loss rate \mdot $\times {f_{cl}^{0.5}}$ \\
\hline
O9Iab  &    W24   & <1.16e-04       &      <6.04e-06 \\
       &    W35   & <9.16e-05       &      <5.08e-06 \\
       &    W38   & <1.23e-04       &      <6.33e-06 \\
       &    W41   & <1.22e-04       &      <6.28e-06 \\
O9Ib   &    W15   & <1.04e-04       &      <5.68e-06 \\
       &    W29   & <9.38e-05       &      <5.26e-06 \\
       &    W37   & <1.61e-04       &      <7.89e-06 \\
       &    W43c   & <6.92e-05       &      <4.19e-06 \\
O9.5Iab &   W61b   & <6.13e-05       &      <3.75e-06 \\
       &    W74   & <6.23e-05       &      <3.80e-06 \\
       &    W2019   & <1.65e-04       &      <7.89e-06 \\
O9.5Ib &   W46b   & <7.20e-05       &      <4.13e-06 \\
       &    W56b   & <6.06e-05       &      <3.63e-06 \\
       &    W65   & <6.11e-05       &      <3.65e-06 \\
       &    W84   & <6.91e-05       &      <4.00e-06 \\
       &    W86   & <1.43e-04       &      <7.10e-06 \\
       &    W2017   & <6.40e-05       &      <3.78e-06 \\
       &    W2028   & <1.16e-04       &      <5.92e-06 \\
       &    W3005   & <1.32e-04       &      <6.50e-06 \\
B0Ia &      W31  &   <1.02e-04   &  <4.00e-06 \\
 &          W34   &  <8.76e-05  &      <3.57e-06 \\
 &          W43a  &  <1.91e-04  &      <5.90e-06 \\
 &          W43b  &  <1.46e-04     &      <4.82e-06 \\
 &          W55   &  <7.38e-05     &      <3.14e-06 \\
B0Iab  &    W49   &  <6.24e-05     &      <2.77e-06 \\
  &         W60   &  <8.15e-05     &      <3.38e-06  \\
  &         W63a  &  <6.66e-05      &     <2.90e-06 \\
  &         W232  &   <1.28e-04     &      <4.74e-06 \\
B0Ib  &    W2002   &  <7.82e-05     &      <3.28e-06 \\
  &        W2011 &  <5.22e-05      &     <2.42e-05 \\
  &        W3024 &  <5.75e-05      &     <2.60e-05 \\
B0.5Ia   &  W10  &  <1.50e-04       &     <4.83e-06 \\
  &         W18  &  <1.27e-04       &     <4.25e-06 \\      
  &         W21  &  <8.33e-06       &     <3.11e-06 \\
B0.5Iab  &  W6a  &  <1.08e-04       &     <3.78e-06 \\
  &         W54   & <4.45e-05       &     <1.94e-06 \\
  &         62a  & <8.32e-05        &     <3.19e-06 \\
B0.5Ib &  W62a  &  <8.29e-05       &     <3.10e-06 \\
B1Ia  &     W19  & <1.04e-04         &    <2.02e-06 \\
  &         W21   & <5.80e-05        &     <1.31e-06 \\
  &         W43b   & <6.76e-05        &     <1.46e-06 \\
  &         W78  & <1.15e-04        &     <2.18e-06 \\
  &         W3019  & <3.26e-04      &     <4.77e-06 \\
B1Iab  &    W49  & <1.26e-04        &     <3.26e-06 \\
   &        W238  & <4.91e-05       &      <1.15e-06 \\
B1.5Ia   &  W8b  & <2.76e-04        &     <2.91e-06 \\
   &        W52  & <1.32e-04        &      <1.68e-06 \\
B2Ia   &    W2a  & <9.82e-05         &     <1.50e-06 \\
   &        W11  & <2.14e-04        &       <2.69e-06 \\
B3Ia   &    W70  & <1.21e-04        &      <2.18e-06 \\
B4Ia   &    W57a  & <9.39e-05       &      <1.11e-06 \\
\hline
\end{tabular}
\tablefoot{Samples from each spectral type have been included. The upper limit flux density is taken to be 3$\sigma$ where $\sigma$ is the rms noise level measured in the primary-beam corrected image using a 2$"$ circular region centred on the optical position. The mass-loss rate calculation was performed as outlined in Sect. \ref{sect:mdotrates}.}
\label{tab:obupperlist}
\end{table*}


\clearpage

\newpage



\counterwithin{figure}{section}

\section{Figures}

\begin{figure*}[ht!]
\begin{center}
\includegraphics[width=0.95\textwidth]{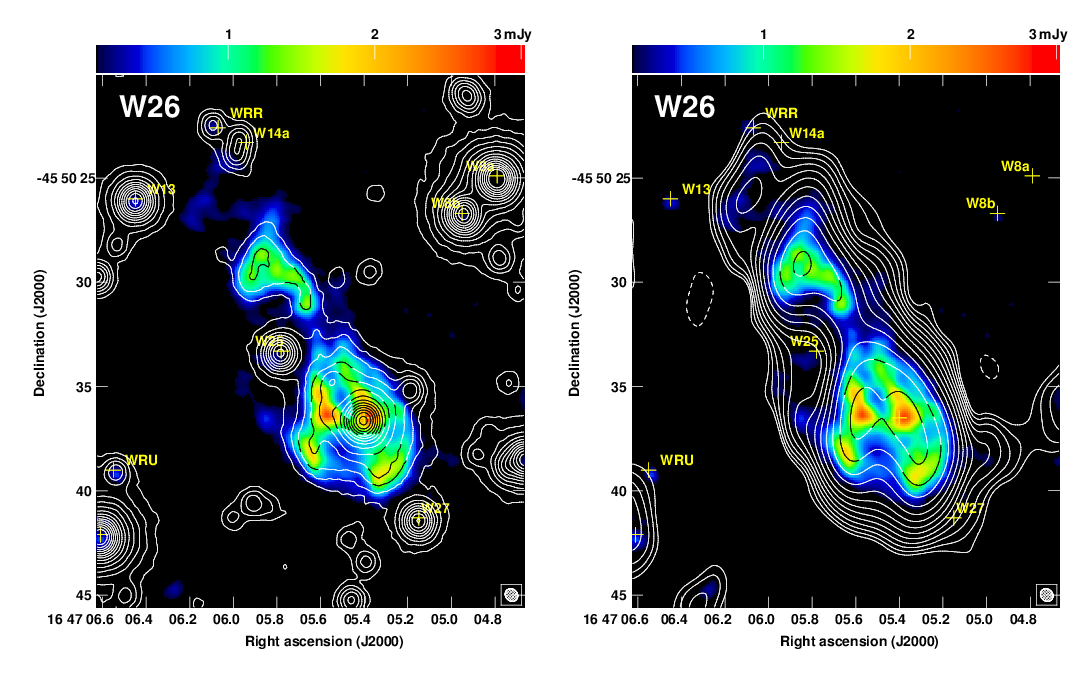}
\includegraphics[width=0.95\textwidth]{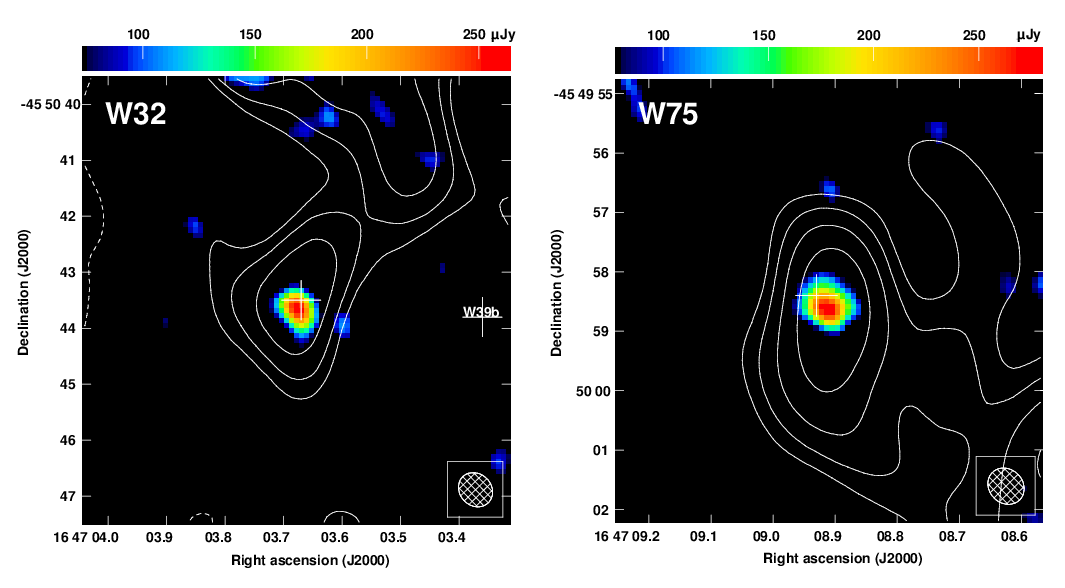}
\captionof{figure}{ALMA image of the red supergiant star Wd1-26 shown in colour-scale with H$\alpha$ (top-left) and radio (top-right) contours overlaid. The H$\alpha$ image is that presented in \cite{wright} from the VPHAS survey of the region and the contours are plotted at $\times$ 3$\sigma$ where $\sigma$ is 64 \mujybm. The radio contours are from recent ATCA observations of Westerlund 1 at 8.6\,GHz with contours plotted at -1,1,1.41,2,2.82,4,5.66,8,11.31,16,22.62,32,45.25,64,90.50\,$\times$\,0.304\,\mjybm \citep[see][for further details]{andrews18}. Also shown are the YHG Wd1-32 (bottom-left) and the RSG Wd1-75 (bottom-right) with 3-mm emission in colour-scale and radio contours plotted as for Wd1-26 at multiples of 0.112 \& 0.087 \mjybm.}
\label{fig:w26halpha}
\end{center}
\end{figure*}

\begin{figure*}
\includegraphics[width=18cm]{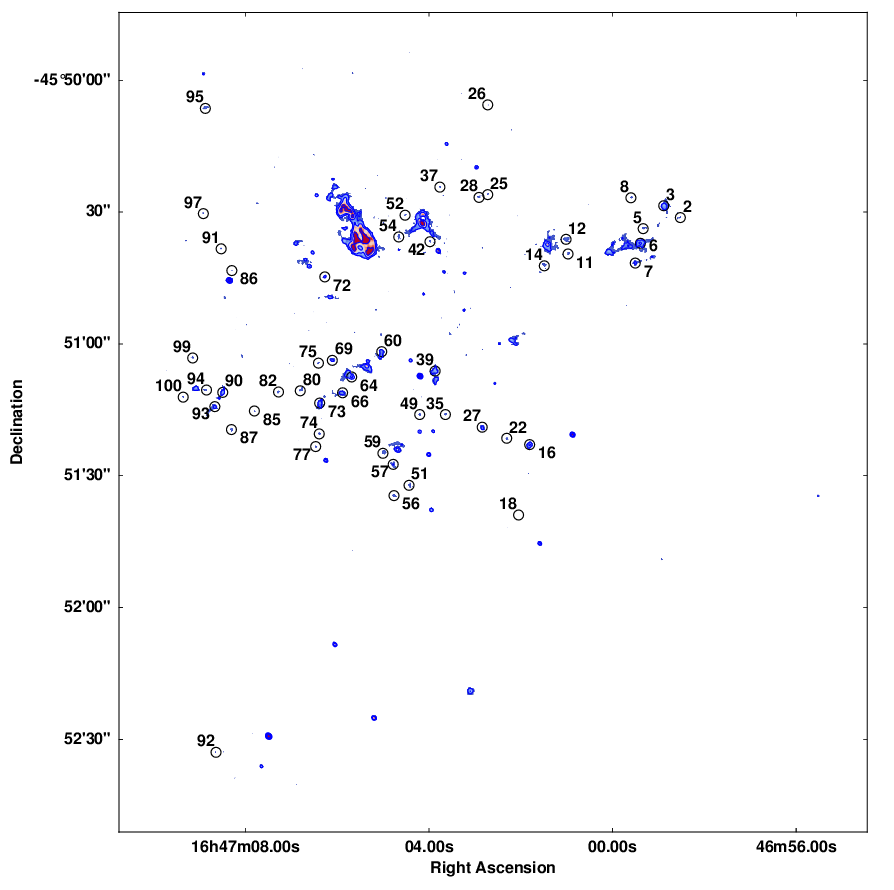}
\caption{ALMA 3-mm colour-scale map with overlaid contours. The colour-scale ranges from 0.10 to 1.2\,\mjybm and the contours are plotted at levels of 0.16,0.29,0.52,0.96,1.74,3.16\,\mjybm. The unknown sources identified in this study (see Table \ref{tab:fcp18_list} for details) are labelled at the peak positions observed in the ALMA data. }
\label{fig:unknownfinderimage}
\end{figure*}

\newpage

\begin{figure*}[ht!]
\begin{center}
\includegraphics[width=0.9\textwidth]{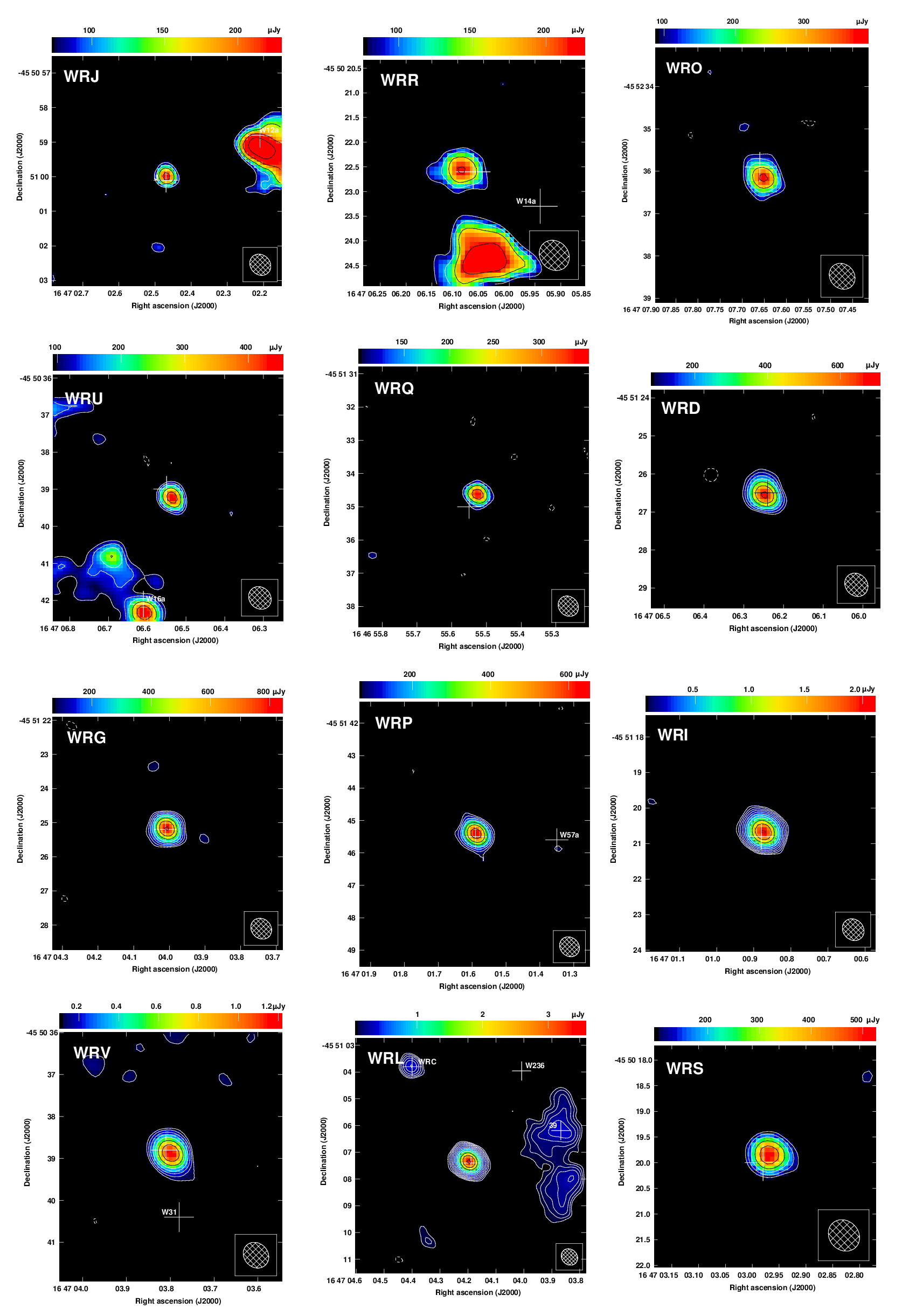}
\caption{ALMA images of the remaining detected stars (see Table \ref{tab:srcelist} for details). Contours are plotted at -1,1,1.41,2,2.82,4,5.66,8,11.31,16,22.62,32,45.25,64\,$\times$3$\sigma$. Unlabelled white crosses show the catalogue positions of each source. Other stars in the vicinity are also labelled as well as the unidentified ALMA FCP18 sources (see Table \ref{tab:fcp18_list} for further information).}
\label{fig:stargroups}
\end{center}
\end{figure*}

\begin{figure*}[ht!]
\ContinuedFloat
\begin{center}
\includegraphics[width=0.9\textwidth]{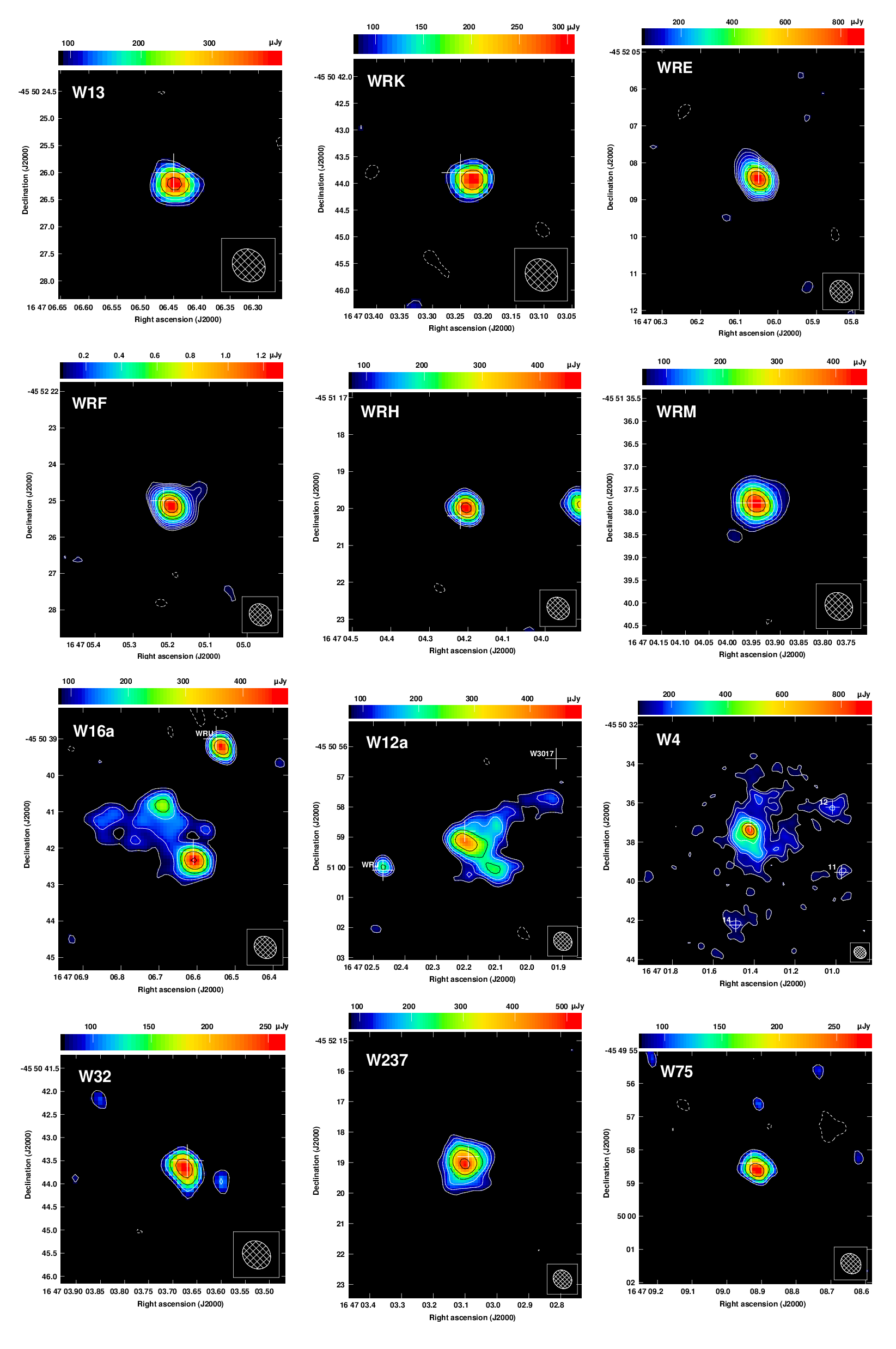}
\caption*{Fig. \ref{fig:stargroups} continued}
\end{center}
\end{figure*}

\begin{figure*}[ht!]
\ContinuedFloat
\begin{center}
\includegraphics[width=0.9\textwidth]{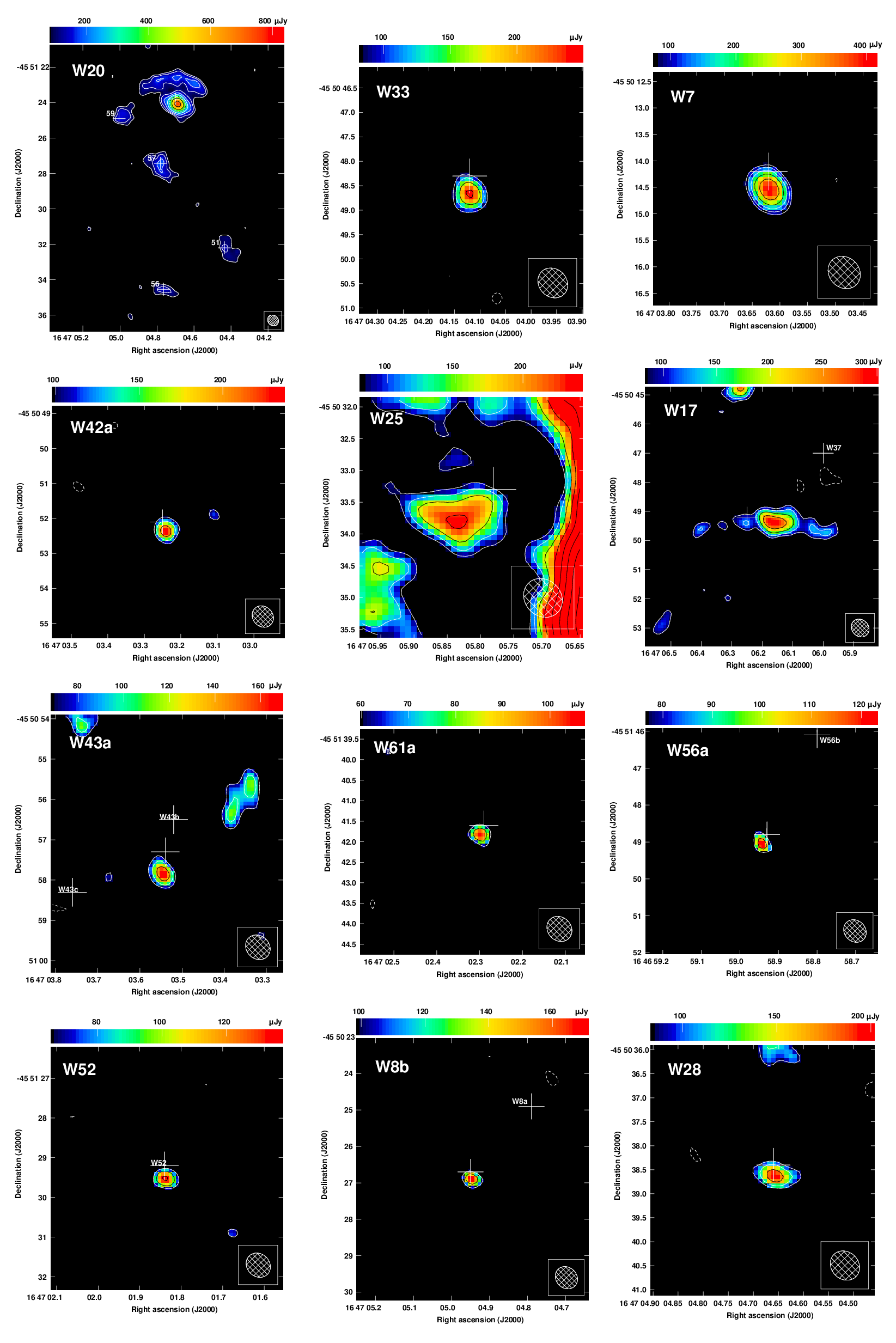}
\caption*{Fig. \ref{fig:stargroups} continued}
\end{center}
\end{figure*}

\begin{figure*}[ht!]
\ContinuedFloat
\begin{center}
\includegraphics[width=0.9\textwidth]{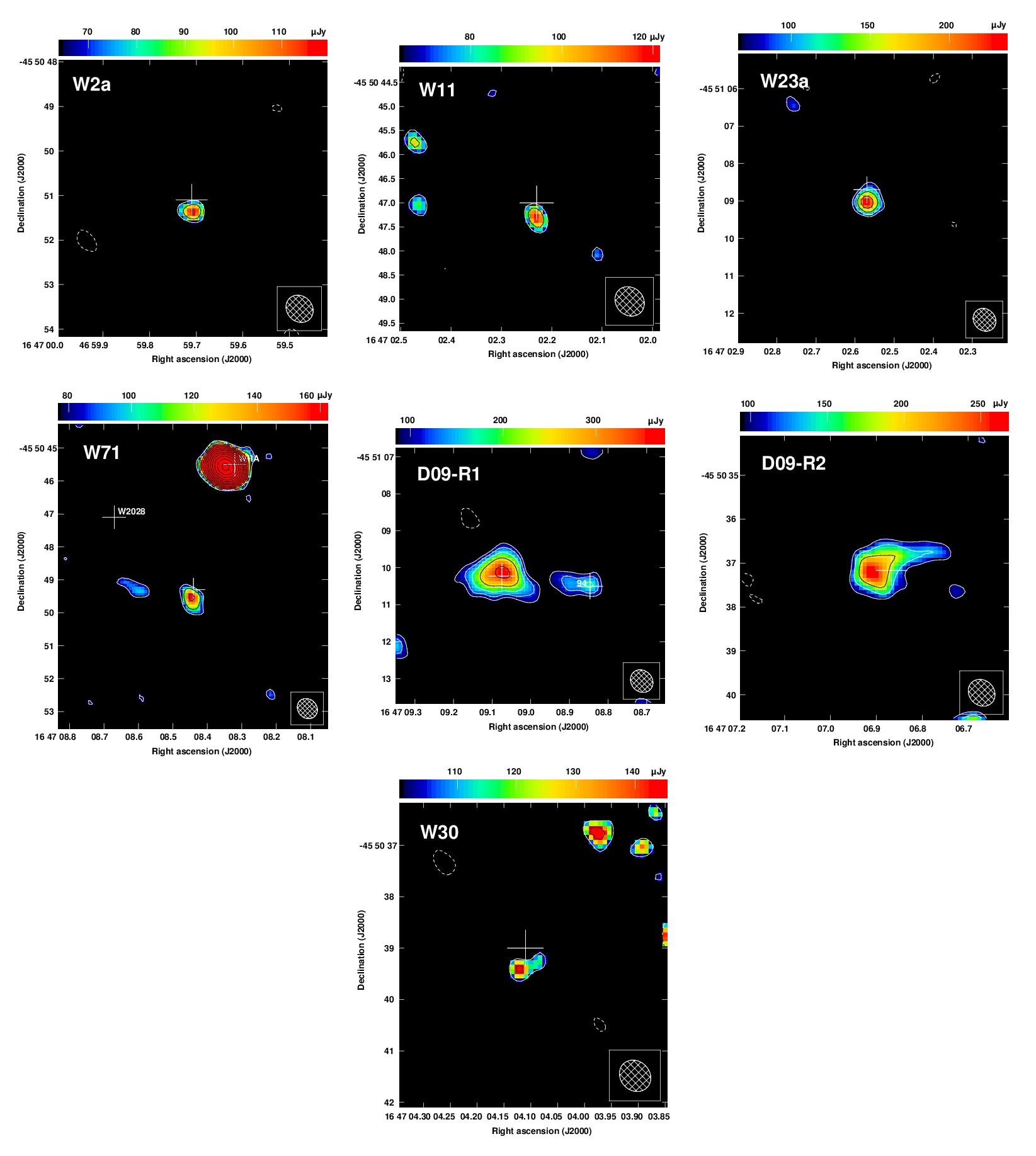}
\caption*{Fig. \ref{fig:stargroups} continued}
\end{center}
\end{figure*}


\clearpage

\newpage


\end{document}